\documentclass[a4paper,oldversion]{aa} 
\usepackage{times,mathptm}
\usepackage{txfonts}
\usepackage{natbib}
\usepackage{graphicx}
\setcitestyle{authoryear}
\bibpunct{(}{)}{;}{a}{}{,} 

\def\cgs{erg~cm$^{-2}$~s$^{-1}$}
\def\cm{cm$^{-2}$}
\def\ergs{erg~s$^{-1}$}

\def\nh{{N$_{\rm H}$}}
\def\neix{\hbox{Ne\ {\sc ix}}}
\def\xmm{{XMM-{\it Newton\/}}}
\def\asca{{\it ASCA\/}}
\def\chandra{{\it Chandra}}
\def\xray{\hbox{X-ray}}

\begin{document}

\title{The X-ray to [Ne V]3426 flux ratio:\\ discovering heavily
obscured AGN in the distant Universe.}

\author{
R.~Gilli\inst{1}, 
C.~Vignali\inst{2},
M.~Mignoli\inst{1},
K.~Iwasawa\inst{1,3},
A.~Comastri\inst{1},
G.~Zamorani\inst{1}
}

\authorrunning{R. Gilli et al.}  
\titlerunning{[Ne~V] selected obscured QSOs}

   \offprints{R. Gilli \\email:{\tt roberto.gilli@oabo.inaf.it}}

   \date{Received ... ; accepted ...}

\institute{ 
INAF -- Osservatorio Astronomico di Bologna, Via Ranzani 1, 40127
Bologna, Italy 
\and 
Dipartimento di Astronomia, Universit\`a degli Studi di Bologna, 
Via Ranzani 1, 40127 Bologna, Italy
\and
Institut de Ci\`encies del Cosmos, Universitat de Barcelona,
Mart\'i y Franqu\`es 1, 08028 Barcelona, Spain}

\abstract{We investigate the possibility of using the ratio between
the 2-10 keV flux and the [Ne~V]3426 emission line flux (X/NeV) as a
diagnostic diagram to discover heavily obscured, possibly Compton-Thick Active Galactic Nuclei (AGN) in the distant Universe. While
being on average about one order of magnitude fainter than the more commonly 
used [O~III]5007 emission line, the [Ne~V]3426 line can be observed with
optical spectroscopy up to $z\sim 1.5$, whereas the [O III]5007 line
is redshifted out of the optical bands already at $z\sim0.8$. First,
we calibrate a relation between X/NeV and the cold absorbing column
density $N_H$ using a sample of 74 bright, nearby Seyferts
with both X-ray and [Ne V] data available in the literature, and for
which the column density is determined unambiguously. Similarly to
what is found for the X-ray to [O III]5007 flux ratio (X/OIII), we
found that the X/NeV ratio decreases towards large column densities, as expected
if [Ne V]3426 emission is a good tracer of the AGN intrinsic
power. Essentially all local Seyferts with X/NeV values below 15 are
found to be Compton-Thick objects. At X/NeV values below 100, the
percentage of Compton-Thick nuclei decreases to $\sim 50\%$, but still
$\sim 80\%$ of the considered sample is absorbed with $N_H>10^{23}$
cm$^{-2}$. Second, we apply this diagnostic diagram to different
samples of distant obscured and unobscured QSOs in the Sloan Digital
Sky Survey (SDSS). SDSS blue, unobscured, type-1 QSOs in the redshift
range $z=[0.1-1.5]$ indeed show X/NeV values typical of unobscured
Seyfert 1s in the local Universe. Conversely, SDSS type-2 QSOs at
$z\sim 0.5$ classified either as Compton-Thick or Compton-Thin on the
basis of their X/OIII ratio, would have been mostly classified in the same
way based on the X/NeV ratio. We apply the X/NeV diagnostic
diagram to 9 SDSS obscured QSOs in the redshift range $z=[0.85-1.31]$,
selected by means of their prominent [Ne~V]3426 line (rest $EW>4$\AA) and
observed with \chandra\ ACIS-S for 10ks each (8 of them as part of our
proprietary program). Based on the X/NeV ratio, complemented by X-ray
spectral analysis, 2 objects appear good Compton-Thick QSO
candidates, 4 objects appear as Compton-Thin QSOs, while 3 have an ambiguous classification. 
When excluding from the sample broad lined QSOs with a red continuum 
and thus considering only genuine narrow-line objects, the efficiency in selecting Compton-Thick QSOs 
through the [Ne~V] line is about 50\% (with large errors, though), more similar 
to what is achieved with [O~III] selection. We discuss the possibility of applying 
the X/NeV diagnostic to deep X-ray surveys to search for Compton-Thick Seyferts 
at $z\sim 1$, i.e. those objects which are thought to be responsible for
the ``missing" X-ray background.

Finally, we compare the optical spectral properties of [Ne~V]-selected QSOs with those of
other SDSS populations of obscured and unobscured QSOs. By restricting the analysis to objects in the same 
redshift (and luminosity) range z=[0.4-1.5], we found evidence that, at any given [Ne~V] luminosity, 
increasing obscuration is accompanied by increasing [O~II]3727 emission. This correlation
is interpreted as evidence for enhanced star formation
in obscured QSOs, which is consistent with current popular scenarios of BH-galaxy coevolution.


}

   \maketitle

\section{Introduction} \label{introduction}

While the cosmological evolution of unobscured QSOs has been traced up
to $z\sim 6$, the evolution of obscured AGN is
much more uncertain and is the subject of intense debate. The
{\it observed} number statistics in current AGN samples is dominated
by unobscured objects which are easier to discover (e.g. $>10^5$
QSOs have been identified in the Sloan Digital Sky Survey), but several
arguments suggest that obscured AGN must be intrinsically more
numerous. Deep X-ray surveys (see \citealt{bh05} for a
review) have indeed shown that, towards faint X-ray fluxes, the surface density of obscured AGN overtakes that of 
unobscured AGN. Also, population synthesis models of the cosmic
X-ray background (XRB), suggest that obscured AGN outnumber
unobscured ones by a factor which ranges from $\sim 2$ to $\sim 8$,
depending on the considered luminosity regime (\citealt{gch07}; see \citealt{tuv09} and \citealt{balla06} for a steeper luminosity
dependence). To understand the cosmological history of accretion onto supermassive black holes
(SMBHs) it is therefore necessary to map and understand the population
of obscured AGN and, in particular, of the most obscured and hence
elusive ones, the so-called Compton-Thick (CT) nuclei, i.e. those
obscured by column densities above $\sim 10^{24}$ cm$^{-2}$. The population of moderately obscured AGN in fact 
does not completely account for the XRB peak intensity at 30 keV, to which CT AGN are expected to
contribute from $\sim 10\%$ \citep{tuv09} to $\sim 25-30\%$
\citep{gch07}, depending on the XRB model assumptions. In addition,
the presence of a large population of CT AGN across the cosmic epochs,
would help in reconciling the measured mass function of local
SMBHs with that expected by integrating the
accretion history of seed black holes \citep{marconi04, shankar04}. 
Finally, popular semi-analytic models of galaxy
formation and evolution \citep{kh00, marulli08}, coupled to hydrodynamical simulations of galaxy mergers
\citep{hop06}, propose that nuclear activity is triggered
during major mergers of gas-rich galaxies and that at its early stage,
the AGN is embedded within optically thick gas shrouds. Despite these
theoretical progresses, the cosmological evolution and luminosity
function of CT objects is currently unknown, and in the synthesis models of
the XRB it has been usually assumed to be equal to that of less obscured objects. 
Whether this is the case or not can only be determined by
obtaining statistically significant samples of distant CT objects and by
comparing them with the local samples (see e.g. \citealt{c04} and \citealt{rdc08} for
reviews on nearby, {\it bona-fide} CT objects).

The number of techniques devised to select CT AGN is rapidly growing,
following the technological development of efficient detectors across
the electromagnetic spectrum. These diverse selection techniques have
allowed the first estimates of the space density of CT AGN in
different redshift and luminosity regimes.

Very hard X-ray selection, i.e. at energies above 10 keV, is
unaffected by absorption up to a few $\times 10^{24}$
cm$^{-2}$, but, because of the still limited instrumental sensitivity,
is mainly sampling the nearby Universe. The population of 
CT AGN detected by INTEGRAL/IBIS \citep{tueller08} and Swift/BAT \citep{malizia09}
at $z<0.02$ is indeed producing only a tiny fraction ($<$1\%) of the cosmic XRB.

Deep X-ray surveys in the more accessible 2-10 keV band are revealing
large populations of heavily obscured objects in the redshift range
$z\sim[0.5-4]$. The generally low photon statistics of the
detected sources, however, prevent an accurate spectral analysis and $N_H$
measurement. The CT nature of a faint X-ray source is therefore often
inferred from the characteristics of the reprocessed spectrum, like the
presence of a prominent fluorescence $K\alpha$ iron line over a flat
continuum. Examples of distant CT AGN selected with this technique
have been found by \citet{tozzi06} and \citet{geo07,geo09} in the \chandra\ Deep Fields. The number of CT AGN candidates
detected in the deep X-ray surveys appears to be in rather good
agreement with the XRB models predictions.

A very recent and promising approach to select CT candidates in the
distant Universe is based on their strong mid-IR flux, where most of
the absorbed radiation should be re-emitted \citep{alejo05}. Recently, \citet{daddi07}, \citet{fiore08, fiore09} and
\citet{alex08} located heavily obscured AGN in objects showing 24$\mu$m emission
in excess of that expected from dust heated by stellar processes. By
stacking the \chandra\ data of these mid-IR-excess objects, a very hard
X-ray spectrum was observed, reminiscent of CT obscuration. These
studies span a broad AGN luminosity range ($L_X\sim10^{42-45}$ erg
s$^{-1}$), but mostly sample populations of objects at $z\sim 2$. The measured space
density of CT AGN at these high redshifts is in general as large as expected from
XRB synthesis models or possibly even larger (see eg. \citealt {trei09_ct}). 

Another way to select obscured QSOs is through their high-ionization
narrow optical emission lines, which are thought to be produced on physical scales (from $\sim0.1$ to a few kpc) mostly free from nuclear
obscuration. Recently, the [O IV]26$\mu m$ line has been used 
to select obscured AGN among galaxies observed with {\it Spitzer}/IRS. However, since this line quickly
moves out of the observable IR bands as redshift increases, this selection mostly concerns the nearby
Universe \citep{diamond09, rigby09}. The most commonly used marker of obscured nuclear activity
therefore remains the [O~III]5007 emission line, which
is strong, falls in the optical domain, and allows object selection up to $z\sim 0.8$.

The 2-10 keV to [O~III]5007 flux ratio (X/OIII) has been often used as
a diagnostic for heavy obscuration in sources with poor X-ray photon
statistics \citep{maio98, cappi06, panessa06}, being low X/OIII ratios
($\lesssim 3$, see e.g. Fig.~4 of \citealt{cappi06}) highly
suggestive of heavy nuclear absorption.  Based on the [O~III]5007
emission line, \citet{zak03} and \citet{reyes08} identified in the Sloan Digital
Sky Survey (SDSS) a population of obscured QSOs at a median redshift
of $z\sim0.3$, at least as abundant as that of type-1 QSOs at the same
redshifts \citep{reyes08}. These results have been extended to lower luminosities
by \citet{bongiorno09}, who measured the luminosity function of [O III]-selected type-2 AGN
in the zCOSMOS spectroscopic survey \citep{lilly07}, finding that the fraction of obscured AGN 
is decreasing with luminosity, in agreement with what is observed in X-ray surveys.

X-ray observations of small samples drawn from the \citet{zak03} catalog, 
suggest that about half of luminous type-2 QSOs (log$L_{O III}>9.3\;L_{\odot}$)
could be CT \citep[hereafter V10; see also \citealt{lamastra09} for
X-ray observations of lower luminosity SDSS type-2 AGN]{ptak06, v06, v10}. Since
[O III] selection is likely missing objects in which also the Narrow Line
Region (NLR) is extincted (like e.g. in the prototype CT AGN NGC~4945
and NGC~6240), the estimated type-2 QSO abundances should be considered as lower limits \citep[V10]{reyes08}.

In this work we explore the possibility of using the high-ionization
[Ne~V]3426 emission line, rather than the [O~III]5007 line, as a tracer
of obscured nuclear activity. Despite being on average a factor of
$\sim 9$ weaker than [O~III]5007 \citep{fo86, zak03} and suffering stronger dust extinction,
the [Ne~V]3426 line is commonly observed in nearby
Seyfert galaxies and, given that high energy photons ($\gtrsim 0.1$
keV) are required to further ionize NeIV, it is considered an
unambiguous sign of nuclear activity \citep[e.g.][]{schmidt98}. In addition, the [Ne~V]3426 emission line is observable up
to $z\sim 1.5$ before being redshifted out of the optical bands,
whereas the [O~III]5007 line is observable only up to $z\sim
0.7-0.8$. Indeed, only 13 out of 887 [O~III]-selected QSOs in the \citet{reyes08} sample
lie at $z>0.7$, with only 3 at $z>0.8$. 
[Ne~V]-selection may then be used to reveal nuclear activity in obscured
sources at $z\sim 1$, i.e. at the epoch where most of the XRB light is thought to be produced.

The structure of the paper is the following: in Section 2 we present and discuss
the sample of nearby Seyfert galaxies used to calibrate the relation between
[Ne~V] and X-ray emission (the details of the sample are given in the Appendix). 
In Section 3 we present the X/NeV diagnostic diagram
and apply it to obscured and unobscured QSO population drawn from the SDSS.
In Section 4 we present \chandra\ observations of a sample of 9 [Ne~V]-selected obscured QSOs at $z\sim 1$
in the SDSS and use the X/NeV diagnostic to estimate the fraction of CT objects among them. 
In Section 5 we discuss efficiency and biases of [Ne~V] selection together with its application to 
sky areas with deep optical spectroscopy and X-ray coverage. In the same Section, the evidence
of enhanced star formation in obscured QSOs at z=0.4-1.5 is also highlighted. The conclusions
are drawn in Section 6. 

\begin{figure*}[t]
\begin{center}
\includegraphics[width=12cm]{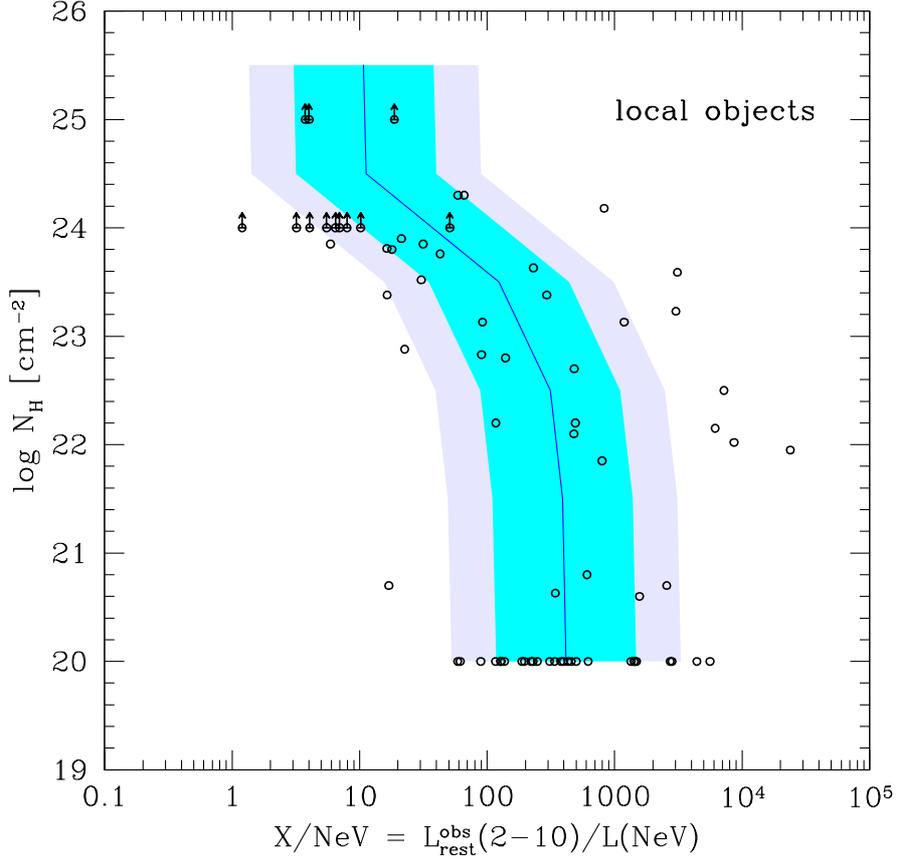}
\caption{Observed 2-10 keV to [Ne~V]3426 luminosity ratio (X/NeV) vs
absorption column density for a sample of 74 Seyfert galaxies
in the local Universe. The solid line shows the expected trend obtained by
starting from the mean X/NeV ratio $\langle$X/NeV$\rangle$ observed
in unobscured objects (i.e. those plotted at log$N_H=20$) and
progressively obscuring the X-ray emission with increasing $N_H$ (up
to log$N_H=25.5$) as plotted on the y-axis. Lower limits at log$N_H=24$ refer to Compton-Thick
objects observed only below 10 keV, while datapoints at log$N_H>24$ and lower limits at log$N_H=25$ refer to CT
objects observed also above 10 keV, for which a more stringent determination of the column density is possible.
The darker (lighter) shaded region is obtained by
making the same computation as above but starting at $\pm1\sigma$ ($\pm
90\%$) around $ \langle$X/NeV$\rangle$ (see text for details).
The region at low ($<15$) X/NeV values is essentially populated only by CT objects.}
\label{xnev_local}
\end{center}
\end{figure*}

\section{The local Seyfert sample}

We searched in the literature for nearby ($z<0.1$) AGN for which both [Ne~V] and
X-ray data are available. A total sample of 74 objects
were found with measured [Ne~V] flux, 2-10 keV flux and X-ray column
density $N_H$. The most difficult information to obtain was that on
[Ne~V] flux, since optical spectra are often limited at wavelengths
$>3700\AA$. On the contrary, the $N_H$ and the 2-10 keV flux values for
bright nearby objects are more easily obtained either from published
papers or from archival X-ray data. The main catalogs of
optical/near-UV Seyfert spectra providing the [Ne~V] fluxes used in
this work are those published by \citet{malkan86}, \citet{mw88}, \citet{storchi95}, \citet{erkens97}, as well as the
compilation by \citet{schmitt98}. A public catalog of HST/STIS spectra of
nearby AGN has been released by \citet{spinelli06}. We
note that the HST/STIS spectra have been extracted in apertures of
$0.2''\times0.2''$, thus sampling physical scales of a few tens of
parsecs at the typical source redshift. This physical scale is often
too small to fully encompass the NLR, which indeed may extend up to a
few kpc. We verified that the [O~III] and [Ne~V] fluxes as measured on
the \citet{spinelli06} spectra are often significantly lower (an
order of magnitude or more) than those measured on larger apertures,
especially in those objects in which an extended NLR has been revealed
\citep{bianchi06}. Therefore, we avoided using
measurements obtained with HST/STIS and considered only those taken
with larger apertures. All the [Ne~V] fluxes used in this work come
from apertures $\gtrsim1.5''$. Whenever more than one measurement is
available for the [Ne~V] flux, we preferred the one obtained with the
largest aperture. For 6 objects we measured the [Ne~V] line flux directly on the calibrated spectrum 
drawn from the catalog of 99 UV-optical spectra of nearby galaxies released
by Storchi-Bergmann et al.\footnote{http://www.if.ufrgs.br/$\sim$thaisa/}.

Since our main aim is to calibrate a diagnostic which can be applied to
distant objects, we did not attempt to correct the [Ne~V] fluxes for the reddening which
may be intrinsic to the NLR. Indeed, while for local objects the
extinction to the NLR can be easily measured through e.g. the ratio
between the narrow components of $H\alpha$ and $H\beta$, in distant
AGN either $H\alpha$ or both $H\alpha$ and $H\beta$ are not observable
and measuring the ratio between the Balmer lines of higher order (e.g.
$H\beta/H\gamma$, $H\gamma/H\delta$) is often unfeasible due to their
weakness over noisy spectra.

As for X-ray data, we generally preferred to use the values obtained
with the most recent and sensitive satellites \chandra\ and \xmm. When
dealing with heavily obscured objects, though, we considered data obtained with {\it Suzaku} and {\it BeppoSAX} in order to
map the energy range above 10 keV and get either a measurement or a tighter lower limit
on the absorbing column (see Fig.~\ref{xnev_local}). Sometimes an observed 2-10 keV flux
measurement is not directly quoted in the considered literature
papers: in those cases we estimated it using the published best-fit
spectral parameters and/or 2-10 keV luminosity. For a few objects we
analyzed archival X-ray data which were still unpublished. The full
sample, along with [Ne~V] fluxes, 2-10 keV fluxes, $N_H$ values and
relative references is presented in the Appendix.

\subsection{Sample biases}

Although an effort has been done to find as many sources as possible,
with both [Ne~V] and X-ray information, the sample we built is by no
means complete and cannot be considered suitable for statistical
studies. Objects in the sample have been targeted for observations for several
different reasons. Moreover, since we require [Ne~V] detection, the
sample might be biased towards objects with stronger [Ne~V]
emission. Indeed, the absence of a [Ne~V] measurement may well be due
to the lack of spectral coverage at 3426\AA, but also to excessively faint
[Ne~V] emission. Given that sometimes the information on the spectral coverage is
missing and that not all optical spectra on which line
measurements are derived have been published, it is often
difficult to judge what is the reason for the lack of [Ne~V] data. On
the contrary, whenever a source is observed in the X-rays, the fine
sensivity of current X-observatories and the brightness of the
considered nearby objects generally allow a measurement of both 2-10
keV flux and column density. Therefore, the sample should not be
biased towards X-ray bright objects. \footnote{We note that we found only very few local objects with [Ne V] emission and observed in the X-rays 
which show poor X-ray photon statistics, despite having similar X-ray fluxes to those analyzed in the previous Section.
These objects were observed with very short exposures (e.g. by the XMM-Slew survey) and have not been 
considered here since it is not possible to have reliable information on their column density and hence place
them in Fig.~\ref{xnev_local}.} 
Also, those objects in which the
NLR is relatively free from obscuration should preferentially appear
in the sample. Indeed, even a modest reddening of E(B-V)=0.5
corresponds to a flux dumping by a factor of $\sim 10$ at
3426$\AA$ assuming standard extinction curves. Compton-Thick objects like NGC~4945 and NGC~6240, in which the
NLR is affected by strong extinction, do not indeed appear in the
sample. Because of the issues discussed above, the
present sample is the one to be used when building a diagnostic
diagram to be applied to samples of distant sources selected on the basis of
their [Ne~V] emission, which indeed would suffer from similar biases.

\section{The X/NeV diagnostic}\label{diag}

We computed the ratio between the flux observed in the 2-10 keV band
and that in the [Ne~V]3426 emission line (X/NeV) for the local Seyfert
sample described in the previous Section and plotted the X/NeV value
against the measured X-ray absorption (Fig.\ref{xnev_local}). This is
essentially the same diagram worked out by \citet[][see also \citealt{cappi06} and \citealt{panessa06}]{maio98}
obtained by replacing the
[O~III] with the [Ne~V] line. Working with local objects allows to deal
with good optical and X-ray spectra and in turn provides a precise
measurement of the X-ray absorption. Assuming that the [Ne~V]3426 flux
is mostly produced in the NLR, far from nuclear obscuration which
affects the X-ray emission, one would expect that the X/NeV ratio on
average decreases with increasing column density. This is indeed what
is observed (see Fig.\ref{xnev_local}): the median X/NeV values for
unabsorbed and for CT Seyferts are about 400 and 7, respectively. For
X/NeV$<15$ almost all objects are CT. For X/NeV$<100$ about 50\%
of the objects are CT, while 80\% are still obscured by columns larger
than $10^{23}$ cm$^{-2}$. The CT outlier with X/NeV$\sim$800, NGC~3281,
likely suffers from extinction in the NLR. NGC~6240 and NGC~4945 do not
even appear in the diagram. Therefore, low X/NeV ratios would select 
clean samples of CT AGN, i.e. not significantly contaminated by low-obscuration 
sources, but not complete.

We note that, for unabsorbed Seyferts, the
mean and median X/NeV values are almost identical and very
similar to what is obtained by scaling down the average X/OIII value
obtained by \citet{panessa06} for Seyfert 1s by a factor of 9,
i.e. the average OIII/NeV emission line ratio measured for local
Seyferts \citep[e.g.][]{so81}. The solid line in
Fig.~\ref{xnev_local} has been obtained starting from the mean X/NeV
ratio observed in unobscured Seyfert 1s (i.e. those with log$N_H$
fixed to 20 in Fig.~\ref{xnev_local}) and progressively absorbing the
2-10 keV flux of unobscured AGN using the spectral templates
of \citet{gch07}. In particular, the 2-10 keV emission in
CT AGN with log$N_H>24.5$, mostly dominated by the Compton-reflected
continuum, has been assumed to be about $\sim2\%$ of the intrinsic one in the same band. The
darker (lighter) shaded region has been obtained using the
1$\sigma$ (90\%) limits of the X/NeV distribution of unabsorbed
Seyferts. Most objects lie within the shaded regions, in agreement with
the expectations that nuclear absorption affects the X-ray emission
only. However a few notable exceptions are present.
For instance, the object with the highest X/NeV ratio in
Fig.~\ref{xnev_local} is NGC~2992, a Seyfert 1.9 galaxy in which the
X-ray flux has been observed to vary by a factor of 20 over the years,
likely as a consequence of switching on and off of its nuclear
activity \citep{gilli00} and/or flares in the inner accretion disk
\citep{murphy07}. Using the average X-ray flux, rather than the
high-state X-ray flux as done in Fig.~\ref{xnev_local}, would
shift NGC~2992 in the shaded area where most of the sources lie. This
highlights the problem of dealing with variable sources and
non-simultaneous optical and X-ray observations, which 
increases the dispersion of the X/NeV distribution. Absorption
variations are also important. Indeed, the lightly obscured object
(log$N_H<21$) with the lowest X/NeV ratio (X/NeV$\sim20$) is the dwarf AGN
POX~52 \citep{barth04}, which shows rapid X-ray flux and
absorption variability \citep{thornton08}. Based on the observed
variability, the position of POX~52 on the diagram can shift from the
plotted position to X/NeV$\sim10$ and log$N_H=22.8$, in the region
populated by obscured objects. Finally, it is worth noting that in
radio-loud objects an additional X-ray component coming from the jet
emission, produces a shift towards higher X/NeV ratios than
radio-quiet objects: indeed $\sim 90\%$ of the radio galaxies in
our samples show X/NeV ratios larger than 400, including some of the objects 
falling on the right of the shaded region in Fig.~\ref{xnev_local}, while the same
fraction is $\sim 25\%$ for the radio-quiet galaxies.

\begin{figure*}[t]
\begin{center}
\includegraphics[width=12cm]{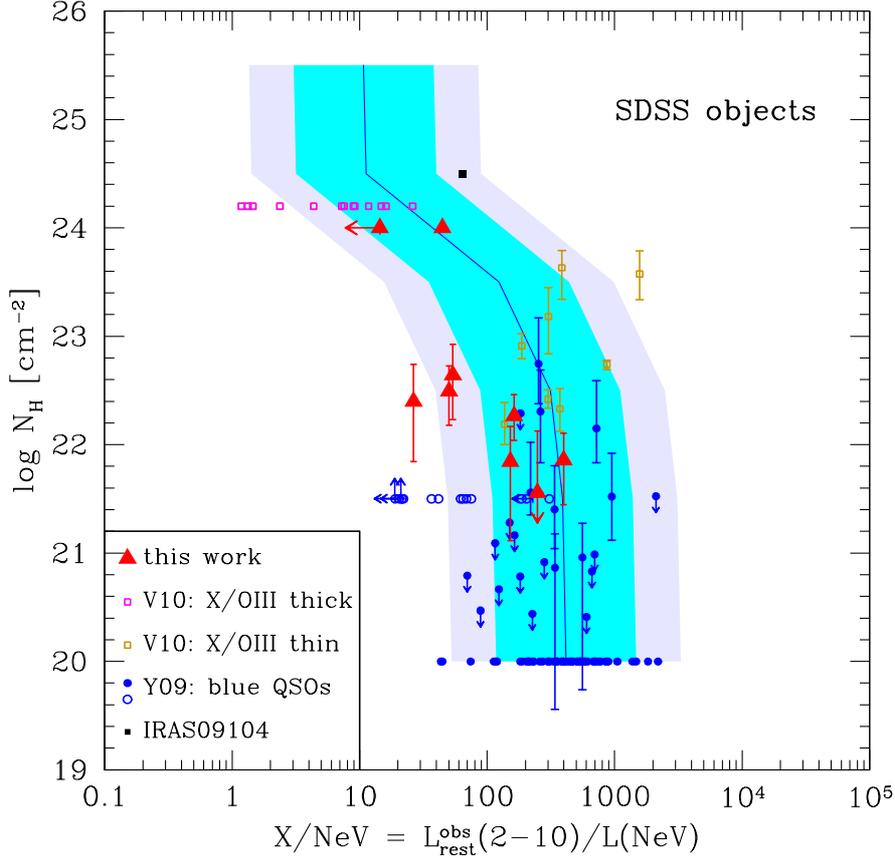}
\caption{The X/NeV vs $N_H$ diagram (see Fig.~\ref{xnev_local})
applied to different samples of SDSS QSOs. The blue SDSS QSOs in the
\citet[][Y09]{y09} catalog populate the same region of
unobscured Seyfert 1s in the local Universe: filled circles are high-significance X-ray
detections for which the $N_H$ values have been measured from the spectral fits by Y09; open circles
are low-significance or non X-ray detections in Y09, which have been plotted at log$N_H=21.5$ (see Section 3.1 for details).
The obscured SDSS QSOs in
the V10 sample are plotted as open squares
(Compton-Thick candidates are placed at log$N_H=24.2$). Objects which
have been classified as Compton-Thick (-Thin) by \citet[][V10]{v10} on the basis of
the their low (high) X/OIII ratio, show correspondingly low (high) X/NeV
ratios, and hence would be also classified as Compton-Thick (-Thin) based on
this new diagnostic. The [Ne~V]-selected obscured QSOs investigated in
this work are shown as filled triangles. The two objects classified in
Table~\ref{xspec} as likely CT QSO candidates based on their X/NeV
ratio and/or X-ray spectral properties, are plotted at log$N_H=24$ (see
also Section~\ref{cxo}). For comparison, the position of the
SDSS-observed, well-known CT QSO IRAS09104 is also shown with a filled
square.}
\label{xnevsdss}
\end{center}
\end{figure*}

\subsection{SDSS type-1 QSOs}

We applied the X/NeV diagnostic ratio derived in the local Universe to
distant obscured and unobscured QSOs in the SDSS. It is important to
note that what is needed here is the {\it 2-10 keV rest frame flux (or
luminosity) prior to absorption correction}, to make a meaningful
comparison with the local sample. We first considered the catalog of
\citet[][hereafter Y09]{y09}, who collected a sample
of 792 QSOs in the SDSS DR5 selected primarily by their blue optical
colors and serendipitously observed by \xmm \footnote{We considered the catalog tables as
updated in Young et al. 2009, ApJS, 185, 250}. By matching the Y09 sample
with the SDSS spectroscopic tables we found 94 objects in the redshift
range $z=[0.12-1.50]$ for which i) the [Ne~V]3426 emission line is
detected at $>3\sigma$ and ii) have been detected in the X-rays at
$>6\sigma$, which is the threshold adopted by Y09 to perform X-ray
spectral fits and measure both the observed 2-10 keV rest frame flux
and the column density $N_H$. To check for possible problems in the
determination of the [Ne~V]3426 flux related to the SDSS automatic
procedures, we visually inspected the 94 SDSS spectra above and found
11 objects in which the listed detection of [Ne~V] emission is seriously affected by instrumental 
features and sky residuals. These objects were then
removed from the sample. We also noticed that the line parameters
listed in the SDSS spectroscopic tables are not accurate for weak
emission lines over a steep continuum. For instance, in about one
quarter of SDSS QSOs, the [O~II]3727 emission line (see the Discussion)
appears as an absorption line due to an overestimate of the
continuum. We then retrieved the SDSS spectra and performed a Gaussian
fit to the emission lines we were interested in, deriving fluxes,
luminosities, and rest frame equivalent widths. All the emission line
parameters used in this work for SDSS objects are derived from our
direct fits. Whenever we compared our line measurements with those
performed by other Authors on the same SDSS spectra (e.g. for a
subsample of type-2 QSOs in \citealt{zak03}), we found an
excellent agreement.

The X/NeV vs $N_H$ relation for the 83 SDSS QSOs selected above is
shown in Fig.~\ref{xnevsdss}. Most objects are unobscured and
populate the same region of local Seyfert 1 galaxies. The median
X/NeV ratio for SDSS QSOs is 370, almost identical to the value of 400
found for local Seyfert 1s. The few objects showing
significant intrinsic X-ray absorption appear in the optical as
intermediate type Seyferts (i.e. 1.8-1.9) rather than pure type-1
AGN.\footnote{One of them is indeed also included in the type-2 QSOs
catalog by \citet{zak03} and in the \citet{v06,v10} samples.}

To check for possible outliers, e.g. blue SDSS QSOs with very low X/NeV ratios
($<15$, as observed for local CT Seyferts), we also considered those objects in the
Y09 catalog showing significant [Ne~V] emission and which were either just detected
(13 objects) or completely undetected (3 objects) in the X-rays.  We verified that only
two objects - appearing in the optical as classical, blue, broad line objects - based on the
[Ne~V] flux we measure and on the 2-10 keV rest frame flux in the Y09 catalog, would
nominally show X/NeV$<15$. We double checked the X-ray fluxes of these sources based on
literature data (i.e. a \chandra\ 50ks exposure available for one of them, providing a good quality spectrum 
with 200 photons,  and the 2XMM catalog results for the other source) and found that, while the soft X-ray fluxes 
are in agreement with the Y09 values, the 2-10 keV fluxes appear significantly (i.e. a factor of 5-10) higher than
those in Y09. Both objects are hard sources and, in particular, the one falling in the \chandra\ ACIS-I field
appears to be absorbed by $N_H >10^{22}$cm$^{-2}$. We speculate that the likely reason for the
mismatch in the 2-10 keV fluxes is that in Y09 the X-ray fluxes of objects with only a few photons are obtained by 
using a photon index fixed to 1.9, which, for hard sources may severely underestimate the
X-ray flux above 2 keV.  Using the correct 2-10 keV fluxes, the two objects have X/NeV values between 40 and 60. 
The X/NeV ratios of the 13 low-significance \xmm\ detections and X/NeV upper limits to the 3 \xmm\ 
non-detections are shown in Fig.~\ref{xnevsdss} as open circles placed at log$N_H$=21.5, since for the majority
of these objects there is evidence of absorption in their optical SDSS spectra. Their X/NeV values are distributed
from $\sim 20$ to $\sim 300$, i.e. they are shifted towards smaller values than those of local Seyfert 1s and 
of high-significance \xmm\ detections in Y09. In principle, if broad line QSOs reach X/NeV ratios as low as
$\sim 20$, then one might suspect that more obscured sources, e.g. those absorbed by columns around $\sim10^{23}$\cm\, could easily contaminate the
CT regime (X/NeV$<15$) defined in the previous Section. We note however that, as discussed above, the X/NeV value could have been underestimated
for many of the low-significance \xmm\ detections in Y09. Furthermore, the two \xmm\ non-detections with X/NeV$\sim20$ appear in the optical as pure
narrow-line Seyferts and are in fact also listed in the \citet{reyes08} type-2 QSO catalog. Likely, both objects are significantly absorbed, in principle even CT,
and we show them as lower limits on $N_H$ in Fig.~\ref{xnevsdss}. Based on these checks we conclude that the CT region defined in the previous Section is 
not contaminated by SDSS blue, unabsorbed QSOs although some mild contamination by moderately absorbed sources might still be possible.

\subsection{SDSS type-2 QSOs at $z\sim 0.5$}

We then considered the objects presented by V10 (see also \citealt{v06}), who
have combined proprietary and archival X-ray observations of a sample
of 25 objects out of 291 [O~III]5007-selected type-2
QSOs \citep{zak03}. The objects in the V10 sample populate
the redshift interval $z\sim[0.3-0.7]$ and have [O~III] luminosities in
excess of $1.9\times10^9\;L_{\odot}$. Based on the observed X/OIII
ratio, V10 divided their sample into 8 Compton-Thin QSOs and 17 CT
QSOs candidates. We considered here the 21 objects in which the [Ne~V]
line is detected at $>3\sigma$ level. Visual inspection of these 21
objects does not show any fake [Ne~V] emission. We placed the V10
objects in the X/NeV vs $N_H$ diagnostic diagram (see
Fig.~\ref{xnevsdss}): an almost perfect correspondence is found between the
X/OIII and X/NeV classifications: all the 8 objects classified as
Compton-Thin by V10 are also classified as Compton-Thin based on their
X/NeV ratio. Conversely, 11 out of 13 objects which appear as CT candidates in
V10, are also classified as CT using the X/NeV$<15$ threshold. This may not be
surprising given that both [O~III] and [Ne~V] lines should trace the
intrinsic AGN power. However, this is a proof that the X/NeV ratio can
be used as a diagnostic for revealing heavy obscuration. In
Fig.~\ref{xnevsdss} we also show the position of the well known
distant ($z=0.44$) CT QSO IRAS09104+4109 \citep{frances00, iwa01iras, pico07iras}, 
whose SDSS optical spectrum
reveals a clear [Ne~V]3426 emission line. IRAS09104 falls in a region
of the X/NeV diagram which, based on the local Seyfert sample
presented in Section \ref{diag}, is equally populated by CT and
Compton-Thin objects.\footnote{The CT nature of IRAS09104 has been questioned by
\citet{pico07iras}, who nonetheless measured heavy absorption (log$N_H\sim5\times 10^{23}$\cm) in this source.}

It is worth noting that the source distribution of SDSS objects in Fig.~\ref{xnevsdss}
appears to be less dispersed that that of local Seyferts in Fig.~\ref{xnev_local}. This may
be related to the fact that i) SDSS objects are intrinsically more luminous, hence less variable,
than local Seyferts and ii) they are located at higher redshifts, which allows the SDSS 3'' fibers 
to encompass the whole [Ne~V]-emitting region, thus reducing aperture issues.

\section{SDSS obscured QSOs at $z\sim 1$}

\subsection{Sample selection}

To directly apply the X/NeV obscuration diagnostic to distant AGN, we
tried to select SDSS obscured QSOs at $z\sim 1$ on the basis of their
[Ne~V] emission. We searched in the SDSS DR7 entire spectroscopic database (table {\tt SpecPhotoAll} 
on the SDSS site) for objects with high [Ne~V] $EW$, since nuclear obscuration should suppress the AGN
continuum but not the flux of the lines produced in the NLR. 

\begin{figure}
\includegraphics[width=8cm]{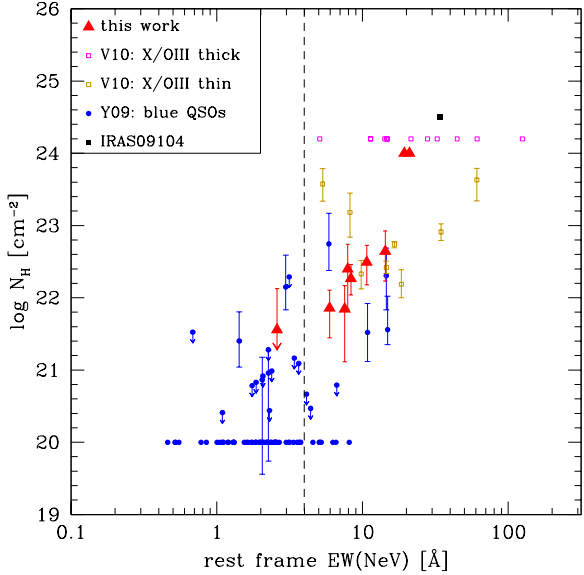}
\caption{Relation between rest frame equivalent width of the [Ne~V]
line and X-ray obscuration for the various samples of SDSS QSOs
considered in this work. Symbols are as in Fig.~\ref{xnevsdss}. The
equivalent width of the [Ne~V] line clearly correlates with X-ray
obscuration. The dashed line shows the $EW_{NeV}>4\AA$ limit used to
select the \chandra\ targets analyzed in this paper.}
\label{ewnh_sdss}
\end{figure}

We concentrated on objects with $0.8<z<1.4$, i.e. those not accessible
through [O~III] selection (thus excluded from the \citealt{zak03}
and \citealt{reyes08} samples), and in which the [Ne~V] line is
observed at $\lambda<8200$\AA, where the SDSS spectral efficiency is
highest. To avoid spurious detections, we conservatively considered
those objects in which the [Ne~V] line is detected at $\geq 5\sigma$ in the
SDSS spectral tables and, to avoid strong contamination from unobscured, broad line (type-1) AGN, we
restricted the search to those spectra with MgII~2800 emission with FWHM$\lesssim3000$
km s$^{-1}$. The cut in the MgII~2800 line width is not stringent: this leaves in the sample
obscured QSO candidates with broad optical lines and a red continuum, such as the red-QSO
population described e.g. by \citet{wilkes02, wilkes05}. The resulting median, rest frame [Ne~V] $EW$ of the
sample is 2\AA. We considered the high $EW$ tail of the distribution, i.e. the objects with $EW>$4\AA~in the SDSS spectroscopic 
tables (89 objects). In Fig.~\ref{ewnh_sdss} we show
the correlation between the [Ne~V] $EW$ and the X-ray absorption as
measured for the SDSS QSO and type-2 QSO samples discussed in the previous
Sections: most objects with $EW>$4\AA~do show significant
X-ray absorption, validating the use of this $EW$ threshold. \footnote{In Fig.~\ref{ewnh_sdss} the 9 [Ne~V]-selected targets discussed in the next Section are also shown. 
After re-fitting the SDSS spectra one object shows $EW\sim 2.6\AA$, i.e. below the $EW>$4\AA~threshold chosen in the SDSS spectroscopic tables. The re-fitted [Ne~V] fluxes and $EW$s of the remaining 8 objects are instead found to differ by only a few percent from those resulting from the SDSS automated fit procedure.}

We visually inspected all the 89 selected spectra to verify possible
problems related to the SDSS automated procedure and to discard QSOs
showing a blue continuum. 
At the end of this process, we selected 27 objects
which, based on their optical spectrum, appear to be good obscured QSO
candidates at $z \sim 1$.


\begin{table}
\caption{Basic sample properties}
\begin{tabular}{lcccr}
\hline \hline
SDSS Name& $z$& R& $f_{NeV}$&  $EW_{NeV}$\\
(1)& (2)& (3)& (4)& (5)\\ 
\hline
J105951.36+301817.4& 0.887& 19.667& 8.12&  8.3\\
J092640.67+023628.7& 1.306& 20.658& 7.76& 14.3\\
J034222.54--055727.9& 0.882& 19.846& 6.83&  7.5\\
J104603.17+071907.2& 0.888& 20.191& 5.73& 10.7\\
J165158.61+432508.6& 0.853& 20.093& 4.20&  5.9 \\
J080859.33+204711.8& 0.908& 21.059& 4.98& 21.0\\
J125848.58+120531.1& 1.035& 20.013& 5.13&  7.9\\
J145503.94+515539.9& 1.277& 19.466& 2.00&  2.6\\
J085600.88+371345.5& 1.022& 21.278& 4.05& 19.3\\
\hline		 	      			   
\end{tabular}

Column description: (1) Source name. (2) Redshift. (3) SDSS PSF
magnitude in the R-band. (4) Measured [Ne~V]3426 flux in units of $10^{-16}$\cgs.
(5) [Ne~V]3426 rest frame equivalent width in units of Angstrom.
\label{bop}
\end{table}

\subsection{\chandra\ observations}\label{cxo}

In \chandra\ Cycle 10 we were awarded 10ks ACIS-S observations of 8 targets which, 
based on their optical spectrum, were considered as good obscured QSOs targets (i.e. they show either narrow emission lines only or broad, 
2000$<$FWHM$<$3000 km s$^{-1}$, lines over a red continuum). Observations have been performed between December 2008 and
July 2009, with the S3 chip at the aimpoint and using the very faint
telemetry mode. We performed standard data reduction and extraction of
spectra using CIAO v3.2.2. A 10ks ACIS-S observation of an archival
object (SDSSJ085600, PI Strauss) satisfying our selection criteria has been also analyzed.

The basic optical properties of these 9 objects (redshift, R-band magnitude,
[Ne~V] flux and rest-frame $EW$) are shown in Table~\ref{bop}, while the
journal of \chandra\ observations is shown in Table~\ref{bxp}.

Apart from the archival object, which shows only 2 X-ray photons
within 1.5'' of the optical source position, all the other targets
were detected, with photon statistics ranging from 18 to 313 counts in
the 0.5-8 keV band. Spectral fitting was performed using XSPEC
v11.3.1. Standard $\chi^2$ minimization has been adopted for spectra
showing more than 100 counts in the 0.5-8 keV band, binned to a
minimum of 10 counts per energy bin. C-statistics \citep{cash79} has been
used for spectra with lower photon statistics, binned to a minimum of
3 counts per bin. Model fits consist of simple power-laws modified by
Galactic absorption and absorption at the source redshift. Errors on
the best fit spectral parameters are quoted at 90\% confidence level
for one interesting parameter. 

\begin{table}
\caption{Log of X-ray observations}
\begin{tabular}{lrrc}
\hline \hline
SDSS Name& ObsId& Date& Exp.\\
(1)& (2)& (3)& (4)\\ 
\hline
J105951.36+301817.4& 10325&  2009-01-29& 10.0\\
J092640.67+023628.7& 10326&  2009-03-23& 10.0\\
J034222.54--055727.9& 10327&  2008-12-02&  9.8\\
J104603.17+071907.2& 10328&  2009-02-10& 10.0\\
J165158.61+432508.6& 10329&  2009-06-30& 10.0\\
J080859.33+204711.8& 10330&  2008-12-22& 10.0\\
J125848.58+120531.1& 10331&  2009-02-23&  9.2\\
J145503.94+515539.9& 10332&  2009-07-27& 10.0\\
J085600.88+371345.5&  6807&  2006-02-17& 10.3\\
\hline		 	      			   
\end{tabular}

Column description: (1) Source name. (2) \chandra\ observation
identifier. (3) Observing date (yyyy-mm-dd). (4) Exposure time in
ks.
\label{bxp}
\end{table} 


\begin{table*}[t]
\caption{Results from X-ray data analysis and spectral fitting}
\begin{center}
\begin{tabular}{lrrcccrrlll}
\hline \hline
SDSS Name& Cts& HR& $N_H^{gal}$& $N_H$& $\Gamma$& $\chi^2/dof$& $f_{2-10}$& $L^{abs}_{2-10}$& $L^{int}_{2-10}$& class\\
(1)& (2)& (3)& (4)& (5)& (6)& (7)& (8)& (9)& (10)& (11)\\ 
\hline
J105951.36+301817.4&  237& -0.30& 2.0& 1.84$^{+1.05}_{-0.75}$& 1.90$^{+0.40}_{-0.35}$& 25.6/20&    16&    44.70&    44.80&      thin\\
J092640.67+023628.7&   65& -0.06& 3.8& 4.40$^{+4.00}_{-2.70}$& 1.67$^{+0.75}_{-0.65}$&     ...&   6.8&    44.61&    44.75&      thin?\\
J034222.54--055727.9&  175& -0.50& 5.3& 0.70$^{+0.77}_{-0.57}$& 1.83$^{+0.56}_{-0.45}$& 10.4/13&    13&    44.60&    44.61&      thin\\
J104603.17+071907.2&   31& -0.16& 2.8& 3.10$^{+2.22}_{-1.59}$&          $1.8^{fixed}$&     ...&   3.7&    44.04&    44.15&     thin?\\
J165158.61+432508.6&  313& -0.38& 1.9& 0.72$^{+0.55}_{-0.44}$& 1.76$^{+0.33}_{-0.30}$& 20.7/27&    19&    44.76&    44.79~&      thin\\
J080859.33+204711.8&   18&  0.80& 4.0&                    ...&   -1.2$^{+0.9}_{-1.4}$&     ...&   14&    43.95&    45.65$^*$& thick\\
J125848.58+120531.1&   18& -0.05& 2.2&    2.5$^{+3.0}_{-1.8}$&          $1.8^{fixed}$&     ...&   1.8&    43.88&    43.97&     thin?\\
J145503.94+515539.9&  125& -0.43& 1.8& 0.36$^{+0.98}_{-0.36}$& 1.82$^{+0.58}_{-0.41}$&  14.4/9&   5.7&    44.67&    44.69&      thin\\
J085600.88+371345.5& $<4$& .....& 2.9&                    ...&                    ...&     ...&$<$2.0& $<$43.50& $<$45.20$^*$& thick\\
\hline		
\end{tabular}
\end{center} 	      			   

Column description: (1) Source name. (2) Net counts in the 0.5-8 keV
band. (3) Hardness ratio, defined as (H-S)/(H+S), where H and S are
the number of photons observed in the 0.5-2 keV and 2-8 keV bands,
respectively. (4) Galactic column density in units of $10^{20}$
cm$^{-2}$ \citep{dl90}. (5) Absorbing column density at the source redshift in units of $10^{22}$
cm$^{-2}$. (6) Photon index. (7) Chi-square value over number of
degrees of freedom: for sources with less than 100 net counts
C-statistics has been used. (8)
Observed 2-10 keV flux in units of $10^{-14}$ erg cm$^{-2}$ s$^{-1}$.
(9) Logarithm of the 2-10 keV rest frame luminosity, not corrected for
absorption, in units of erg s$^{-1}$. (10) Logarithm of the intrinsic
2-10 keV rest frame luminosity, i.e. corrected for absorption, in
units of erg s$^{-1}$; $^*$for the two candidate Compton-Thick objects
it is assumed $L^{int} = 50 \times L^{obs}$ (see text). (11) Object
classification based on the X-ray spectral analysis and the X/NeV
ratio.
\label{xspec}
\end{table*}

Four sources have been detected with sufficient photon statistics
($>100$ counts) to perform a rough spectral analysis and classify them
as Compton-Thin, as also suggested by their soft hardness ratios (HR$\leq-0.3$). One object, namely SDSSJ080859, shows a
faint (18 counts), very hard (HR=0.8) spectrum with a hint of excess
emission at $\sim 3.3$ keV. If the excess is interpreted as a
redshifted 6.4 keV iron $K\alpha$ emission line superimposed to a hard
power-law continuum ($\Gamma=-1.2$), this would imply a rest frame
equivalent width of $EW=2.8^{+3.3}_{-2.1}$ keV. The X-ray spectral
features of SDSSJ080859 are therefore highly suggestive of CT
absorption. Three objects (SDSSJ092640, SDSSJ104603 and SDSSJ125848) are intermediate
cases in terms of both photon statistics (from 18 to 65 counts) and X-ray hardness (from HR=-0.16 to -0.05). By fitting the data with an absorbed power-law
(fixing $\Gamma$ to 1.8 for two of them),
these sources appear to be Compton-Thin. All
the detected sources have relatively bright observed 2-10 keV fluxes
($>3\times 10^{-14}$ \cgs) and rest-frame 2-10 keV luminosities
$\gtrsim 10^{44}$ erg s$^{-1}$ (prior to absorption corrections),
placing them among luminous obscured QSOs.  As for the archival
undetected object, SDSSJ085600, by assuming a 2-10 keV band detection
limit of 4 photons and a pure reflection spectrum ({\tt pexrav} in XSPEC), the corresponding upper limits to the observed
2-10 keV flux and rest frame luminosity, are $2\times10^{-14}$ erg cm$^{-2}$
s$^{-1}$ and $3\times10^{43}$ erg~s$^{-1}$, respectively. A summary of the results
obtained from the X-ray analysis of these 9 objects is given in
Table~\ref{xspec}. Optical and X-ray spectra are shown in
Fig.~\ref{oxspec}

\subsection {Position on the X/NeV diagram}

We placed the new 9 SDSS obscured QSOs at $z\sim 1$ on the X/NeV diagram
(see Fig.~\ref{xnevsdss}). The archival undetected X-ray object
SDSSJ085600 shows X/NeV$<15$, which is strongly suggestive of CT
absorption. Indeed, most SDSS QSOs classified as CT by
V10 do show X/NeV ratio below this value, while
all those classified as Compton-Thin by V10 lie at X/NeV$>100$.

The faint (18 X-ray counts) object SDSSJ0808 has X/NeV$=45$. This value has been observed in both CT and Compton-Thin
AGN (see Fig.~\ref{xnev_local}), but, based on the X-ray spectral
analysis, SDSSJ0808 appears to have all the signatures of CT
absorption. We then classify SDSSJ085600 and SDSSJ0808 as likely CT
candidates. 

The four objects with highest photon statistics have also the lowest
absorption and the highest X/NeV ratios (X/NeV$>150$), and occupy a
region of the diagram which is populated by the Compton-Thin QSOs in
V10 and by the obscured tail of the broad line SDSS QSOs studied by
\citet{y09}. 

Finally, three objects do show X/NeV ratios in the range 20-60,
which are typical of heavily obscured AGN, but are still consistent
with Compton-Thin absorption. Both the X-ray and optical spectra of
these sources would favor mild absorption, hence a Compton-Thin interpretation, although no
clear cut classification can be made with the current data.

We note that the actual degree of obscuration of our two CT candidates SDSSJ085600 and SDSSJ0808 can only be
confirmed by obtaining good-quality X-ray spectra through deeper observations. Indeed, any selection
method which simply relies on X-ray hardness ratios, or on the comparison between the measured, obscured X-ray emission (if any) with some other 
indicator of the intrinsic nuclear power (e.g. dust-reprocessed IR-emission, or high-ionization, narrow optical emission lines) can only
provide an indirect way to select CT AGN. Different caveats and limitations affect these different selection methods, like e.g. reddening in line-selected sources
and contamination from star formation in IR-selected sources.

\section{Discussion}

\subsection{[Ne~V] as a good tracer of nuclear luminosity: effects of reddening in the Narrow Line Region and anisotropy}

The X/NeV vs X-ray absorption diagnostic diagram presented in Section 3 has been 
derived directly from observed values, without applying any corrections to either [Ne~V] or X-ray emission.
The rationale behind is that X-rays come from the innermost nuclear regions and 
can be depressed by small scale ($<$1 pc) absorption, while the [Ne~V] is instead a good indicator of the 
intrinsic nuclear luminosity being i) emitted on larger (kpc) scales, free from nuclear obscuration and ii) being
isotropic. We now discuss whether this two hypotheses are satisfied
and what happens if they are not.

As for the first point, we note that significant extinction towards
the NLR is commonly observed in local AGN \citep[e.g.][]{dd88}.
Under the simplest version of the AGN
unification schemes, one would expect to observe similar properties in
the NLR of both type-1 and type-2 AGN. Therefore, if the dust content
in the NLR of both AGN types is the same, the same
extinction correction should be applied to the measured values of the
[Ne~V] lines, which would simply shift
towards lower X/NeV values all the datapoints in Fig.~\ref{xnev_local}
and \ref{xnevsdss}, without altering the X/NeV vs $N_H$ trend. However,
based on the results by \citet{dd88}, there is some
evidence that the extinction to the NLR in local Seyfert 2s is
somewhat higher than in Seyfert 1s, the following relation holding for
the median values: $A_V^{Sy2}-A_V^{Sy1}\sim1.5-1.0=0.5$. Using
standard extinction curves \citep[e.g.][]{gb07}, this
translates into an average correction in the [Ne~V] flux a factor of $\sim 2.3$ larger in Seyfert 2s
than in Seyfert 1s. Applying some extinction correction to both the observed data and model curves would then 
produce a stronger shift towards lower X/NeV ratios in Seyfert 2s than in Seyfert 1s, making the
anti-correlation between X/NeV ratio and obscuration even more evident.

At any rate, our main interest is on the possibility of applying the
X/NeV diagnostic ratio to $z\sim 1$ objects, for which it is often
impossible to measure the extinction to the NLR because of the lack of
strong Balmer lines in the optical spectrum. Therefore, we simply do
not apply any reddening correction to the measured [Ne~V] fluxes.

As for the second point, the isotropy of [Ne~V] emission, we note that there is some evidence that the
[Ne~V] lines in SDSS type-1 QSOs are broader than in type-2 QSOs. As an example,
in the composite spectrum of QSOs from the 2dF survey \citep{colless01} the [Ne~V] width (FWHM$\sim1200$ km s$^{-1}$) is found to be 
in-between that of broad lines (e.g. H$\beta$ and MgII, FWHM$\sim4000$ km s$^{-1}$) and that of narrow lines 
(e.g. [O~II] and[O~III], FWHM$\sim600$ km s$^{-1}$; see Fig.2 in \citealt{corbett03}).


According to the standard Unified Model, this would
suggest that, whenever a direct look towards the nucleus is available,
an additional broader [Ne~V] component is observed, coming
from low density and fast moving clouds which cannot be observed in
type-2 AGN, possibly being located in-between
the BLR and the NLR. If this is true, then the total [Ne~V] emission
line in not a good isotropic indicator of the intrinsic AGN power,
since in type-2 objects only a portion of the [Ne~V] emitting region
can be observed. If anything, this would go in the direction of
underestimating the total [Ne~V] luminosity in type-2 AGN. Correcting
for this effect, would make the X/NeV ratio even lower in type-2 AGN.
To summarize, if corrections for reddening and anisotropy were applied, 
the X/NeV vs $N_H$ trend shown in Fig.~\ref{xnevsdss} would be even 
stronger.

\subsection{Comparison between [O~III] and [Ne~V] selection}

As discussed in Section 4.2, 2 out of 9 [Ne~V]-selected type-2 QSOs
($\approx 22\%$) appear to be likely CT candidates based on their X-ray
spectral properties and/or X/NeV ratio. At face value, the fraction of [Ne~V]-selected
CT QSO candidates appears lower than that based on
[O~III] selection, which is around 60-70\% (V10). However, we note
that the difference between the CT detection rate can be mainly
ascribed to intrinsic differences in the two parent samples of
obscured QSOs. Indeed, our sample also includes obscured QSOs
which have a red continuum but still do show broad optical
lines. These objects (see Fig.~\ref{oxspec}), appear to be part of the
population of red QSOs discussed e.g. by \citet{wilkes02, wilkes05}
and \citet{urrutia05}, which on average are absorbed by columns
below $N_H=10^{23}$ \cm. On the contrary, V10 selected their objects
from the sample of \citet{zak03}, in which strict
criteria on the line width have been adopted, producing an
ensamble of pure type-2 objects. As a simple check, when considering
only those [Ne~V]-selected objects which look like pure type-2 spectra
(e.g. MgII FWHM $\lesssim2000$ km s$^{-1}$) the fraction of CT
candidates increases to $\approx$50\% (2/4). Despite the very low statistics,
this fraction is consistent with what has been found by V10, suggesting
that, when the search is restricted to pure type-2 objects, [Ne~V]
selection may be an efficient way to pick up CT AGN at $z\sim 1$.

\subsection {Application of the X/NeV diagnostic to spectroscopic surveys
with deep X-ray coverage}

Synthesis models predict that from 10\% to about 30\% of the XRB at 30 keV
is not accounted for by the integrated emission of Compton-Thin AGN. This ``missing''
background is expected to be produced by Compton-Thick AGN, and most of it
is expected to be produced by CT AGN with Seyfert-like intrinsic luminosities at a redshift
of $z\sim1$. The [Ne~V]-selected CT QSOs in the SDSS represent the high-luminosity, low space density tail
of the distribution of CT AGN at $z\sim1$, and are expected  to provide
only a minor contribution to the missing XRB. Selection of lower-luminosity CT AGN at 
$z\sim 1$ is therefore needed, which can in principle be done by applying the X/NeV diagnostics to objects
in sky areas with deep spectroscopic surveys and deep X-ray coverage. 
As an example, the combination between the zCOSMOS-bright 
spectroscopic survey \citep{lilly07, lilly09} and the \chandra-COSMOS X-ray survey \citep{elvis09, puccetti09} in the COSMOS field \citep{scoville07} would be able to identify 
CT AGN at $z\sim 1$ down to intrinsic $L_X\sim10^{43}$\ergs, i.e. the population which is thought to produce a large fraction of the missing XRB (Vignali et al. in preparation).
In the GOODS-S and -N fields, the population of CT AGN at $z\sim 1$ can be tracked further down to intrinsic $L_X$ of $10^{42}$\ergs.
Furthermore, since the CT samples obtained using the X/NeV diagnostic appear to be relatively free from contamination by less obscured objects, by using spectroscopic surveys with well defined selection functions it would be possible to estimate the space density of this missing AGN population.

It has to be noted that selection of obscured AGN through the [Ne~V]3426 line is most likely a lower limit, since it misses objects with dusty Narrow Line Regions, 
which can instead be picked up by mid-IR selection \citep{daddi07,fiore08}. However, when sufficiently deep
mid-IR coverage is not available for spectroscopic survey fields, [Ne~V]-selection may then represent a promising 
and ready-to-use method to get large samples of $z\sim 1$ CT AGN. In addition, the comparison between the space density 
of [Ne~V] and mid-IR selected objects with similar redshifts and bolometric luminosities in fields with full multiwavelength coverage, 
may indicate the fraction of objects
in which the NLR is free from obscuration, thus constraining the physical scale on which absorption arises. 

\begin{figure}[t]
\includegraphics[width=8cm]{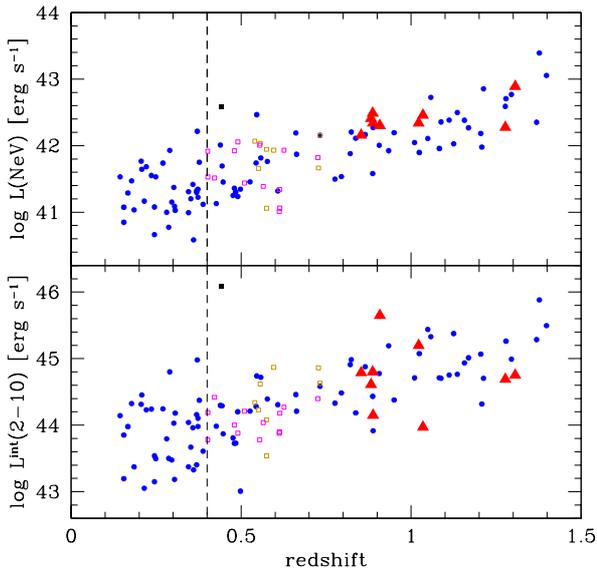}
\caption{[Ne~V] luminosity ({\it upper panel}) and {\it intrinsic} 2-10 keV rest
frame luminosity ({\it lower panel}) vs redshift for the different SDSS
QSO samples presented in this work. Symbols are as in the previous
Figures. For Compton-Thick QSO candidates, we assumed that the
intrinsic 2-10 keV rest frame luminosity is a factor of 50 larger than
the observed one. The dashed line shows the $z>0.4$ threshold adopted
to compare the [O~II], [Ne~V] and X-ray properties of QSOs of similar
intrinsic luminosity (see text).}
\label{lnexz_sdss}
\end{figure}

\begin{figure}[t]
\includegraphics[width=8cm]{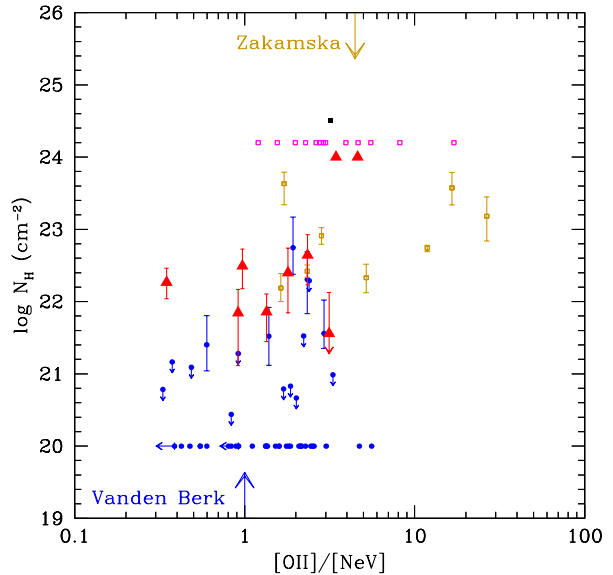}
\caption{[O~II]/[Ne~V] luminosity ratio vs X-ray column density for SDSS
QSOs and type-2 QSOs at $z>0.4$. Symbols are as in the previous Figures. A
trend is visible in which progressively more obscured objects
have enhanced [O~II] emission relative to [Ne~V]. We interpret this
trend as due to increased star formation in obscured QSOs, which
boosts the [O~II] line emission. The [O~II]/[Ne~V] ratio as measured on
the SDSS type-1 QSO composite spectrum of \citet{vanden01} and on the
SDSS type-2 QSO composite spectrum of \citet{zak03} are shown with
the two labeled arrows.}
\label{oiinenh_sdss}
\end{figure}

\subsection{Enhanced star formation in type-2 QSOs?}

Popular semi-analytic models of galaxy formation and evolution
\citep{kh00, marulli08, hop08}
propose that, at least for the most massive and luminous objects,
nuclear activity and star formation are both triggered by major
mergers of gas rich galaxies, and that at the early stages of the
merger, when star formation is more vigorous, the AGN is embedded
within optically thick gas shrouds. According to this scenario one can
therefore expect to observe strong star formation in obscured QSOs.

At redshift $z\sim 2$, the concurrent obscured black hole growth and
star formation has been observed in the population of bright
submillimeter sources \citep[e.g.][]{alex05}.
At low redshifts ($z<0.3$), \citet{kim06} noted that the ratio
between the luminosity of the [O~II]3727 line and the [O~III]5007
line is a factor of $\sim 4$ higher in the type-2 QSO composite
spectrum of \citet{zak03} than in the average spectrum of
type-1 QSOs. Because of the similar [O~III] luminosity of the two
populations, the excess of low-ionized oxygen in type-2 QSOs could not
be explained in terms of different ionization parameters of the Narrow
Line Regions, and was then interpreted by \citet{kim06} as due to enhanced
star formation in type-2 QSOs.

We then investigated the star formation in the samples presented in
this work by measuring the [O~II]3727 flux on the SDSS spectra and
considered the [O~II] over [Ne~V] ratio as a function of the nuclear
obscuration. To compare objects in the same redshift range we
restricted our analysis to $z>0.4$. As shown in Fig.\ref{lnexz_sdss},
this cut removes only low-redshifts type-1 QSOs and ensures that both
type-1 and type-2 QSOs span a similar luminosity range.\footnote{The intrinsic X-ray luminosity of CT QSOs has been assumed
to be 50$\times$ the observed one, while V10 defined as CT candidates those objectes in which the 
expected intrinsic X-ray luminosity was at least a factor of 100 larger than the observed one. 
We note that, even with a factor of 2 larger correction, the intrinsic
luminosity of the CT QSOs considered in this work would still be
comparable with that of unobscured QSOs in the same redshift range.}
We note that, for QSOs in the redshift range $z$=0.4-1.5, the 3'' size of 
the SDSS fibers encloses regions as large as 16-26~kpc diameter, and therefore 
samples a significant portion of the host galaxy in which star formation
can take place.

In Fig.~\ref{oiinenh_sdss} we plot the [O~II]/[Ne~V]
ratio as a function of the measured X-ray column density. The blue
SDSS QSOs in the Y09 sample do show [O~II]/[Ne~V] ratios on average
lower than type-2 QSOs. Some positive correlation, albeit with a large
scatter, is indeed seen between the [O~II]/[Ne~V] ratio and the absorbing column
density $N_H$. When considering those objects with observed [O~II]/[Ne~V]
$>4$, we found that only 2 out of 12 are not obscured, and
half of them (6 objects) are likely obscured by CT absorption. Conversely, there
are no objects with $N_H>10^{23}$\cm among those with [O~II]/[Ne~V] $<1$. The [O~II]/[Ne~V]
ratios measured on the SDSS type-2 and type-1 QSO composites by \citet{zak03}
and \citet{vanden01} were also considered and found
to be in good agreement with the averages measured in this work for obscured and unobscured
QSOs, respectively (see Fig.~\ref{oiinenh_sdss}). 
We tried to compute the significance of the correlation between the  [O~II]/[Ne~V] ratio and the logarithm of the 
column density. We note that it is difficult to deal with objects which are either unobscured or CT candidates, because they cannot 
be treated statistically as proper upper or lower limits on $N_H$, since the gas column density can plausibly vary only within a bounded range
(i.e. it cannot be zero or infinite). For simplicity we therefore assumed log$N_H$=20 for unobscured objects and log$N_H$=24 for CT candidates, 
respectively  (see Fig.~\ref{oiinenh_sdss}). The presence of a correlation has been estimated through the {\sc asurv} 
software package \citep{lavalley92}, using the generalized Kendall's $\tau$ and the Spearman's $\rho$ correlation tests. We found that the 
probability that the correlation is not present is only $2\times10^{-4}$ and $1\times10^{-4}$, respectively.
If the [O II] emission measured in type-2 QSOs is
interpreted as entirely due to star formation, the median [O II] luminosities of the [O III]- and
[Ne V]-selected samples would correspond to star formation rates of $\approx 100$ and $\approx 200\; M_\odot$/yr, 
respectively (using the relation by \citealt{kewley04}). These values could decrease by up to a factor of $\sim 2$ if the AGN contribution
to the [OII] emission is significant \citep{silverman09}. 
This finding is consistent with the 
expectations from the AGN evolutionary sequence outlined above.

\section{Conclusions}

We have presented a diagnostic diagram to identify heavily obscured, 
Compton-Thick AGN candidates at $z\sim 1$ based on the ratio between the 2-10 keV flux
and the [Ne~V]3426 emission line flux (X/NeV). The diagnostic was calibrated on a sample of 74 
local Seyfert galaxies and then applied to populations of type-1 and type-2 QSOs at different
redshifts (from $z\sim 0.1$ to $z=1.5$) selected from the SDSS. The main results obtained in this work can be summarized as follows.
\\

\noindent
$\bullet$
The observed X/NeV ratio is found to decrease with increasing absorption: the mean X/NeV ratio for
unobscured Seyferts is about 400, about 80\% of local Seyferts with X/NeV$<100$ are obscured by column 
densities above $10^{23}$\cm\, and essentially all objects with 
observed X/NeV $<15$ are Compton-Thick.
\\

\noindent
$\bullet$
We considered a sample of 83 blue type-1 QSOs and 21 [O III]-selected type-2 QSOs in the SDSS
which have been observed in the X-rays and show significant [Ne V] detection. It was verified that they 
follow the same X/NeV vs X-ray absorption trend which is observed for local Seyferts. Furthermore, SDSS type-2 QSOs 
classified either as Compton-Thick or Compton-Thin on the basis of their X/OIII ratio, would have been mostly classified in the same
way based on the X/NeV ratio. 
\\

\noindent
$\bullet$
The X/NeV diagnostic was used to investigate the obscuration of 
9 SDSS obscured QSOs in the redshift range $z=[0.85-1.31]$, which is not accessible through
[O III] selection. The 9 objects were selected by means of their prominent [Ne~V]3426 line 
($EW>4$\AA), and \chandra\ snapshot observations for 8 of them were obtained (one object is from the archive). 
Based on the X/NeV ratio, complemented by X-ray
spectral analysis, only 2 objects appear good Compton-Thick QSO
candidates. However, when considering the 4 genuine narrow-line objects only (FWHM of the MgII line $\lesssim 2000$ km $s^{-1}$), 
the efficiency in selecting Compton-Thick QSOs through the [Ne~V] line is about 50\% (2/4), which is more similar, 
despite the large uncertainties, to what is achieved with [O~III] selection (60-70\%; \citealt{v10}). 
\\

\noindent
$\bullet$
We verified that neither extinction nor anisotropy corrections on the [Ne~V] emission 
would affect our conclusions and that the X/NeV diagnostic is therefore a good method to identify clean, despite not complete, samples of
heavily obscured AGN.
We discussed the possibility of applying the X/NeV diagnostic to objects in sky areas with deep optical spectroscopy and X-ray coverage.
This will allow to identify Compton-Thick Seyferts at $z\sim 1$, i.e. those objects which are thought to be responsible for
a large fraction of the ``missing" X-ray background.
\\

\noindent
$\bullet$
Finally, the optical emission line properties of [Ne~V]-selected QSOs were compared with those of
other SDSS populations of obscured and unobscured QSOs. By restricting the analysis to objects in the same 
redshift (and luminosity) range $z$=[0.4-1.5], we found evidence that the ratio between the [O~II]3727 and [Ne~V]3426 luminosity
increases with obscuration. This correlation is interpreted as evidence of enhanced star formation
in obscured QSOs, which is consistent with current popular scenarios of BH-galaxy coevolution.

{
\acknowledgements

We thank Martin Elvis, Alessandro Marconi, Lucia Pozzetti, Guido Risaliti, Marco Salvati
and Pilar Esquej for stimulating discussions and Monica Young for kindly providing
the catalog of SDSS QSOs for which \xmm\ data were available. We thank the referee for a timely and useful report. 
We acknowledge partial support from ASI-INAF and PRIN/MIUR under grants
I/023/05/00, I/088/06/00 and 2006-02-5203.
}

\appendix
\section{Master table for the local sample}

In the Tables at the end of the Appendix we present the sample of local objects which has
been used to build the X/NeV vs X-ray absorption diagram shown in
Fig.~\ref{xnev_local}. The X-ray absorption, X-ray flux and [Ne~V] flux
have been derived from the literature.  As for X-ray data, we
generally preferred to use the values obtained with the most recent
and sensitive satellites, \chandra\ and \xmm. When dealing with heavily
obscured objects, though, we considered
measurements obtained with {\it Suzaku} and {\it BeppoSAX} in order to map the
energy range above 10 keV and get a more robust measurements of the
absorbing column density. Sometimes an observed 2-10 keV flux measurement is
not directly quoted in the considered literature papers: in those
cases we estimated it using the published best fit spectral parameters
and/or 2-10 keV luminosity. We analyzed the archival, unpublished X-ray data
of 4 objects (Mrk~78, Tol~1506.3-00, NGC~4074, IIIZw77) and re-analyzed the X-ray spectrum 
of NGC~1320 since the 2-10 keV flux we derived from the best-fit parameters in \citet{greenhill08} 
appears to differ significantly from that listed in the 2XMM catalog. The X-ray spectra and best fit
parameters for these objects are presented below.

\begin{figure*}
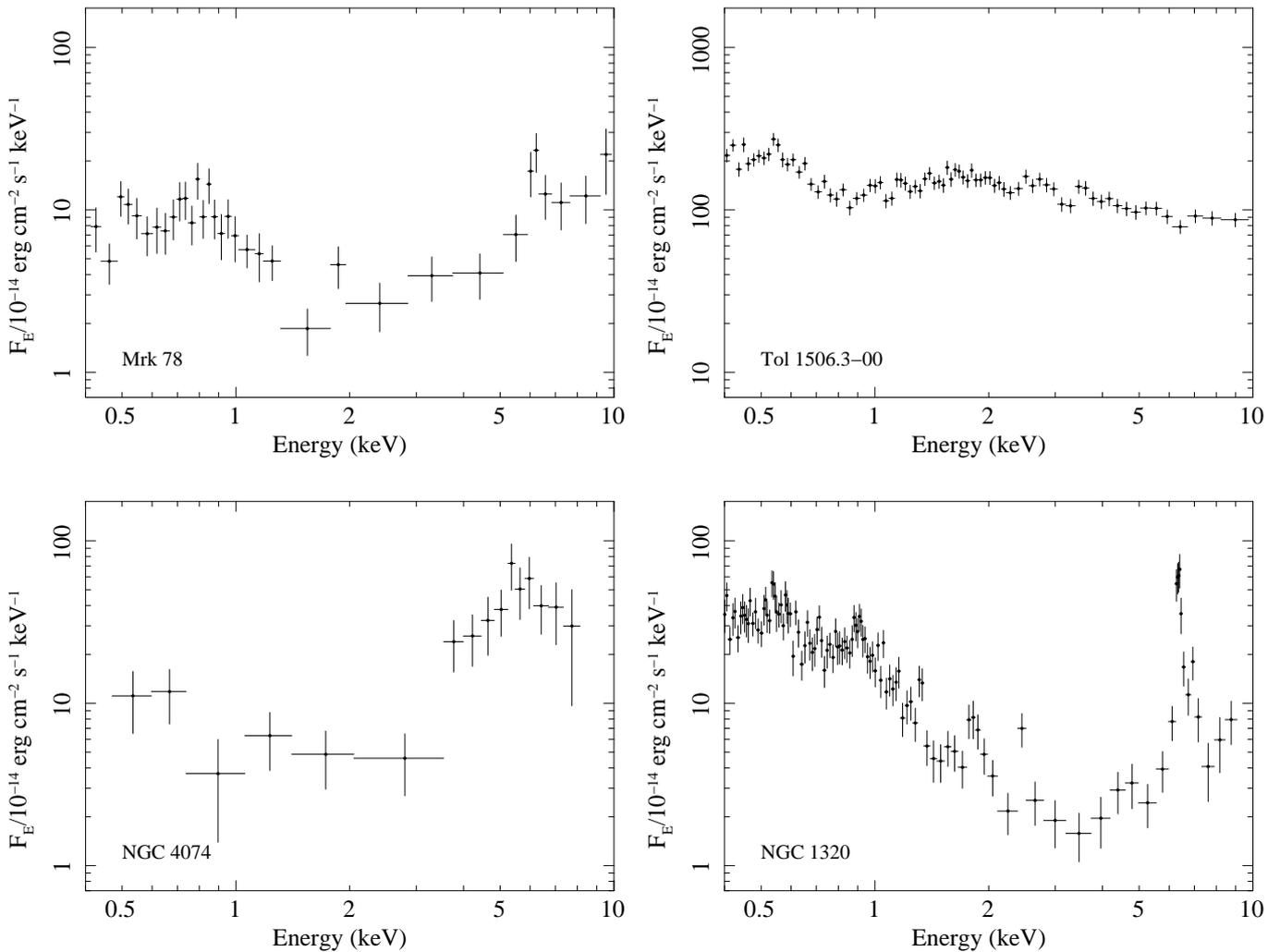

\includegraphics[angle=-90,width=0.49\textwidth]{4039fa1a.ps}
\includegraphics[angle=-90,width=0.49\textwidth]{4039fa1b.ps}
\vglue0.6cm
\includegraphics[angle=-90,width=0.49\textwidth]{4039fa1c.ps}
\includegraphics[angle=-90,width=0.49\textwidth]{4039fa1d.ps}

\caption{
\xmm\ EPIC-pn data of the four archival observations presented here 
(from top-left to bottom-right: Mrk~78, Tol~1506.3$-$00, NGC~4074, and 
NGC~1320).  
The data have been corrected for the effective area and further converted to 
flux density in units of $10^{-14}$ erg~cm$^{-2}$~s$^{-1}$~keV$^{-1}$ 
for displaying purposes.}
\label{epic_spectra}
\end{figure*}

\subsection{X-ray spectra of individual objects analyzed in this work}

For the AGN observed by \xmm\ (Mrk~78, Tol~1506.3-00, NGC~4074, NGC~1320), EPIC pn spectral results are presented, 
while for \asca\ data (IIIZw77), we refer to SIS spectra. 
All of the sources presented below were pointed as targets, with the exception 
of NGC~4074, which was observed at an off-axis angle of $\approx$~7.3\arcmin. 

\subsection{\xmm\ data}
\subsubsection{Mrk~78}
The \xray\ spectrum of Mrk~78 (top-left panel in Fig.~A.1; effective 
exposure time in EPIC-pn of $\approx$~7.2~ks) is well fitted with a model 
including, in the soft \xray\ energy band, a thermal ($kT\approx$~0.6~keV) 
and a power-law component (with a photon index $\Gamma\approx$~2.3). 
The emission at energies above 2~keV is well reproduced by an absorbed 
($N_H\approx~5.7\times10^{23}$~cm$^{-2}$) power-law (with photon index 
fixed to 1.8 because of the limited photon statistics) plus an iron 
emission line. The line energy (E=6.31--6.45~keV) indicates emission from 
neutral or mildly ionized iron, while its $EW$ ($\approx$~340~eV) is 
consistent, within the errors, with being produced by transmission through 
the same matter responsible for the absorption of the nuclear component. 
The 0.5--2~keV and 2--10~keV fluxes are $\approx8.1\times10^{-14}$~\cgs\
and $\approx5.9\times10^{-13}$~\cgs, respectively; 
the de-absorbed, rest-frame 2--10~keV luminosity for this source is 
$\approx8.5\times10^{42}$~\ergs.

\subsubsection{Tol~1506.3$-$00}
Tol~1506.3$-$00 was observed for $\approx$~3.7~ks with \xmm\ 
(top-right panel in Fig.~A.1). Its spectrum is characterized by 
a thermal ($kT\approx$~0.2~keV) plus a power-law component at soft energies, 
while the emission above $\approx$~2~keV is well parameterized by a power-law 
with $\Gamma\approx1.6$ and mild absorption 
($N_H\approx~7.1\times10^{21}$~cm$^{-2}$). No iron line is 
present, the upper limit on its $EW$, in the case of neutral iron, is 
$\approx$~40~eV (90\% confidence level). 
The 0.5--2~keV and 2--10~keV fluxes are $\approx2.35\times10^{-12}$~\cgs\ 
and $\approx7.71\times10^{-12}$~\cgs, respectively. 
The intrinsic, rest-frame 2--10~keV luminosity of Tol~1506.3$-$00 
is $\approx5.5\times10^{43}$~\ergs.

\subsubsection{NGC~4074}
The \xmm\ spectrum of NGC~4074 (bottom-left panel in Fig.~A.1; exposure time 
of $\approx$~2.6~ks) requires a double power-law model, with the 
component at energies above $\approx$~3~keV being absorbed by a column density 
\nh~$\approx~2.4\times10^{23}$~cm$^{-2}$ (assuming $\Gamma=1.8$). 
No iron line is present, with a 90\% upper limit on the $EW$ of a neutral 
iron line of $\approx$~190~eV. 
The measured 0.5--2~keV (2--10~keV) flux is $\approx8.6\times10^{-14}$~\cgs\ 
($\approx2.59\times10^{-12}$~\cgs), while the de-absorbed, rest-frame 
2--10~keV luminosity is $\approx8.0\times10^{42}$~\ergs.

\subsubsection{NGC~1320}
Among the sources observed by \xmm\ and presented in this Appendix, 
NGC~1320 is the one with the longest exposure time ($\approx$~11.5~ks) 
and most complex \xray\ spectrum, as shown in Fig.~A.1 (bottom-right panel). 
The \xray\ data for this source are highly suggestive of the 
presence of both a transmission and reflection component. 
While the transmitted component is well parameterized by a $\Gamma=1.8$ 
power-law continuum absorbed by thick matter (with a column density of 
$\approx2.4\times10^{24}$~cm$^{-2}$), for the reflection component 
the parameters are basically unconstrained, calling for observations with 
higher photon statistics and data above 10~keV to properly model and 
constrain the two spectral components. 
The best-fitting model requires also the presence of a power-law 
($\Gamma\approx2.9$) at low energies and a \neix\ emission line at 
an energy of $\approx$~930~eV ($EW\approx$~50~eV). 
The measured 0.5--2~keV (2--10~keV) flux is $\approx2.3\times10^{-13}$~\cgs\ 
($\approx5.7\times10^{-13}$~\cgs) \footnote{consistent with the value quoted in the 2XMM catalog.}, and the de-absorbed, rest-frame 
2--10~keV luminosity is $\approx8.4\times10^{42}$~\ergs; we note, however, 
that this value is somehow uncertain, given the complex \xray\ modeling of the 
data presented here. 

\subsection{\asca\ data}

\subsubsection{IIIZw77}

An \asca\ observation with a net exposure time of $\approx 65$~ks has 
detected a faint \xray\ source at this galaxy. The \xray\ spectrum 
(see Fig.~A.2) shows a hard \xray\ excess above 3~keV 
when a simple power-law is fitted. 
This hard \xray\ component probably originates from an obscured 
active nucleus. When modeled by an absorbed power-law with photon 
index $\Gamma = 1.8$, the absorbing column density is estimated to be 
$N_{\rm H}=8^{+5}_{-4}\times 10^{23}$~cm$^{-2}$. The data have no 
sufficient quality to constrain a Fe K emission-line. The soft X-ray
component can be described by a power-law of $\Gamma\approx 2.5$. 
While the origin of the soft \xray\ emission is unclear, assuming that no 
strong star formation is taking place in this E/S0 galaxy, it is 
likely extended photoionized gas. The observed fluxes in the 0.5-2~keV 
and 2-10~keV bands are $1\times 10^{-13}$~\cgs\ and 
$6\times 10^{-13}$~\cgs, respectively. The absorption-corrected 
2--10~keV luminosity derived from the absorbed power-law model is 
$1\times 10^{43}$~\ergs.

\begin{figure}
\includegraphics[angle=-90,width=0.49\textwidth]{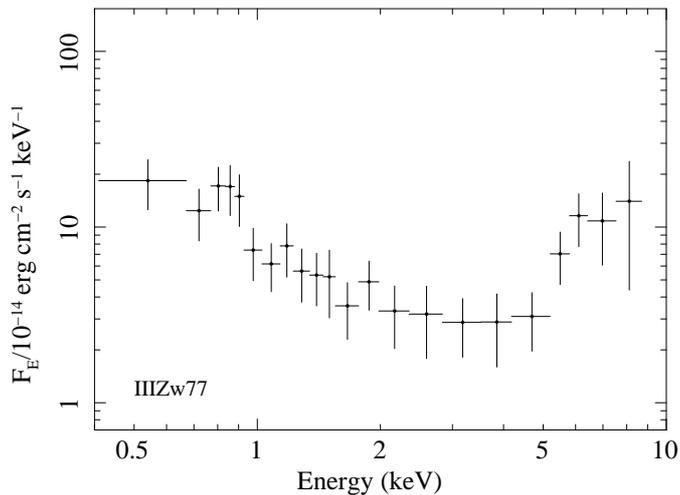}
\caption{
ASCA data of the source IIIZw77, corrected for the effective area and further converted to 
flux density in units of $10^{-14}$ erg~cm$^{-2}$~s$^{-1}$~keV$^{-1}$ for displaying purposes.}
\label{asca_spectrum}
\end{figure}

\begin{table*}
\caption{The local sample: Seyfert 1.0 to 1.5, Narrow Line Seyfert 1s (NLS1) and Broad Line Radio Galaxies (BLRGs).}
\begin{center}
\begin{tabular}{lcccrcrcc}
\hline \hline
Name& Type& $z$& log$N_H$& log$f_{2-10}$& $Ref_X$& log$f_{NeV}$& A& $Ref_{NeV}$\\
(1)& (2)& (3)& (4)& (5)& (6)& (7)& (8)& (9)\\ 
\hline
       NGC~3227&   1.5&  0.0039&   22.83& -11.09&  1&  -13.05&  L&   1\\ 
       NGC~3516&   1.5&  0.0088&   20.63& -10.62&  2&  -13.15&  L&   1\\ 
       NGC~3783&   1.0&  0.0097&   20.00& -10.07&  3&  -12.86&  M&   2\\ 
       NGC~4051&   1.5&  0.0020&   20.00& -10.64&  4&  -12.98&  L&   1\\ 
       NGC~4151&   1.5&  0.0033&   22.88& -10.35&  1&  -11.70&  L&   1\\ 
       NGC~4253&  NLS1&  0.0129&   20.00& -10.80&  5&  -13.49&  M&   3\\ 
       NGC~4395&   1.0&  0.0010&   20.80& -11.30&  6&  -14.10&  S&   4\\ 
       NGC~4593&   1.0&  0.0090&   20.00& -10.43&  7&  -13.60&  M&   2\\ 
       NGC~5548&   1.5&  0.0172&   20.00& -10.28&  7&  -12.57&  L&   1\\ 
       NGC~6814&   1.5&  0.0052&   20.00& -11.82&  8&  -13.77&  M&   2\\ 
       NGC~7469&   1.2&  0.0163&   20.00& -10.64&  7&  -12.74&  L&   1\\ 
         Mrk~42&  NLS1&  0.0246&   20.00& -11.98&  9&  -14.52&  L&   5\\ 
        Mrk~359&  NLS1&  0.0169&   20.00& -11.24&  10&  -13.30&  S&   4\\ 
        Mrk~486&  NLS1&  0.0389&   20.00& -11.55&  11&  -13.68&  S&   4\\ 
        Mrk~493&  NLS1&  0.0313&   20.00& -11.46&  10&  -14.52&  M&   3\\ 
        Mrk~509&   1.2&  0.0344&   20.00& -10.47&  7&  -12.24&  M&   2\\ 
        Mrk~704&   1.5&  0.0291&   23.13& -11.25&  12&  -13.21&  S&   4\\ 
        Mrk~705&   1.2&  0.0285&   20.00& -10.64&  13&  -13.80&  S&   4\\ 
        Mrk~841&   1.5&  0.0364&   20.00& -10.79&  14&  -13.15&  M&   2\\ 
        Mrk~896&   1.0&  0.0264&   20.00& -11.46&  10&  -14.10&  M&   2\\ 
        Mrk~926&   1.5&  0.0469&   20.00& -10.51&  7&  -13.96&  S&  6\\ 
       Mrk~1239&  NLS1&  0.0194&   23.52& -11.82&  15&  -13.31&  S&   4\\ 
         3C~120&  BLRG&  0.0330&   20.70& -10.31&  16&  -13.72&  S&  6\\ 
         3C~227&  BLRG&  0.0858&   22.10& -11.70&  17&  -14.40&  M&  7\\ 
         3C~382&  BLRG&  0.0578&   20.00& -10.21&  18&  -13.85&  M&  7\\ 
       3C~390.3&  BLRG&  0.0561&   20.60& -10.40&  19&  -13.59&  S&   4\\ 
         3C~445&  BLRG&  0.0562&   22.70& -11.17&  20&  -13.85&  M&   2\\ 
      Fairall~9&   1.2&  0.0470&   20.00& -10.92&  7&  -13.03&  M&   8\\ 
     Fairall~51&   1.0&  0.0142&   22.20& -10.60&  21&  -13.29&  M&   2\\ 
   Fairall~1116&   1.0&  0.0582&   20.00& -11.27&  22&  -13.54&  M&   8\\ 
   Tol~1351-375&   1.9&  0.0520&   22.20& -11.42&  23&  -13.48&  M&   2\\ 
  Tol~1506.3-00&   1.5&  0.0543&   21.85& -11.11&  24&  -14.00&  M&   2\\ 
     H1143-182&   1.5&  0.0329&   20.00& -10.55&  4&  -13.19&  M&   8\\ 
     H1846-786&   1.0&  0.0743&   20.00& -11.10&  7&  -13.49&  M&   2\\ 
        Akn~120&   1.0&  0.0330&   20.00& -10.42&  25&  -14.15&  S&   4\\ 
        Akn~564&  NLS1&  0.0247&   20.00& -10.80&  26&  -13.33&  S&   4\\ 
   MCG-6-30-15&   1.2&  0.0077&   20.00& -10.62&  27&  -14.05&  M&   2\\ 
   1ES1615+061&   1.5&  0.0380&   20.00& -11.10&  28&  -14.22&  M&   2\\ 
       IIIZw77&   1.2&  0.0342&   23.90& -12.22&  24&  -13.55&  M&  9\\ 
    ESO~141-G55&   1.2&  0.0360&   20.00& -10.57&  29&  -13.17&  M&  8\\ 
         POX~52&   1.0&  0.0218&   20.70& -12.11&  30&  -13.35&  M&  10\\ 
\hline		 	      			   
\end{tabular}
\end{center}

Column description: (1) Source name. (2) Spectroscopic classification
(1.0, 1.2 and 1.5 refer to Seyfert types). (3) Redshift. (4) Logarithm
of the cold X-ray absorbing column density. For those objects in which
no cold absorption is measured in excess of the Galactic value, the
column has been fixed to log$N_H$=20. (5) Logarithm of the observed
2-10 keV flux in erg cm$^{-2}$ s$^{-1}$. (6) Reference for the X-ray
data. (7) Logarithm of the [Ne~V]3426 flux in erg cm$^{-2}$
s$^{-1}$. (8) Aperture used to measure the [Ne~V] flux, defined
following Schmitt (1998): S (small), M (medium) and L (large)
correspond to apertures in the ranges 1-3'', 3-7'', and  $>$7'',
respectively. (9) Reference for the [Ne~V] data. \\
{\it X-ray references:} 
(1) \citealt{cappi06}; (2) \citealt{bianchi04};
(3) \citealt{blustin02}; (4) \citealt{bianchi09}; 
(5) \citealt{landi05}; (6) \citealt{iwa10}; (7) \citealt{shinozaki06};
(8) \citealt{vf07}; (9) \citealt{vaughan99}; (10) \citealt{gallo06};
(11) \citealt{ballo08}; (12) \citealt{landi07}; (13) \citealt{gallo05};
(14) \citealt{pico05}; (15) \citealt{grupe04}; (16) \citealt{grandi06};
(17) \citealt{hard07};
(18) \citealt{gliozzi07}; (19) \citealt{evans06}; (20) \citealt{sambruna07};
(21) \citealt{jime08}
(22) \citealt{dammando08}; (23) \citealt{guido00}; (24) this work;
 (25) \citealt{vaughan04};
(26) \citealt{vignali04}; (27) \citealt{ponti04}; (28) \citealt{guainazzi98};
(29) \citealt{gondoin03}; (30) \citealt{thornton08}. \\
{\it [Ne~V] references:}
(1) \citealt{anderson70}; (2) \citealt{mw88}; (3) \citealt{op85}; 
(4) \citealt{erkens97}; 
(5) \citealt{malkan86};
(6) \citealt{db88}; 
(7) \citealt{o76}; (8) \citealt{winkler92}; (9) \citealt{o81iii}; (10) \citealt{kunth87}.

\end{table*}

\begin{table*}
\caption{The local sample: Seyfert 1.8 to 2.0 and Narrow Line Radio Galaxies (NLRGs).}
\begin{center}
\begin{tabular}{lcccrcrcc}
\hline \hline
Name& Type& $z$& log$N_H$& $f_{2-10}$& $Ref_X$& $f_{NeV}$& A& $Ref_{NeV}$\\
(1)& (2)& (3)& (4)& (5)& (6)& (7)& (8)& (9)\\ 
\hline
       NGC~1068&  2.0&  0.0038&   25.00&  -11.34&   1& -11.94&   M&   1\\ 
       NGC~1275&  NLRG& 0.0175&   20.00&  -10.91&   2$^a$& -12.70&   L&   2\\ 
       NGC~1320&  2.0&  0.0089&   24.00&  -12.24&   3& -13.15&   S&   3\\ 
       NGC~1667&  2.0&  0.0152&   24.00&  -13.00&   4& -14.00&   L&   4\\ 
       NGC~2992&  1.9&  0.0077&   21.95&  -10.13&   5& -14.52&   S&  5\\ 
       NGC~3081&  2.0&  0.0079&   23.85&  -11.89&   6& -12.66&   L&   6\\ 
       NGC~3281&  2.0&  0.0106&   24.18&  -11.54&   7& -14.52&   S&   5\\ 
       NGC~3393&  2.0&  0.0125&   25.00&  -12.68&   8& -13.25&   L&   4\\ 
       NGC~4074&  2.0&  0.0224&   23.38&  -11.59&   3& -14.05&   S&  7\\ 
       NGC~4507&  2.0&  0.0118&   23.63&  -10.89&   9& -13.26&   L&   6\\ 
       NGC~5135&  2.0&  0.0136&   24.00&  -12.80&   10& -13.54&   L&   4\\ 
       NGC~5506&  1.9&  0.0062&   22.50&  -10.10&   11& -13.96&   M&  8\\ 
       NGC~5643&  2.0&  0.0040&   23.80&  -12.08&   12& -13.33&   M&  8\\ 
       NGC~5728&  2.0&  0.0093&   24.30&  -11.89&   13& -13.66&   L&   4\\ 
       NGC~7130&  2.0&  0.0161&   24.00&  -12.29&   14& -14.00&   S&  9\\ 
       NGC~7314&  1.9&  0.0048&   22.02&  -10.39&   15& -14.30&   M&  8\\ 
       NGC~7582&  2.0&  0.0052&   24.00&  -11.64&   16& -12.48&   M&  10\\ 
       NGC~7674&  2.0&  0.0289&   25.00&  -12.12&   17& -13.40&   L&  11\\ 
          Mrk~1&  2.0&  0.0159&   24.00&  -12.89&   18& -13.49&   L&   11\\ 
         Mrk~34&  2.0&  0.0505&   24.00&  -12.74&   19& -13.55&   L&   11\\ 
         Mrk~78&  2.0&  0.0371&   23.76&  -12.23&   3& -13.85&   L&   11\\ 
        Mrk~348&  2.0&  0.0150&   23.13&  -10.56&   20& -13.64&   L&   11\\ 
       Mrk~463E&  2.0&  0.0500&   23.85&  -12.39&   21& -13.89&   L&   11\\ 
        Mrk~477&  2.0&  0.0377&   23.38&  -11.92&   22& -13.14&   L&   4\\ 
        Mrk~573&  2.0&  0.0171&   24.00&  -12.92&   23& -13.00&   S&   3\\ 
        Mrk~609&  1.8&  0.0344&   20.00&  -11.83&   24& -14.40&   S&  12\\ 
        Mrk~612&  2.0&  0.0204&   23.81&  -12.44&   18& -13.66&   L&   11\\ 
          3C~33&  NLRG& 0.0597&   23.59&  -11.13&   25& -14.70&   M&  1\\ 
   MCG-5-23-16&  2.0&  0.0085&   22.15&  -10.10&   26& -13.89&   S&   5\\ 
        WAS~49b&  2.0&  0.0630&   22.80&  -12.20&   20& -14.30&   S&  13\\ 
       Cygnus~A&  NLRG& 0.0560&   23.23&  -10.89&   25& -14.40&   M&   14\\ 
        IC~3639&  2.0&  0.0109&   24.00&  -13.10&   18& -13.60&   L&   4\\ 
   Tol~0109-383&  2.0&  0.0117&   24.30&  -11.80&   27& -13.62&   S&   15\\ 
\hline		 	      			   
\end{tabular}
\end{center}

Column description: (1) Source name. (2) Spectroscopic classification (1.8, 1.9 and 2.0 refer to Seyfert types). (3)
Redshift. (4) Logarithm of the cold X-ray absorbing column
density. For those objects in which no cold absorption is measured in excess of the Galactic value, 
the column has been fixed to log$N_H$=20. (5) Observed 2-10 keV flux in units of erg
cm$^{-2}$ s$^{-1}$. (6) Reference for the X-ray data. (7) [Ne~V]3426
flux in units of erg cm$^{-2}$ s$^{-1}$. (8) Aperture used to measure the [Ne~V]
flux, defined following Schmitt (1998): S (small), M (medium) and L (large) correspond 
to apertures in the ranges 1-3'', 3-7'', and $>$7'', respectively. (9) Reference for the [Ne~V] data.\\
{\it X-ray references:} 
(1) \citealt{cappi06}; (2) \citealt{panessa06}; (3) this work; (4) \citealt{bianchi05};
(5) \citealt{gilli00}; (6) \citealt{bassani99}; (7) \citealt{vignali02}; (8) \citealt{maio98};
(9) \citealt{matt04}; (10) \citealt{guainazzi05fabian}; (11) \citealt{bianchi03};
(12) \citealt{guainazzi04}; (13) \citealt{comastri10}; (14) \citealt{guido99}; (15) \citealt{dg05};
(16) \citealt{pico07ngc}; (17) \citealt{pino98}; (18) \citealt{guainazzi05matt}; (19) \citealt{greenhill08}; (20) \citealt{awaki00}; 
(21) \citealt{bianchi08}; (22) \citealt{levenson01}; (23) \citealt{shu07}; 
(24) \citealt{gallo06lehmann}; (25) \citealt{evans06}; (26) \citealt{balestra04};
(27) \citealt{iwa01tol}\\
Notes: $^a$: The column density of NGC 1275, which is at the center within the Perseus cluster, is difficult to measure: \citet{evans06}
do not find absorption in excess of the Galactic one based on \chandra\ data.

{\it [Ne~V] references:} 
(1) \citealt{koski78}; (2) \citealt{anderson70}; (3) \citealt{erkens97}; 
(4) This work (based on the catalog of UV-optical spectra of nearby galaxies by Storchi-Bergmann et al., see text);
(5) \citealt{db88}; (6) \citealt{db86}; (7) \citealt{so81};
(8) \citealt{mw88}; (9) \citealt{sf90}; (10) \citealt{ward80}; (11) \citealt{malkan86}; (12) \citealt{o81};
(13) \citealt{moran92}; (14) \citealt{om75}; (15) \citealt{fs83}.

\end{table*}

\begin{figure*}[t]
\begin{center}
\includegraphics[keepaspectratio=false, height=7cm, width=4.0cm, angle=270]{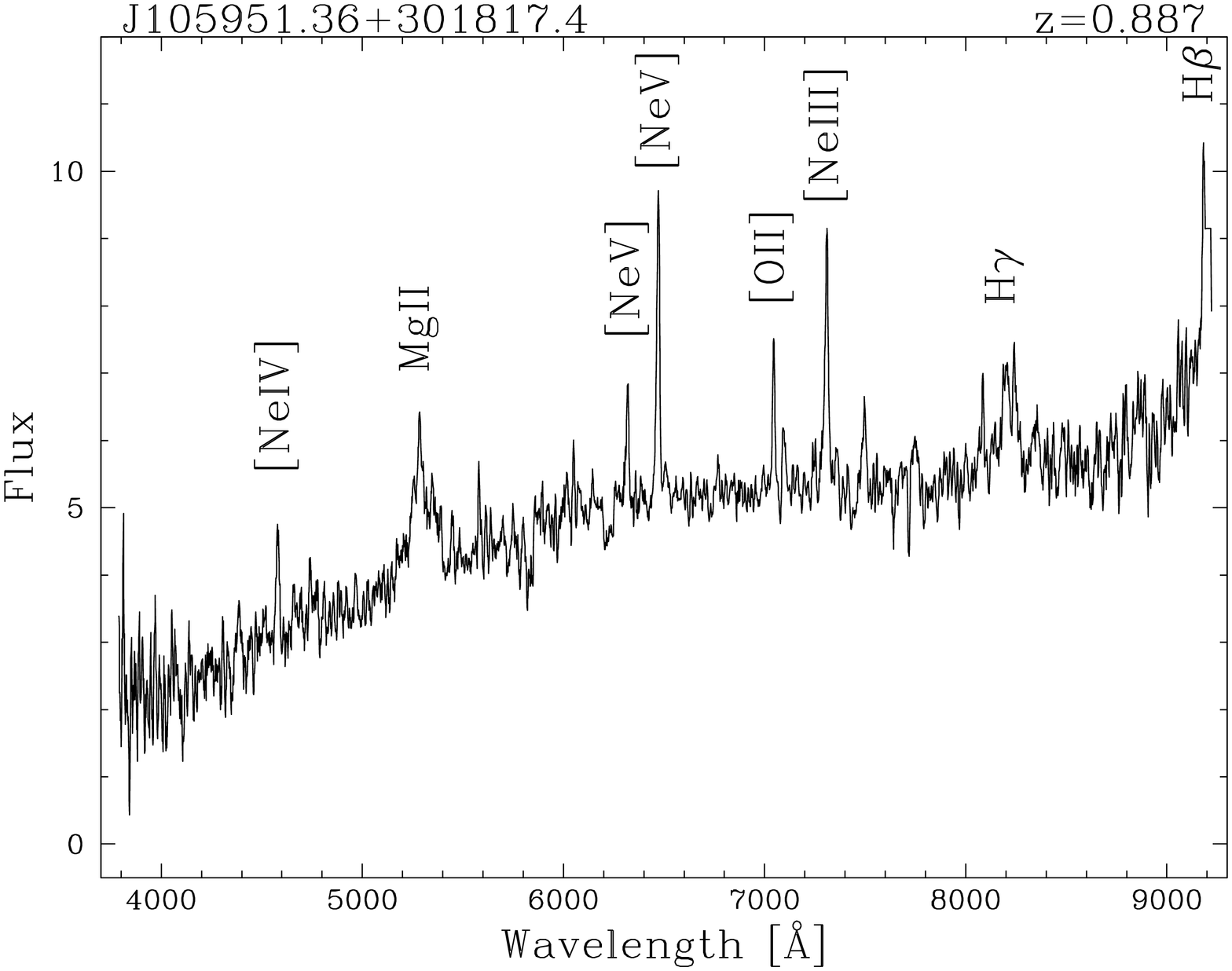}
\includegraphics[keepaspectratio=false, height=8cm, width=4.0cm, angle=270]{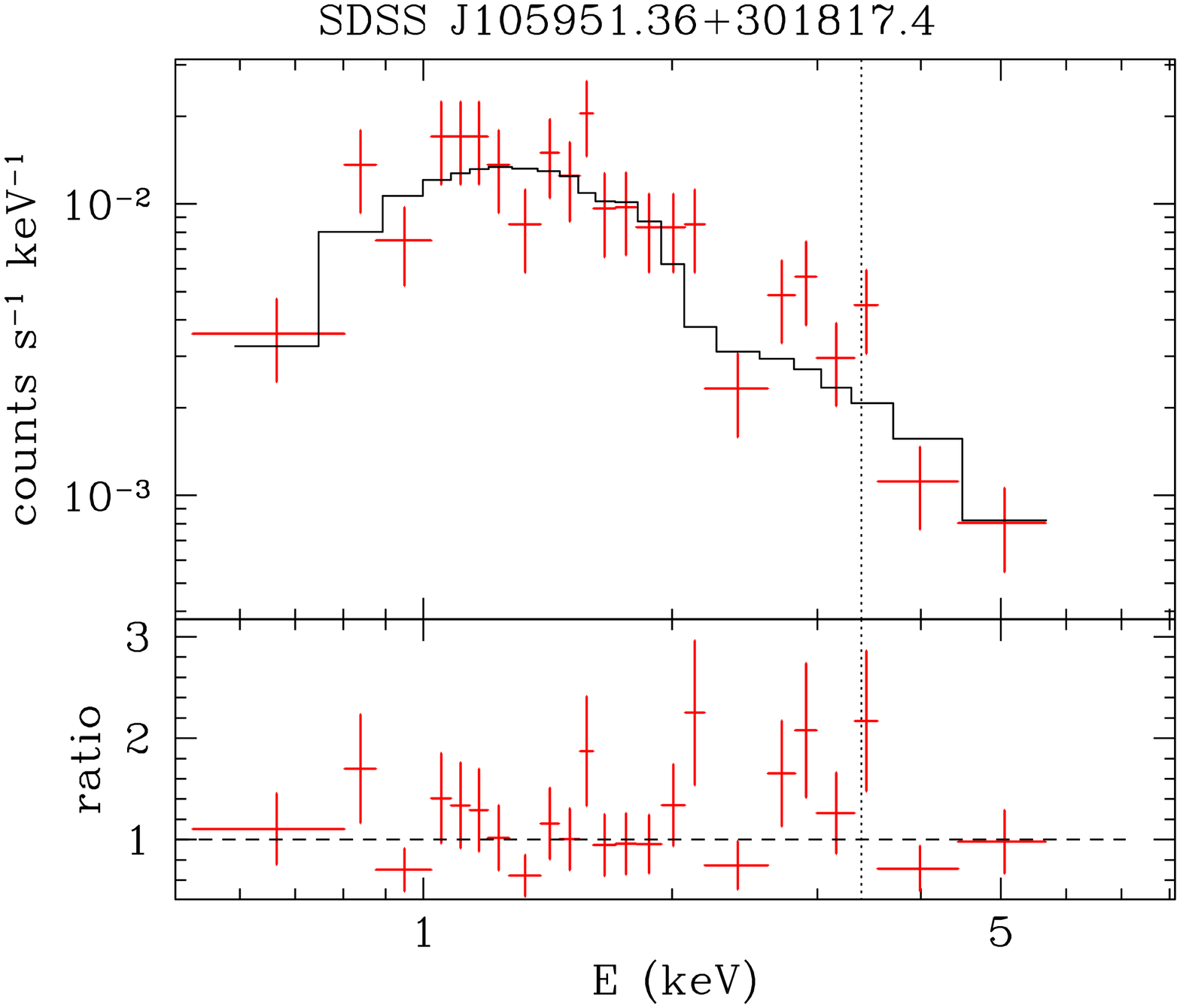}
\includegraphics[keepaspectratio=false, height=7cm, width=4.0cm, angle=270]{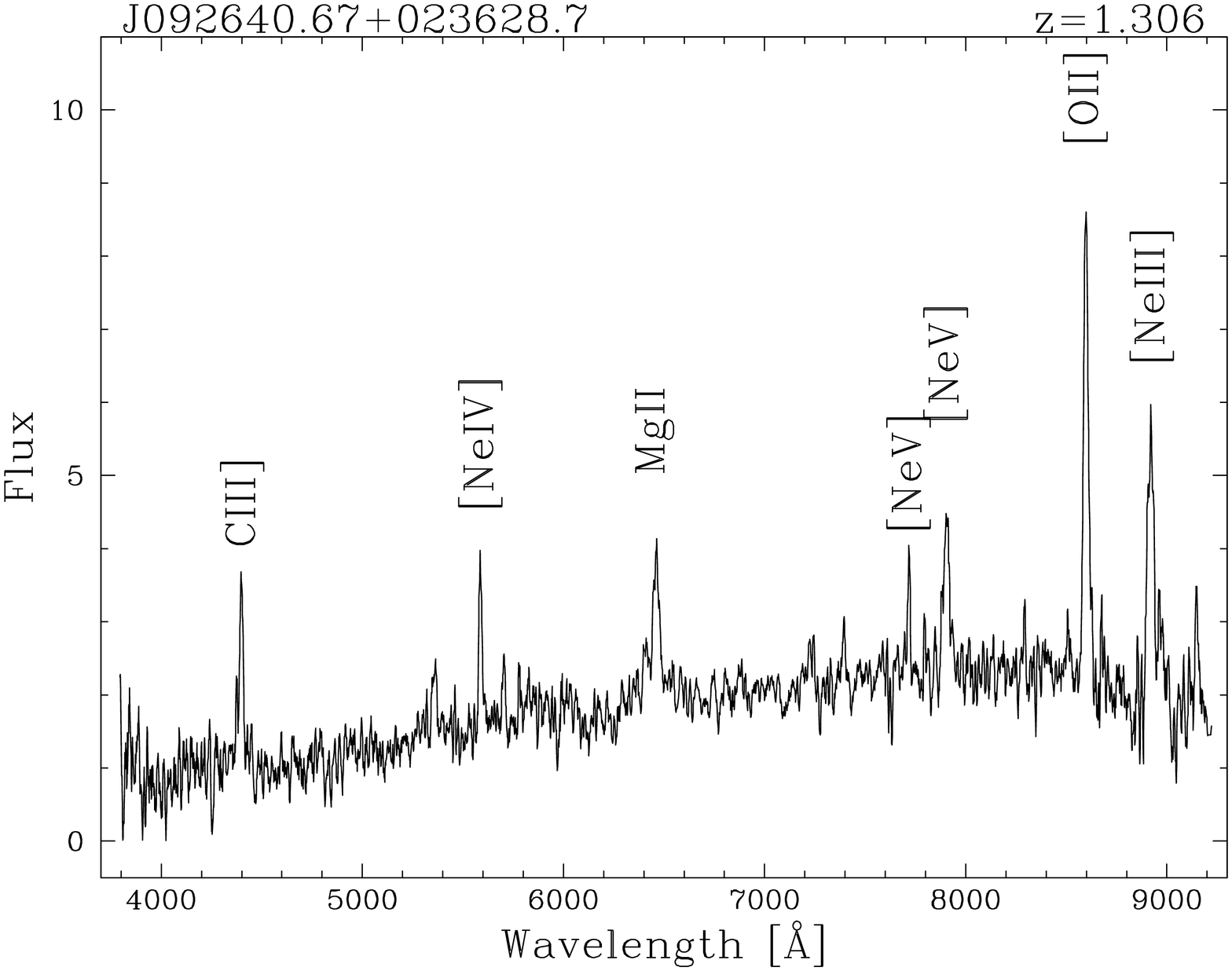}
\includegraphics[keepaspectratio=false, height=8cm, width=4.0cm, angle=270]{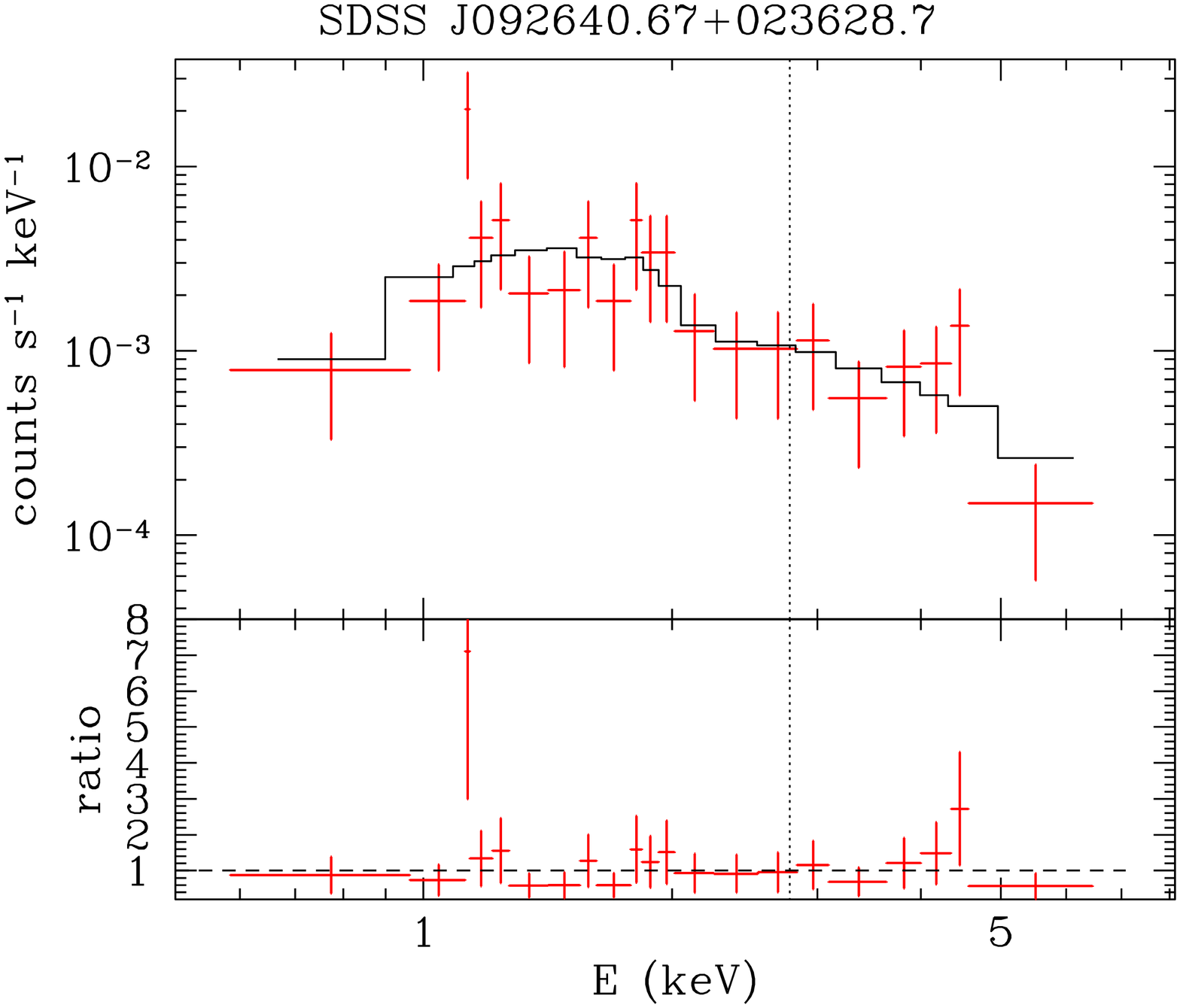}
\includegraphics[keepaspectratio=false, height=7cm, width=4.0cm, angle=270]{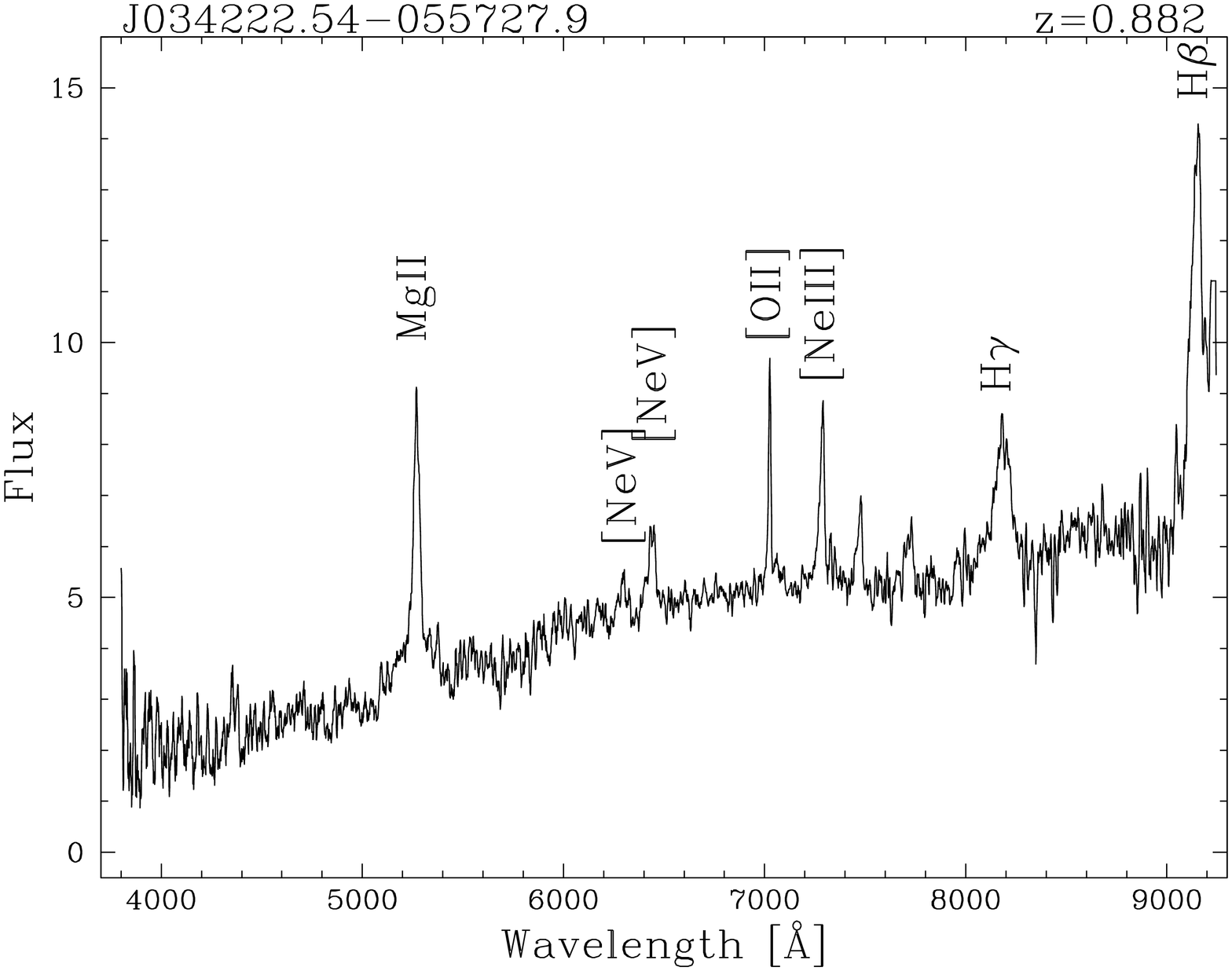}
\includegraphics[keepaspectratio=false, height=8cm, width=4.0cm, angle=270]{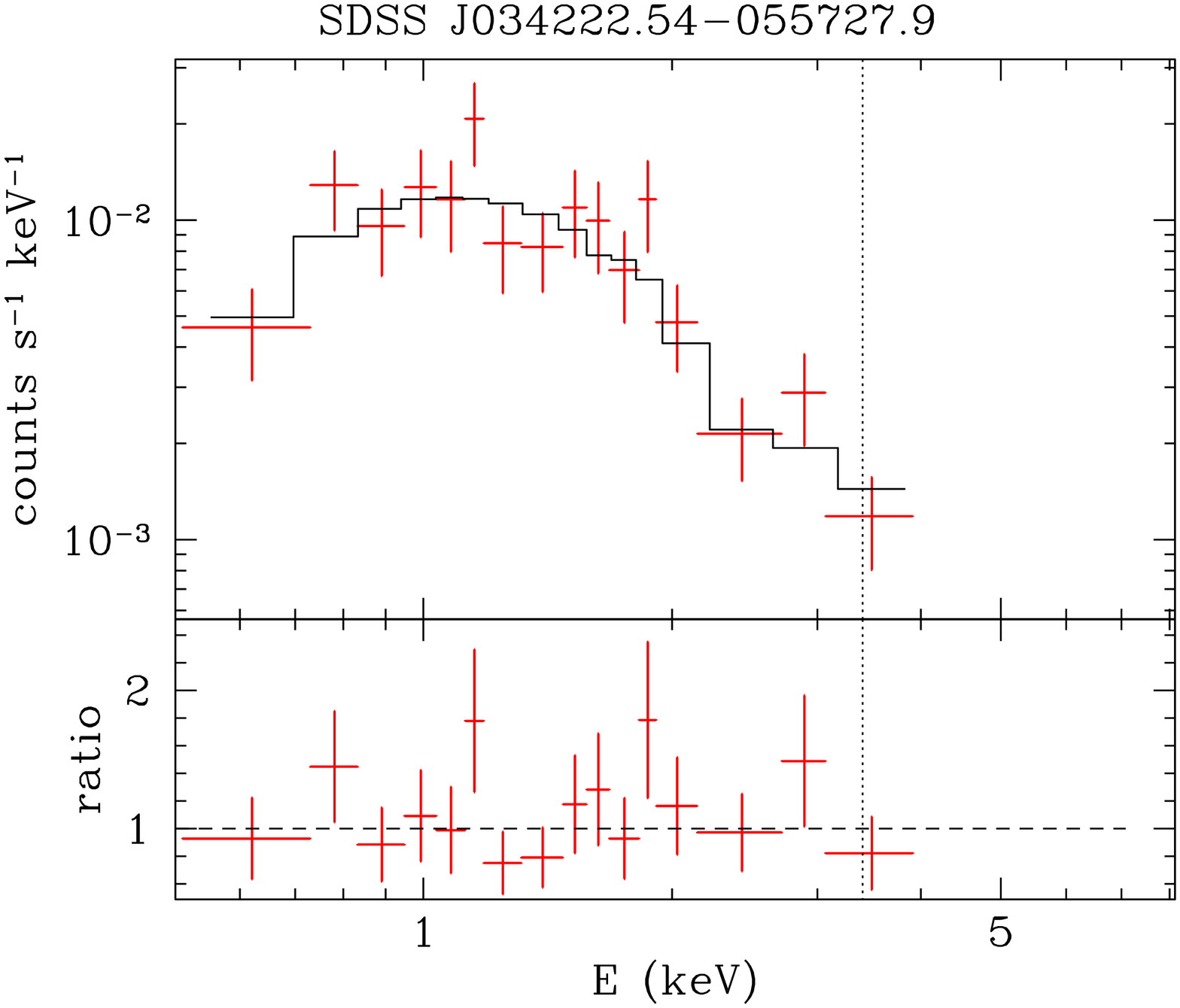}
\includegraphics[keepaspectratio=false, height=7cm, width=4.0cm, angle=270]{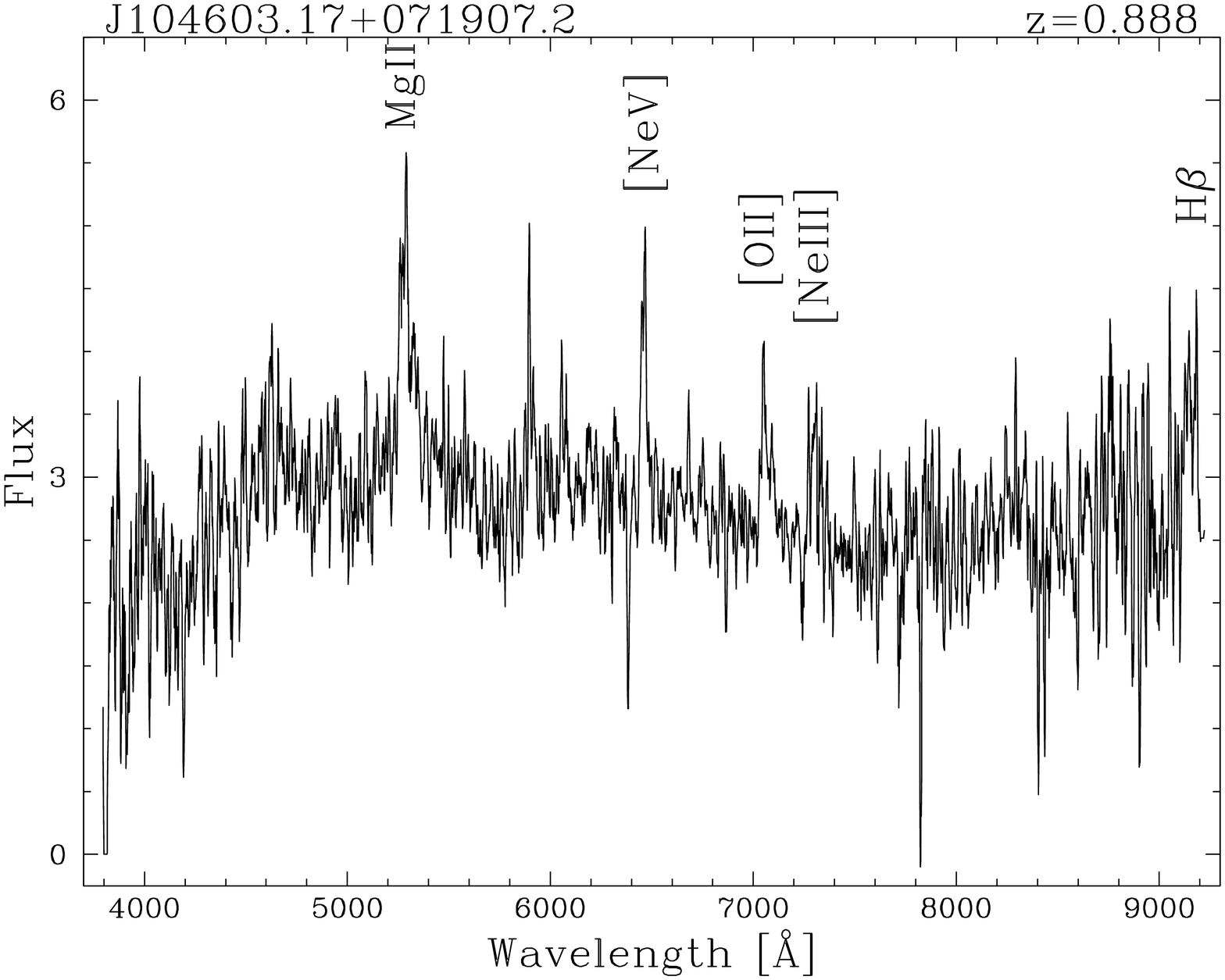}
\includegraphics[keepaspectratio=false, height=8cm, width=4.0cm, angle=270]{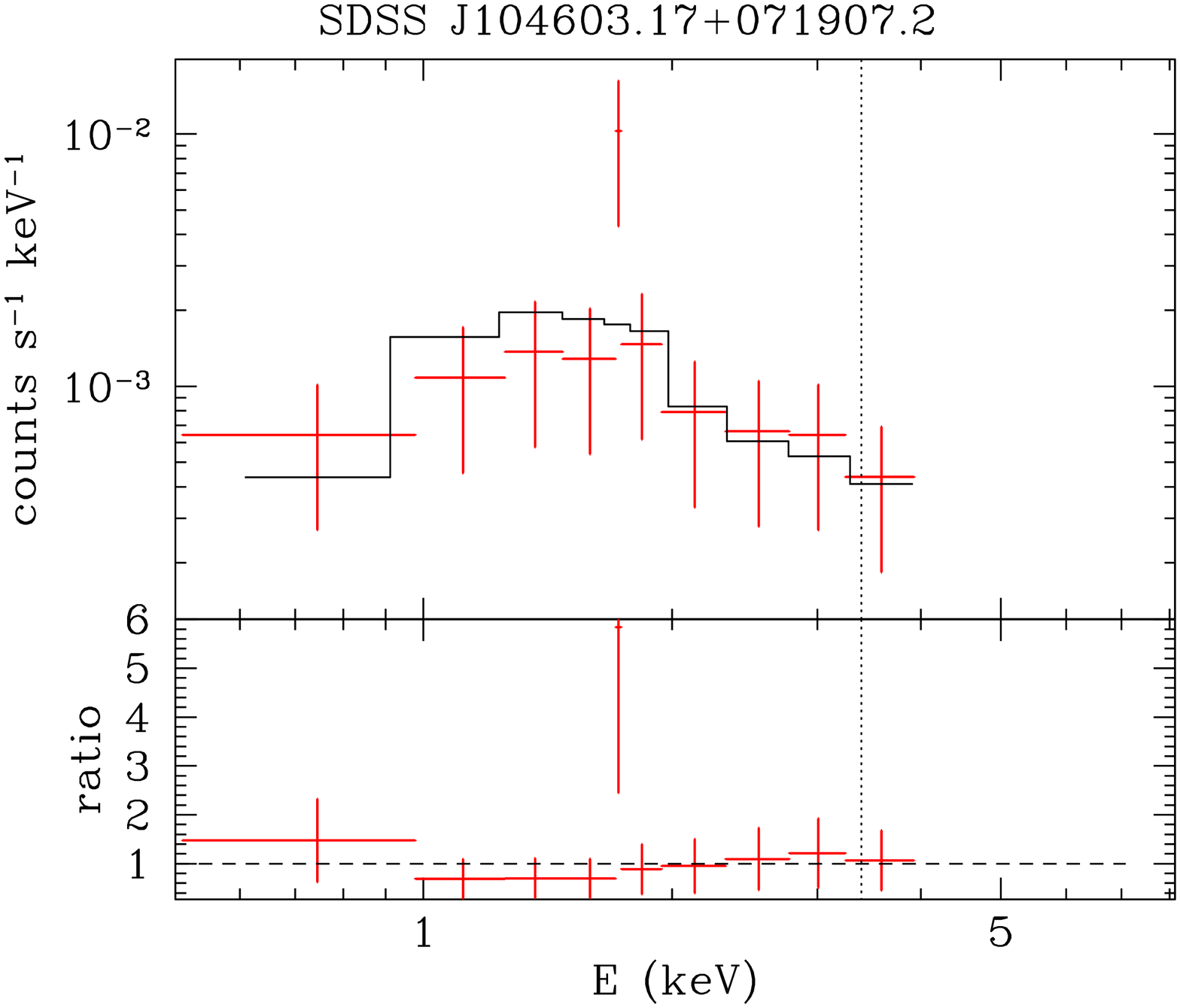}
\includegraphics[keepaspectratio=false, height=7cm, width=4.0cm, angle=270]{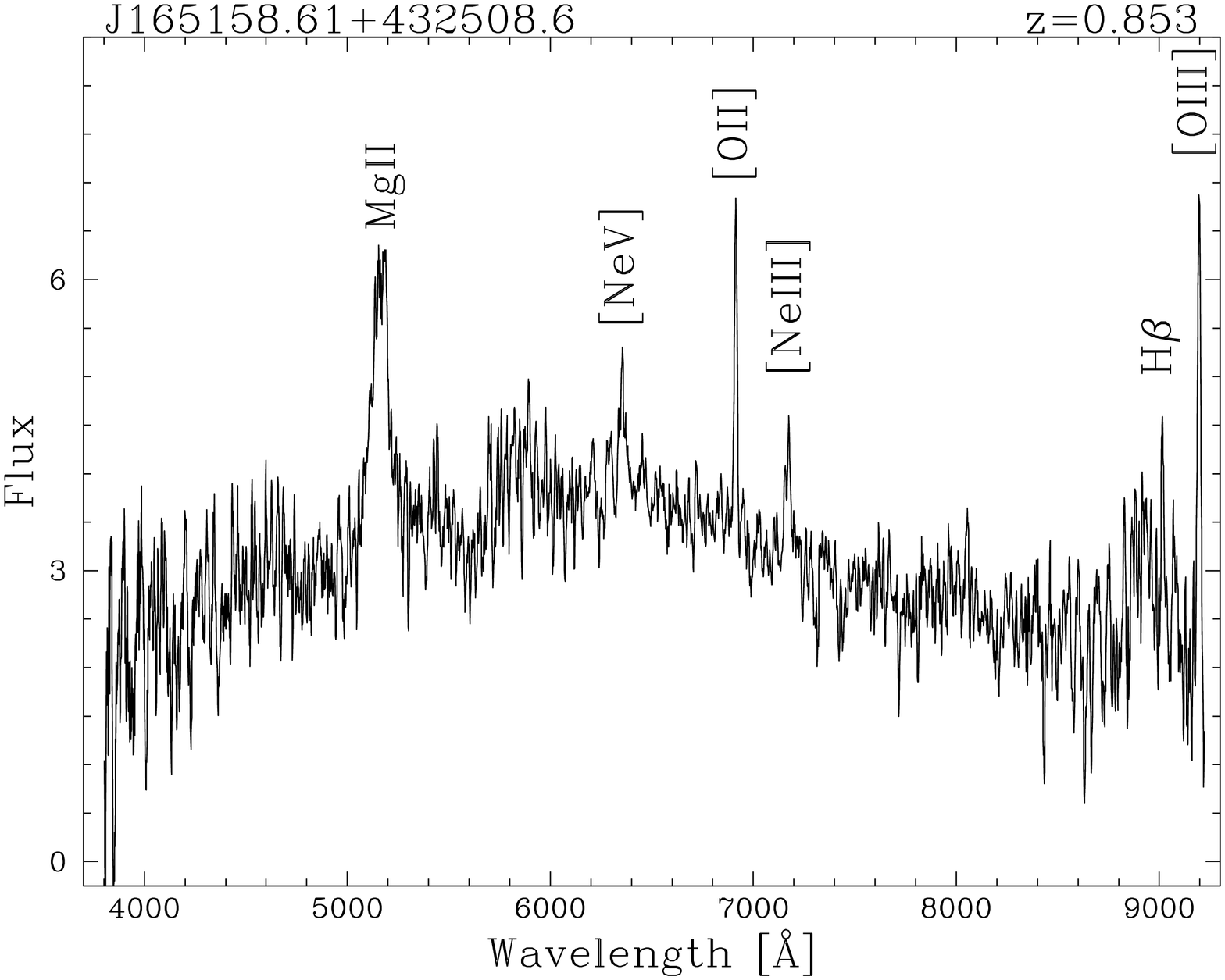}
\includegraphics[keepaspectratio=false, height=8cm, width=4.0cm, angle=270]{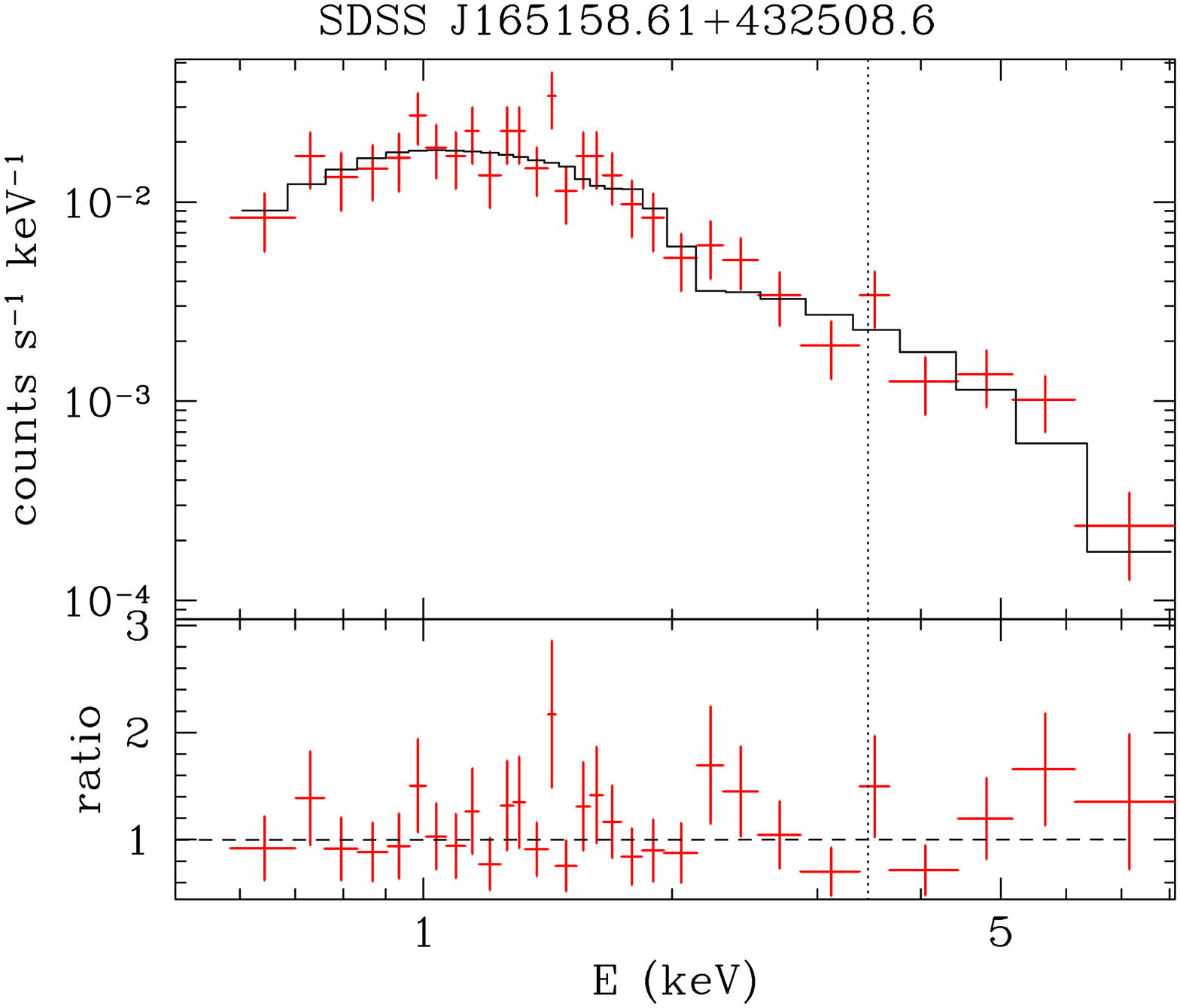}
\end{center}
\caption{The SDSS optical spectra (left) and \chandra\ X-ray spectra (right) of the 9
  obscured QSOs at $z\sim1$ presented in this work. Both optical and X-ray spectra are plotted in the
observed frame. Flux units for the SDSS spectra are $10^{-17}$\cgs $\AA^{-1}$. The dotted vertical lines in the right panels mark
the expected position in the observed frame of the 6.4 keV Fe K$\alpha$ line. The archival object SDSS
  J085600 is not detected in the X-rays: we show a 30''$\times30"$ \chandra\ image in the 0.5-8 keV band around the source position
(marked with a 1.5'' radius circle).}
\label{oxspec}
\end{figure*}

\begin{figure*}[t]
\begin{center}
\includegraphics[keepaspectratio=false, height=7cm, width=4.0cm, angle=270]{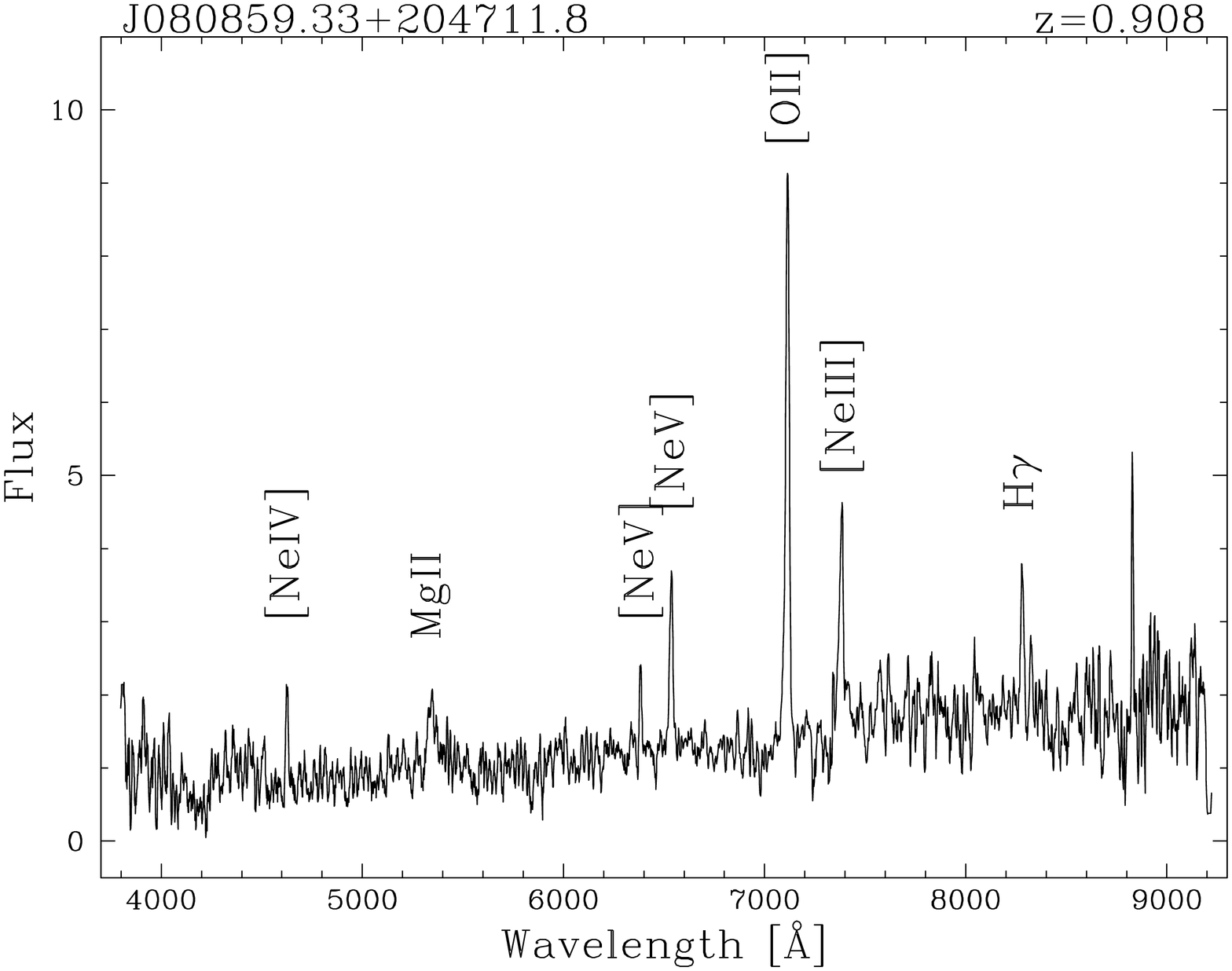}
\includegraphics[keepaspectratio=false, height=8cm, width=4.0cm, angle=270]{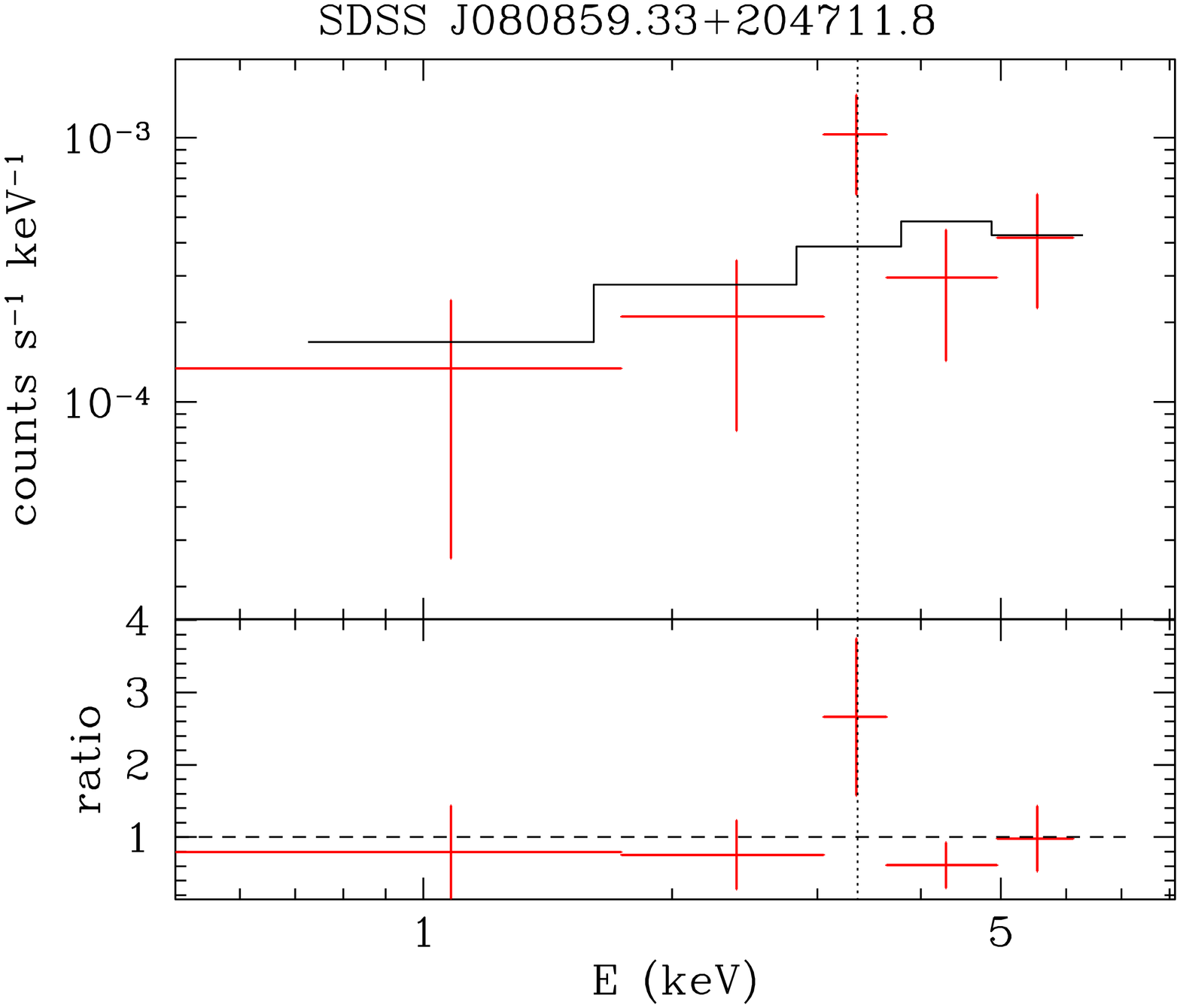}
\includegraphics[keepaspectratio=false, height=7cm, width=4.0cm, angle=270]{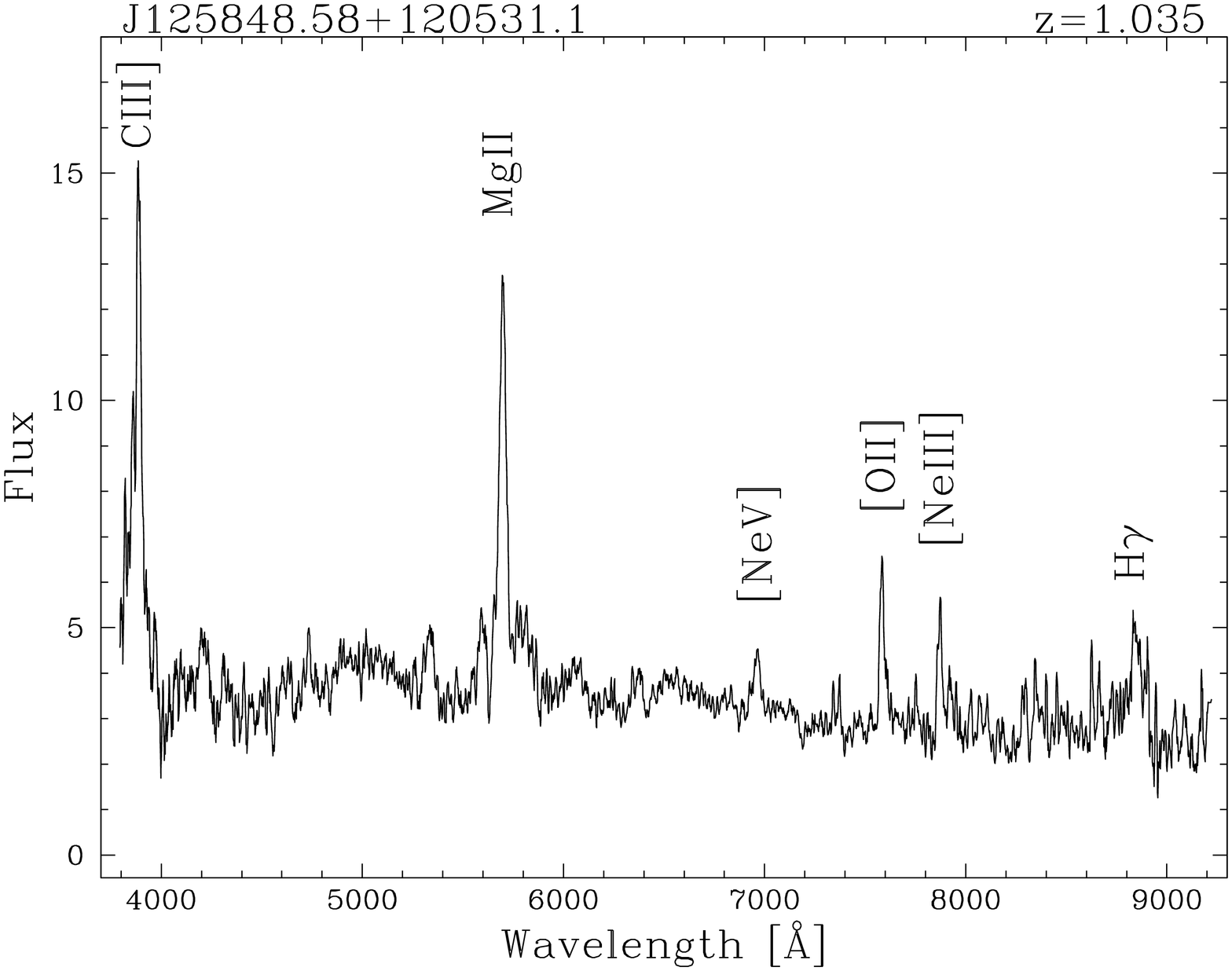}
\includegraphics[keepaspectratio=false, height=8cm, width=4.0cm, angle=270]{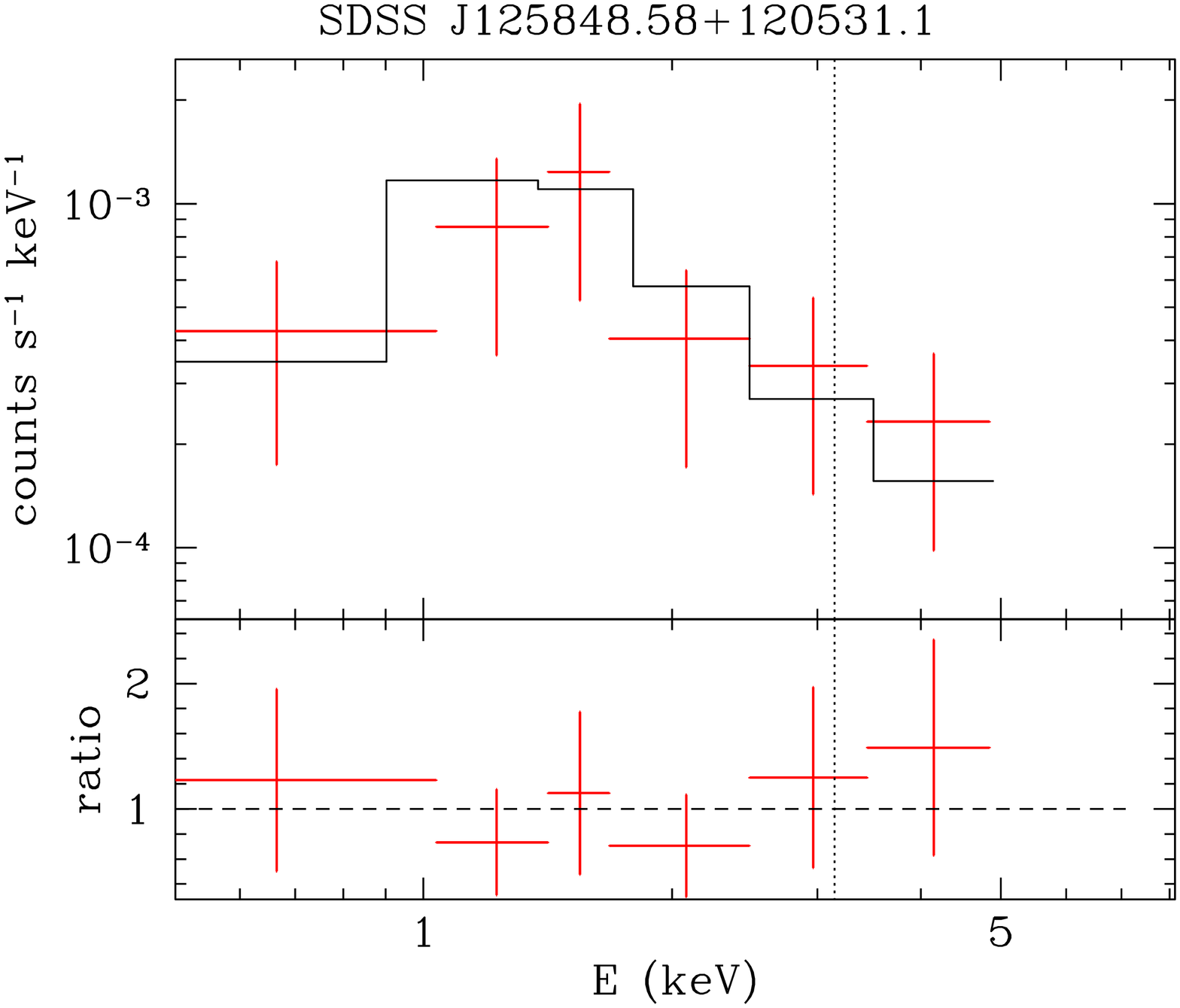}
\includegraphics[keepaspectratio=false, height=7cm, width=4.0cm, angle=270]{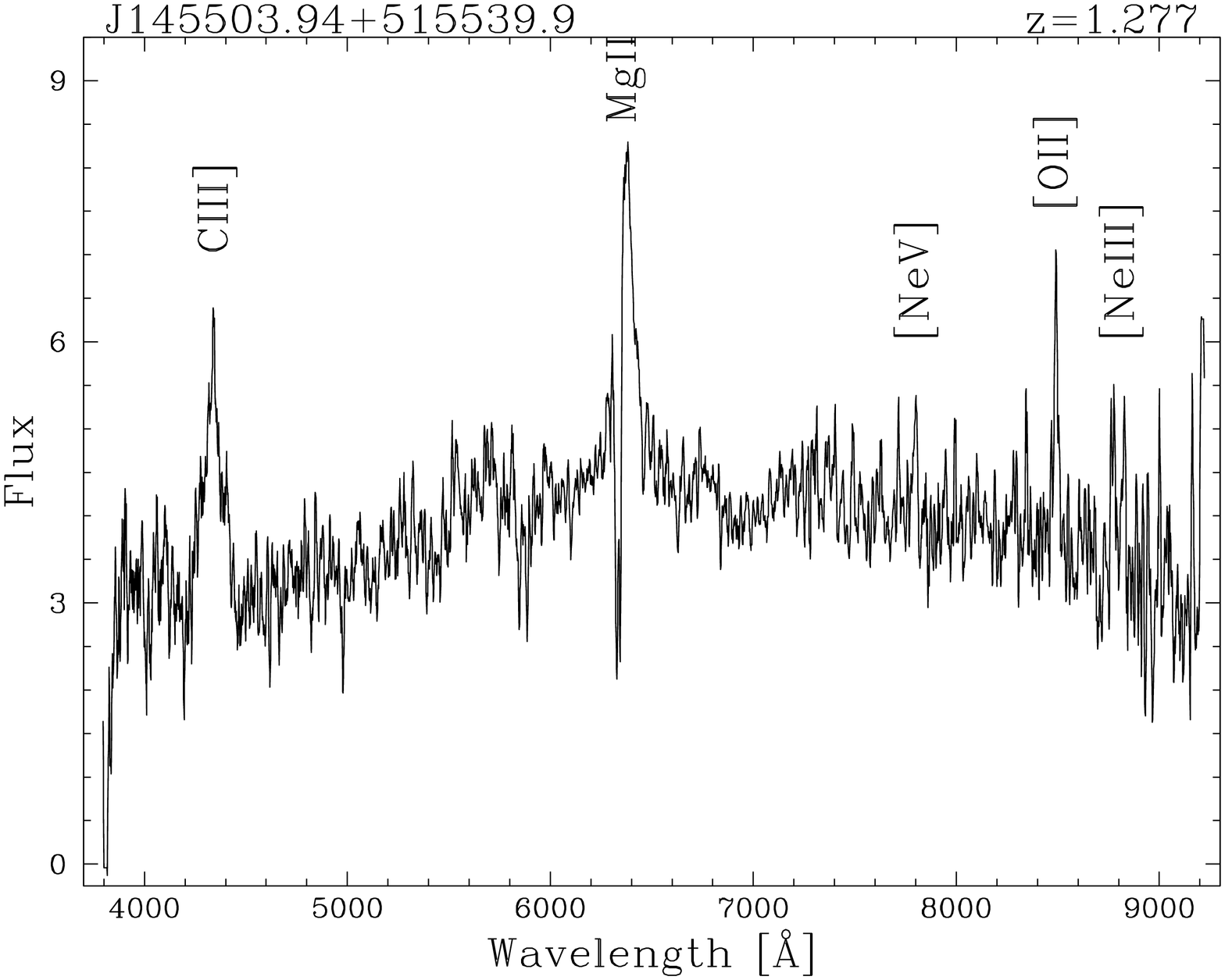}
\includegraphics[keepaspectratio=false, height=8cm, width=4.0cm, angle=270]{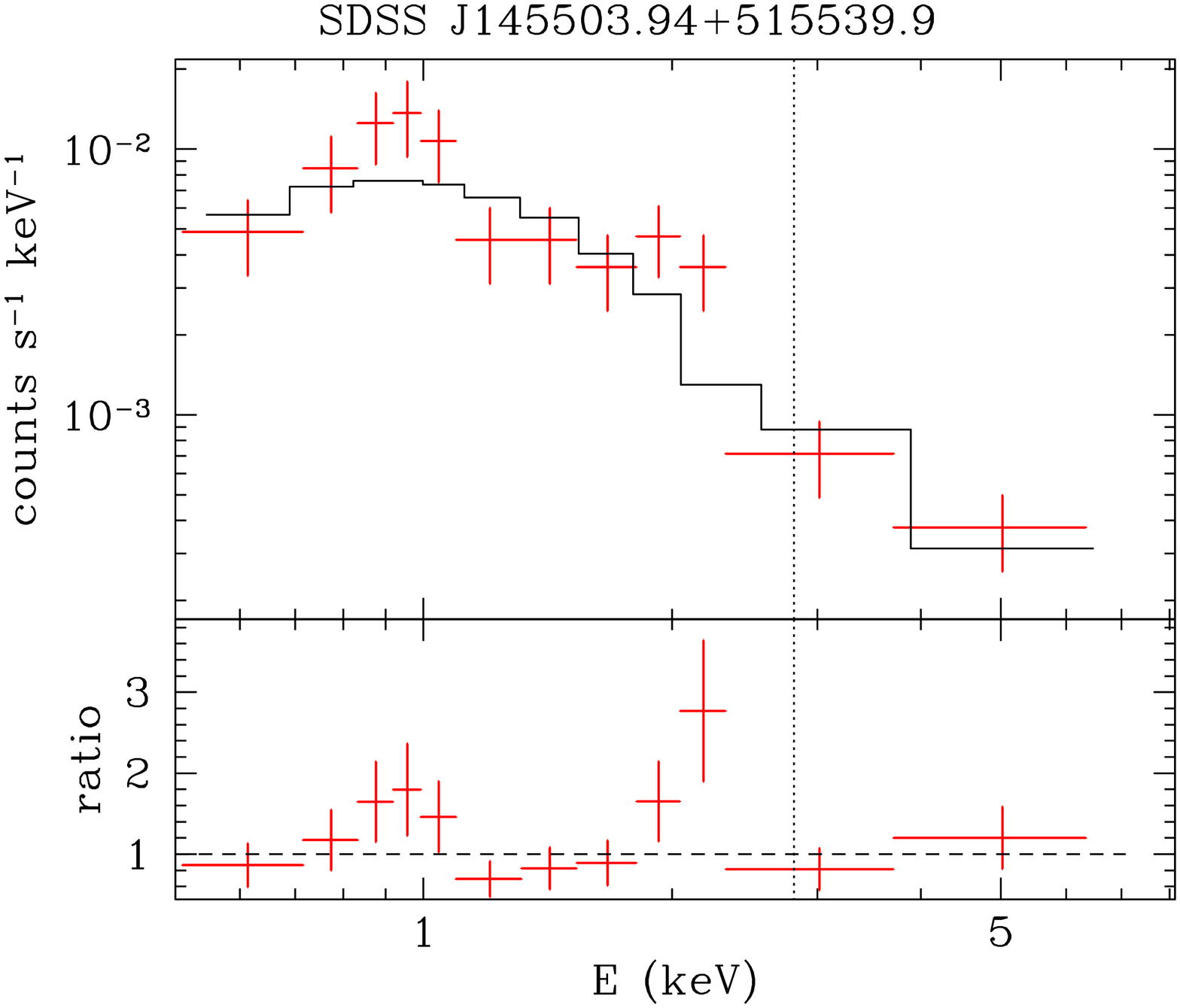}
\includegraphics[keepaspectratio=false, height=7cm, width=4.0cm, angle=270]{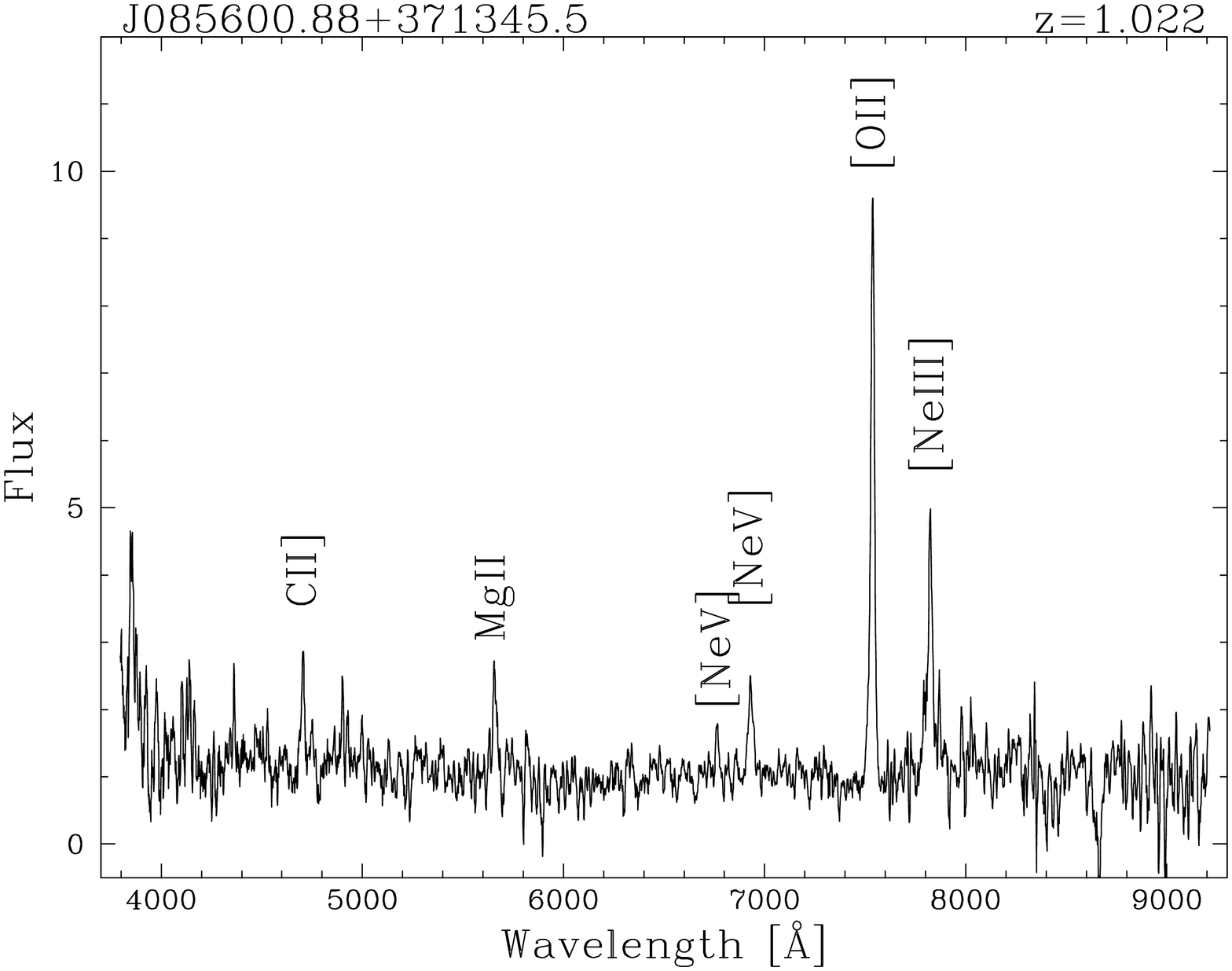}
\includegraphics[keepaspectratio=false, height=8cm, width=4.0cm, angle=270]{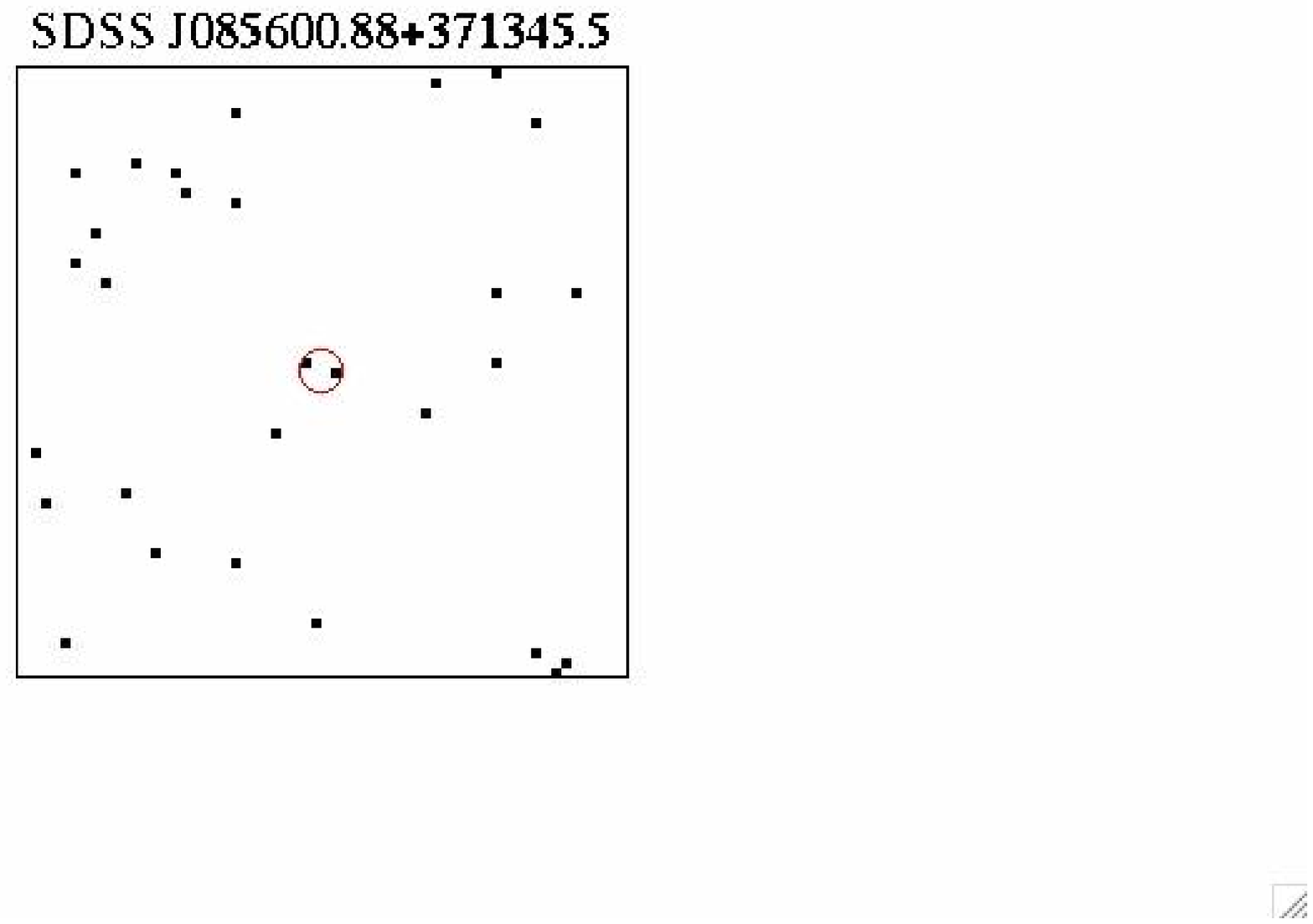}
\end{center}
{\bf Fig.\ref{oxspec}.} - continued.
\end{figure*}

\bibliographystyle{aa} 
\bibliography{../biblio} 

\begin{thebibliography}{136}
\expandafter\ifx\csname natexlab\endcsname\relax\def\natexlab#1{#1}\fi

\bibitem[{{Alexander} {et~al.}(2008){Alexander}, {Chary}, {Pope}, {Bauer},
  {Brandt}, {Daddi}, {Dickinson}, {Elbaz}, \& {Reddy}}]{alex08}
{Alexander}, D.~M., {Chary}, R., {Pope}, A., {et~al.} 2008, \apj, 687, 835

\bibitem[{{Alexander} {et~al.}(2005){Alexander}, {Smail}, {Bauer}, {Chapman},
  {Blain}, {Brandt}, \& {Ivison}}]{alex05}
{Alexander}, D.~M., {Smail}, I., {Bauer}, F.~E., {et~al.} 2005, \nat, 434, 738

\bibitem[{{Anderson}(1970)}]{anderson70}
{Anderson}, K.~S. 1970, \apj, 162, 743

\bibitem[{{Awaki} {et~al.}(2000){Awaki}, {Ueno}, {Taniguchi}, \&
  {Weaver}}]{awaki00}
{Awaki}, H., {Ueno}, S., {Taniguchi}, Y., \& {Weaver}, K.~A. 2000, \apj, 542,
  175

\bibitem[{{Balestra} {et~al.}(2004){Balestra}, {Bianchi}, \&
  {Matt}}]{balestra04}
{Balestra}, I., {Bianchi}, S., \& {Matt}, G. 2004, \aap, 415, 437

\bibitem[{{Ballantyne} {et~al.}(2006){Ballantyne}, {Everett}, \&
  {Murray}}]{balla06}
{Ballantyne}, D.~R., {Everett}, J.~E., \& {Murray}, N. 2006, \apj, 639, 740

\bibitem[{{Ballo} {et~al.}(2008){Ballo}, {Giustini}, {Schartel}, {Cappi},
  {Jim{\'e}nez-Bail{\'o}n}, {Piconcelli}, {Santos-Lle{\'o}}, \&
  {Vignali}}]{ballo08}
{Ballo}, L., {Giustini}, M., {Schartel}, N., {et~al.} 2008, \aap, 483, 137

\bibitem[{{Barth} {et~al.}(2004){Barth}, {Ho}, {Rutledge}, \&
  {Sargent}}]{barth04}
{Barth}, A.~J., {Ho}, L.~C., {Rutledge}, R.~E., \& {Sargent}, W.~L.~W. 2004,
  \apj, 607, 90

\bibitem[{{Bassani} {et~al.}(1999){Bassani}, {Dadina}, {Maiolino}, {Salvati},
  {Risaliti}, {della Ceca}, {Matt}, \& {Zamorani}}]{bassani99}
{Bassani}, L., {Dadina}, M., {Maiolino}, R., {et~al.} 1999, \apjs, 121, 473

\bibitem[{{Bianchi} {et~al.}(2003){Bianchi}, {Balestra}, {Matt}, {Guainazzi},
  \& {Perola}}]{bianchi03}
{Bianchi}, S., {Balestra}, I., {Matt}, G., {Guainazzi}, M., \& {Perola}, G.~C.
  2003, \aap, 402, 141

\bibitem[{{Bianchi} {et~al.}(2008){Bianchi}, {Chiaberge}, {Piconcelli},
  {Guainazzi}, \& {Matt}}]{bianchi08}
{Bianchi}, S., {Chiaberge}, M., {Piconcelli}, E., {Guainazzi}, M., \& {Matt},
  G. 2008, \mnras, 386, 105

\bibitem[{{Bianchi} {et~al.}(2006){Bianchi}, {Guainazzi}, \&
  {Chiaberge}}]{bianchi06}
{Bianchi}, S., {Guainazzi}, M., \& {Chiaberge}, M. 2006, \aap, 448, 499

\bibitem[{{Bianchi} {et~al.}(2005){Bianchi}, {Guainazzi}, {Matt}, {Chiaberge},
  {Iwasawa}, {Fiore}, \& {Maiolino}}]{bianchi05}
{Bianchi}, S., {Guainazzi}, M., {Matt}, G., {et~al.} 2005, \aap, 442, 185

\bibitem[{{Bianchi} {et~al.}(2009){Bianchi}, {Guainazzi}, {Matt}, {Fonseca
  Bonilla}, \& {Ponti}}]{bianchi09}
{Bianchi}, S., {Guainazzi}, M., {Matt}, G., {Fonseca Bonilla}, N., \& {Ponti},
  G. 2009, \aap, 495, 421

\bibitem[{{Bianchi} {et~al.}(2004){Bianchi}, {Matt}, {Balestra}, {Guainazzi},
  \& {Perola}}]{bianchi04}
{Bianchi}, S., {Matt}, G., {Balestra}, I., {Guainazzi}, M., \& {Perola}, G.~C.
  2004, \aap, 422, 65

\bibitem[{{Blustin} {et~al.}(2002){Blustin}, {Branduardi-Raymont}, {Behar},
  {Kaastra}, {Kahn}, {Page}, {Sako}, \& {Steenbrugge}}]{blustin02}
{Blustin}, A.~J., {Branduardi-Raymont}, G., {Behar}, E., {et~al.} 2002, \aap,
  392, 453

\bibitem[{{Bongiorno} {et~al.}(2009){Bongiorno}, {Mignoli}, {Zamorani},
  {Lamareille}, {Lanzuisi}, {Miyaji}, {Bolzonella}, {Carollo}, {Contini},
  {Kneib}, {Le Fevre}, {Lilly}, {Mainieri}, {Renzini}, {Scodeggio}, {Bardelli},
  {Brusa}, {Caputi}, {Civano}, {Coppa}, {Cucciati}, {de la Torre}, {de Ravel},
  {Franzetti}, {Garilli}, {Halliday}, {Hasinger}, {Koekemoer}, {Iovino},
  {Kampczyk}, {Knobel}, {Kovac}, {Le Borgne}, {Le Brun}, {Maier}, {Merloni},
  {Nair}, {Pello}, {Peng}, {Perez Montero}, {Ricciardelli}, {Salvato},
  {Silverman}, {Tanaka}, {Tasca}, {Tresse}, {Vergani}, {Zucca}, {Abbas},
  {Bottini}, {Cappi}, {Cassata}, {Cimatti}, {Guzzo}, {Leauthaud}, {Maccagni},
  {Marinoni}, {McCracken}, {Memeo}, {Meneux}, {Oesch}, {Porciani}, {Pozzetti},
  \& {Scaramella}}]{bongiorno09}
{Bongiorno}, A., {Mignoli}, M., {Zamorani}, G., {et~al.} 2009, ArXiv e-prints

\bibitem[{{Brandt} \& {Hasinger}(2005)}]{bh05}
{Brandt}, W.~N. \& {Hasinger}, G. 2005, \araa, 43, 827

\bibitem[{{Cappi} {et~al.}(2006){Cappi}, {Panessa}, {Bassani}, {Dadina},
  {Dicocco}, {Comastri}, {della Ceca}, {Filippenko}, {Gianotti}, {Ho},
  {Malaguti}, {Mulchaey}, {Palumbo}, {Piconcelli}, {Sargent}, {Stephen},
  {Trifoglio}, \& {Weaver}}]{cappi06}
{Cappi}, M., {Panessa}, F., {Bassani}, L., {et~al.} 2006, \aap, 446, 459

\bibitem[{{Cash}(1979)}]{cash79}
{Cash}, W. 1979, \apj, 228, 939

\bibitem[{{Colless} {et~al.}(2001){Colless}, {Dalton}, {Maddox}, {Sutherland},
  {Norberg}, {Cole}, {Bland-Hawthorn}, {Bridges}, {Cannon}, {Collins}, {Couch},
  {Cross}, {Deeley}, {De Propris}, {Driver}, {Efstathiou}, {Ellis}, {Frenk},
  {Glazebrook}, {Jackson}, {Lahav}, {Lewis}, {Lumsden}, {Madgwick}, {Peacock},
  {Peterson}, {Price}, {Seaborne}, \& {Taylor}}]{colless01}
{Colless}, M., {Dalton}, G., {Maddox}, S., {et~al.} 2001, \mnras, 328, 1039

\bibitem[{{Comastri}(2004)}]{c04}
{Comastri}, A. 2004, in Astrophysics and Space Science Library, Vol. 308,
  Supermassive Black Holes in the Distant Universe, ed. {A.~J.~Barger}, 245--+

\bibitem[{{Comastri} {et~al.}(2010){Comastri}, {Iwasawa}, {Gilli}, {Vignali},
  {Ranalli}, {Matt}, \& {Fiore}}]{comastri10}
{Comastri}, A., {Iwasawa}, K., {Gilli}, R., {et~al.} 2010, \apj, submitted

\bibitem[{{Corbett} {et~al.}(2003){Corbett}, {Croom}, {Boyle}, {Netzer},
  {Miller}, {Outram}, {Shanks}, {Smith}, \& {Rhook}}]{corbett03}
{Corbett}, E.~A., {Croom}, S.~M., {Boyle}, B.~J., {et~al.} 2003, \mnras, 343,
  705

\bibitem[{{Daddi} {et~al.}(2007){Daddi}, {Alexander}, {Dickinson}, {Gilli},
  {Renzini}, {Elbaz}, {Cimatti}, {Chary}, {Frayer}, {Bauer}, {Brandt},
  {Giavalisco}, {Grogin}, {Huynh}, {Kurk}, {Mignoli}, {Morrison}, {Pope}, \&
  {Ravindranath}}]{daddi07}
{Daddi}, E., {Alexander}, D.~M., {Dickinson}, M., {et~al.} 2007, \apj, 670, 173

\bibitem[{{Dahari} \& {De Robertis}(1988)}]{dd88}
{Dahari}, O. \& {De Robertis}, M.~M. 1988, \apj, 331, 727

\bibitem[{{D'Ammando} {et~al.}(2008){D'Ammando}, {Bianchi},
  {Jim{\'e}nez-Bail{\'o}n}, \& {Matt}}]{dammando08}
{D'Ammando}, F., {Bianchi}, S., {Jim{\'e}nez-Bail{\'o}n}, E., \& {Matt}, G.
  2008, \aap, 482, 499

\bibitem[{{Della Ceca} {et~al.}(2008){Della Ceca}, {Severgnini}, {Caccianiga},
  {Comastri}, {Gilli}, {Fiore}, {Piconcelli}, {Malaguti}, \& {Vignali}}]{rdc08}
{Della Ceca}, R., {Severgnini}, P., {Caccianiga}, A., {et~al.} 2008, Memorie
  della Societa Astronomica Italiana, 79, 65

\bibitem[{{Dewangan} \& {Griffiths}(2005)}]{dg05}
{Dewangan}, G.~C. \& {Griffiths}, R.~E. 2005, \apjl, 625, L31

\bibitem[{{Diamond-Stanic} {et~al.}(2009){Diamond-Stanic}, {Rieke}, \&
  {Rigby}}]{diamond09}
{Diamond-Stanic}, A.~M., {Rieke}, G.~H., \& {Rigby}, J.~R. 2009, \apj, 698, 623

\bibitem[{{Dickey} \& {Lockman}(1990)}]{dl90}
{Dickey}, J.~M. \& {Lockman}, F.~J. 1990, \araa, 28, 215

\bibitem[{{Durret} \& {Bergeron}(1986)}]{db86}
{Durret}, F. \& {Bergeron}, J. 1986, \aap, 156, 51

\bibitem[{{Durret} \& {Bergeron}(1988)}]{db88}
{Durret}, F. \& {Bergeron}, J. 1988, \aaps, 75, 273

\bibitem[{{Elvis} {et~al.}(2009){Elvis}, {Civano}, {Vignali}, {Puccetti},
  {Fiore}, {Cappelluti}, {Aldcroft}, {Fruscione}, {Zamorani}, {Comastri},
  {Brusa}, {Gilli}, {Miyaji}, {Damiani}, {Koekemoer}, {Finoguenov}, {Brunner},
  {Urry}, {Silverman}, {Mainieri}, {Hasinger}, {Griffiths}, {Carollo}, {Hao},
  {Guzzo}, {Blain}, {Calzetti}, {Carilli}, {Capak}, {Ettori}, {Fabbiano},
  {Impey}, {Lilly}, {Mobasher}, {Rich}, {Salvato}, {Sanders}, {Schinnerer},
  {Scoville}, {Shopbell}, {Taylor}, {Taniguchi}, \& {Volonteri}}]{elvis09}
{Elvis}, M., {Civano}, F., {Vignali}, C., {et~al.} 2009, \apjs, 184, 158

\bibitem[{{Erkens} {et~al.}(1997){Erkens}, {Appenzeller}, \&
  {Wagner}}]{erkens97}
{Erkens}, U., {Appenzeller}, I., \& {Wagner}, S. 1997, \aap, 323, 707

\bibitem[{{Evans} {et~al.}(2006){Evans}, {Worrall}, {Hardcastle}, {Kraft}, \&
  {Birkinshaw}}]{evans06}
{Evans}, D.~A., {Worrall}, D.~M., {Hardcastle}, M.~J., {Kraft}, R.~P., \&
  {Birkinshaw}, M. 2006, \apj, 642, 96

\bibitem[{{Ferland} \& {Osterbrock}(1986)}]{fo86}
{Ferland}, G.~J. \& {Osterbrock}, D.~E. 1986, \apj, 300, 658

\bibitem[{{Fiore} {et~al.}(2008){Fiore}, {Grazian}, {Santini}, {Puccetti},
  {Brusa}, {Feruglio}, {Fontana}, {Giallongo}, {Comastri}, {Gruppioni},
  {Pozzi}, {Zamorani}, \& {Vignali}}]{fiore08}
{Fiore}, F., {Grazian}, A., {Santini}, P., {et~al.} 2008, \apj, 672, 94

\bibitem[{{Fiore} {et~al.}(2009){Fiore}, {Puccetti}, {Brusa}, {Salvato},
  {Zamorani}, {Aldcroft}, {Aussel}, {Brunner}, {Capak}, {Cappelluti}, {Civano},
  {Comastri}, {Elvis}, {Feruglio}, {Finoguenov}, {Fruscione}, {Gilli},
  {Hasinger}, {Koekemoer}, {Kartaltepe}, {Ilbert}, {Impey}, {LeFloc'h},
  {Lilly}, {Mainieri}, {Martinez-Sansigre}, {McCracken}, {Menci}, {Merloni},
  {Miyaji}, {Sanders}, {Sargent}, {Schinnerer}, {Scoville}, {Silverman},
  {Smolcic}, {Steffen}, {Santini}, {Taniguchi}, {Thompson}, {Trump}, {Vignali},
  {Urry}, \& {Yan}}]{fiore09}
{Fiore}, F., {Puccetti}, S., {Brusa}, M., {et~al.} 2009, \apj, 693, 447

\bibitem[{{Fosbury} \& {Sansom}(1983)}]{fs83}
{Fosbury}, R.~A.~E. \& {Sansom}, A.~E. 1983, \mnras, 204, 1231

\bibitem[{{Franceschini} {et~al.}(2000){Franceschini}, {Bassani}, {Cappi},
  {Granato}, {Malaguti}, {Palazzi}, \& {Persic}}]{frances00}
{Franceschini}, A., {Bassani}, L., {Cappi}, M., {et~al.} 2000, \aap, 353, 910

\bibitem[{{Gallo}(2006)}]{gallo06}
{Gallo}, L.~C. 2006, \mnras, 368, 479

\bibitem[{{Gallo} {et~al.}(2005){Gallo}, {Balestra}, {Costantini}, {Boller},
  {Burwitz}, {Ferrero}, \& {Mathur}}]{gallo05}
{Gallo}, L.~C., {Balestra}, I., {Costantini}, E., {et~al.} 2005, \aap, 442, 909

\bibitem[{{Gallo} {et~al.}(2006){Gallo}, {Lehmann}, {Pietsch}, {Boller},
  {Brinkmann}, {Friedrich}, \& {Grupe}}]{gallo06lehmann}
{Gallo}, L.~C., {Lehmann}, I., {Pietsch}, W., {et~al.} 2006, \mnras, 365, 688

\bibitem[{{Gaskell} \& {Benker}(2007)}]{gb07}
{Gaskell}, C.~M. \& {Benker}, A.~J. 2007, ArXiv e-prints

\bibitem[{{Georgantopoulos} {et~al.}(2009){Georgantopoulos}, {Akylas},
  {Georgakakis}, \& {Rowan-Robinson}}]{geo09}
{Georgantopoulos}, I., {Akylas}, A., {Georgakakis}, A., \& {Rowan-Robinson}, M.
  2009, \aap, 507, 747

\bibitem[{{Georgantopoulos} {et~al.}(2007){Georgantopoulos}, {Georgakakis}, \&
  {Akylas}}]{geo07}
{Georgantopoulos}, I., {Georgakakis}, A., \& {Akylas}, A. 2007, \aap, 466, 823

\bibitem[{{Gilli} {et~al.}(2007){Gilli}, {Comastri}, \& {Hasinger}}]{gch07}
{Gilli}, R., {Comastri}, A., \& {Hasinger}, G. 2007, \aap, 463, 79

\bibitem[{{Gilli} {et~al.}(2000){Gilli}, {Maiolino}, {Marconi}, {Risaliti},
  {Dadina}, {Weaver}, \& {Colbert}}]{gilli00}
{Gilli}, R., {Maiolino}, R., {Marconi}, A., {et~al.} 2000, \aap, 355, 485

\bibitem[{{Gliozzi} {et~al.}(2007){Gliozzi}, {Sambruna}, {Eracleous}, \&
  {Yaqoob}}]{gliozzi07}
{Gliozzi}, M., {Sambruna}, R.~M., {Eracleous}, M., \& {Yaqoob}, T. 2007, \apj,
  664, 88

\bibitem[{{Gondoin} {et~al.}(2003){Gondoin}, {Orr}, \& {Lumb}}]{gondoin03}
{Gondoin}, P., {Orr}, A., \& {Lumb}, D. 2003, \aap, 398, 967

\bibitem[{{Grandi} {et~al.}(2006){Grandi}, {Malaguti}, \& {Fiocchi}}]{grandi06}
{Grandi}, P., {Malaguti}, G., \& {Fiocchi}, M. 2006, \apj, 642, 113

\bibitem[{{Greenhill} {et~al.}(2008){Greenhill}, {Tilak}, \&
  {Madejski}}]{greenhill08}
{Greenhill}, L.~J., {Tilak}, A., \& {Madejski}, G. 2008, \apjl, 686, L13

\bibitem[{{Grupe} {et~al.}(2004){Grupe}, {Mathur}, \& {Komossa}}]{grupe04}
{Grupe}, D., {Mathur}, S., \& {Komossa}, S. 2004, \aj, 127, 3161

\bibitem[{{Guainazzi} {et~al.}(2005{\natexlab{a}}){Guainazzi}, {Fabian},
  {Iwasawa}, {Matt}, \& {Fiore}}]{guainazzi05fabian}
{Guainazzi}, M., {Fabian}, A.~C., {Iwasawa}, K., {Matt}, G., \& {Fiore}, F.
  2005{\natexlab{a}}, \mnras, 356, 295

\bibitem[{{Guainazzi} {et~al.}(2005{\natexlab{b}}){Guainazzi}, {Matt}, \&
  {Perola}}]{guainazzi05matt}
{Guainazzi}, M., {Matt}, G., \& {Perola}, G.~C. 2005{\natexlab{b}}, \aap, 444,
  119

\bibitem[{{Guainazzi} {et~al.}(1998){Guainazzi}, {Piro}, {Capalbi}, {Parmar},
  {Yamauchi}, \& {Matsuoka}}]{guainazzi98}
{Guainazzi}, M., {Piro}, L., {Capalbi}, M., {et~al.} 1998, \aap, 339, 337

\bibitem[{{Guainazzi} {et~al.}(2004){Guainazzi}, {Rodriguez-Pascual}, {Fabian},
  {Iwasawa}, \& {Matt}}]{guainazzi04}
{Guainazzi}, M., {Rodriguez-Pascual}, P., {Fabian}, A.~C., {Iwasawa}, K., \&
  {Matt}, G. 2004, \mnras, 355, 297

\bibitem[{{Hardcastle} {et~al.}(2007){Hardcastle}, {Croston}, \&
  {Kraft}}]{hard07}
{Hardcastle}, M.~J., {Croston}, J.~H., \& {Kraft}, R.~P. 2007, \apj, 669, 893

\bibitem[{{Hopkins} {et~al.}(2006){Hopkins}, {Hernquist}, {Cox}, {Di Matteo},
  {Robertson}, \& {Springel}}]{hop06}
{Hopkins}, P.~F., {Hernquist}, L., {Cox}, T.~J., {et~al.} 2006, \apjs, 163, 1

\bibitem[{{Hopkins} {et~al.}(2008){Hopkins}, {Hernquist}, {Cox}, \& {Kere{\v
  s}}}]{hop08}
{Hopkins}, P.~F., {Hernquist}, L., {Cox}, T.~J., \& {Kere{\v s}}, D. 2008,
  \apjs, 175, 356

\bibitem[{{Iwasawa} {et~al.}(2001{\natexlab{a}}){Iwasawa}, {Fabian}, \&
  {Ettori}}]{iwa01iras}
{Iwasawa}, K., {Fabian}, A.~C., \& {Ettori}, S. 2001{\natexlab{a}}, \mnras,
  321, L15

\bibitem[{{Iwasawa} {et~al.}(2001{\natexlab{b}}){Iwasawa}, {Matt}, {Fabian},
  {Bianchi}, {Brandt}, {Guainazzi}, {Murayama}, \& {Taniguchi}}]{iwa01tol}
{Iwasawa}, K., {Matt}, G., {Fabian}, A.~C., {et~al.} 2001{\natexlab{b}},
  \mnras, 326, 119

\bibitem[{{Iwasawa} {et~al.}(2010){Iwasawa}, {Tanaka}, \& {Gallo}}]{iwa10}
{Iwasawa}, K., {Tanaka}, Y., \& {Gallo}, L.~C. 2010, ArXiv e-prints

\bibitem[{{Jim{\'e}nez-Bail{\'o}n} {et~al.}(2008){Jim{\'e}nez-Bail{\'o}n},
  {Guainazzi}, {Matt}, {Bianchi}, {Krongold}, {Piconcelli}, {Santos Lle{\'o}},
  \& {Schartel}}]{jime08}
{Jim{\'e}nez-Bail{\'o}n}, E., {Guainazzi}, M., {Matt}, G., {et~al.} 2008, in
  Revista Mexicana de Astronomia y Astrofisica, vol. 27, Vol.~32, Revista
  Mexicana de Astronomia y Astrofisica Conference Series, 131--133

\bibitem[{{Kauffmann} \& {Haehnelt}(2000)}]{kh00}
{Kauffmann}, G. \& {Haehnelt}, M. 2000, \mnras, 311, 576

\bibitem[{{Kewley} {et~al.}(2004){Kewley}, {Geller}, \& {Jansen}}]{kewley04}
{Kewley}, L.~J., {Geller}, M.~J., \& {Jansen}, R.~A. 2004, \aj, 127, 2002

\bibitem[{{Kim} {et~al.}(2006){Kim}, {Ho}, \& {Im}}]{kim06}
{Kim}, M., {Ho}, L.~C., \& {Im}, M. 2006, \apj, 642, 702

\bibitem[{{Koski}(1978)}]{koski78}
{Koski}, A.~T. 1978, \apj, 223, 56

\bibitem[{{Kunth} {et~al.}(1987){Kunth}, {Sargent}, \& {Bothun}}]{kunth87}
{Kunth}, D., {Sargent}, W.~L.~W., \& {Bothun}, G.~D. 1987, \aj, 93, 29

\bibitem[{{Lamastra} {et~al.}(2009){Lamastra}, {Bianchi}, {Matt}, {Perola},
  {Barcons}, \& {Carrera}}]{lamastra09}
{Lamastra}, A., {Bianchi}, S., {Matt}, G., {et~al.} 2009, \aap, 504, 73

\bibitem[{{Landi} {et~al.}(2005){Landi}, {Malizia}, \& {Bassani}}]{landi05}
{Landi}, R., {Malizia}, A., \& {Bassani}, L. 2005, \aap, 441, 69

\bibitem[{{Landi} {et~al.}(2007){Landi}, {Masetti}, {Morelli}, {Palazzi},
  {Bassani}, {Malizia}, {Bazzano}, {Bird}, {Dean}, {Galaz}, {Minniti}, \&
  {Ubertini}}]{landi07}
{Landi}, R., {Masetti}, N., {Morelli}, L., {et~al.} 2007, \apj, 669, 109

\bibitem[{{Lavalley} {et~al.}(1992){Lavalley}, {Isobe}, \&
  {Feigelson}}]{lavalley92}
{Lavalley}, M., {Isobe}, T., \& {Feigelson}, E. 1992, in Astronomical Society
  of the Pacific Conference Series, Vol.~25, Astronomical Data Analysis
  Software and Systems I, ed. {D.~M.~Worrall, C.~Biemesderfer, \& J.~Barnes},
  245--+

\bibitem[{{Levenson} {et~al.}(2001){Levenson}, {Weaver}, \&
  {Heckman}}]{levenson01}
{Levenson}, N.~A., {Weaver}, K.~A., \& {Heckman}, T.~M. 2001, \apj, 550, 230

\bibitem[{{Lilly} {et~al.}(2007){Lilly}, {Le F{\`e}vre}, {Renzini}, {Zamorani},
  {Scodeggio}, {Contini}, {Carollo}, {Hasinger}, {Kneib}, {Iovino}, {Le Brun},
  {Maier}, {Mainieri}, {Mignoli}, {Silverman}, {Tasca}, {Bolzonella},
  {Bongiorno}, {Bottini}, {Capak}, {Caputi}, {Cimatti}, {Cucciati}, {Daddi},
  {Feldmann}, {Franzetti}, {Garilli}, {Guzzo}, {Ilbert}, {Kampczyk}, {Kovac},
  {Lamareille}, {Leauthaud}, {Borgne}, {McCracken}, {Marinoni}, {Pello},
  {Ricciardelli}, {Scarlata}, {Vergani}, {Sanders}, {Schinnerer}, {Scoville},
  {Taniguchi}, {Arnouts}, {Aussel}, {Bardelli}, {Brusa}, {Cappi}, {Ciliegi},
  {Finoguenov}, {Foucaud}, {Franceschini}, {Halliday}, {Impey}, {Knobel},
  {Koekemoer}, {Kurk}, {Maccagni}, {Maddox}, {Marano}, {Marconi}, {Meneux},
  {Mobasher}, {Moreau}, {Peacock}, {Porciani}, {Pozzetti}, {Scaramella},
  {Schiminovich}, {Shopbell}, {Smail}, {Thompson}, {Tresse}, {Vettolani},
  {Zanichelli}, \& {Zucca}}]{lilly07}
{Lilly}, S.~J., {Le F{\`e}vre}, O., {Renzini}, A., {et~al.} 2007, \apjs, 172,
  70

\bibitem[{{Lilly} {et~al.}(2009){Lilly}, {LeBrun}, {Maier}, {Mainieri},
  {Mignoli}, {Scodeggio}, {Zamorani}, {Carollo}, {Contini}, {Kneib},
  {LeF{\`e}vre}, {Renzini}, {Bardelli}, {Bolzonella}, {Bongiorno}, {Caputi},
  {Coppa}, {Cucciati}, {de la Torre}, {de Ravel}, {Franzetti}, {Garilli},
  {Iovino}, {Kampczyk}, {Kovac}, {Knobel}, {Lamareille}, {LeBorgne}, {Pello},
  {Peng}, {P{\'e}rez-Montero}, {Ricciardelli}, {Silverman}, {Tanaka}, {Tasca},
  {Tresse}, {Vergani}, {Zucca}, {Ilbert}, {Salvato}, {Oesch}, {Abbas},
  {Bottini}, {Capak}, {Cappi}, {Cassata}, {Cimatti}, {Elvis}, {Fumana},
  {Guzzo}, {Hasinger}, {Koekemoer}, {Leauthaud}, {Maccagni}, {Marinoni},
  {McCracken}, {Memeo}, {Meneux}, {Porciani}, {Pozzetti}, {Sanders},
  {Scaramella}, {Scarlata}, {Scoville}, {Shopbell}, \& {Taniguchi}}]{lilly09}
{Lilly}, S.~J., {LeBrun}, V., {Maier}, C., {et~al.} 2009, \apjs, 184, 218

\bibitem[{{Maiolino} {et~al.}(1998){Maiolino}, {Salvati}, {Bassani}, {Dadina},
  {della Ceca}, {Matt}, {Risaliti}, \& {Zamorani}}]{maio98}
{Maiolino}, R., {Salvati}, M., {Bassani}, L., {et~al.} 1998, \aap, 338, 781

\bibitem[{{Malaguti} {et~al.}(1998){Malaguti}, {Palumbo}, {Cappi}, {Comastri},
  {Otani}, {Matsuoka}, {Guainazzi}, {Bassani}, \& {Frontera}}]{pino98}
{Malaguti}, G., {Palumbo}, G.~G.~C., {Cappi}, M., {et~al.} 1998, \aap, 331, 519

\bibitem[{{Malizia} {et~al.}(2009){Malizia}, {Stephen}, {Bassani}, {Bird},
  {Panessa}, \& {Ubertini}}]{malizia09}
{Malizia}, A., {Stephen}, J.~B., {Bassani}, L., {et~al.} 2009, \mnras, 399, 944

\bibitem[{{Malkan}(1986)}]{malkan86}
{Malkan}, M.~A. 1986, \apj, 310, 679

\bibitem[{{Marconi} {et~al.}(2004){Marconi}, {Risaliti}, {Gilli}, {Hunt},
  {Maiolino}, \& {Salvati}}]{marconi04}
{Marconi}, A., {Risaliti}, G., {Gilli}, R., {et~al.} 2004, \mnras, 351, 169

\bibitem[{{Mart{\'{\i}}nez-Sansigre} {et~al.}(2005){Mart{\'{\i}}nez-Sansigre},
  {Rawlings}, {Lacy}, {Fadda}, {Marleau}, {Simpson}, {Willott}, \&
  {Jarvis}}]{alejo05}
{Mart{\'{\i}}nez-Sansigre}, A., {Rawlings}, S., {Lacy}, M., {et~al.} 2005,
  \nat, 436, 666

\bibitem[{{Marulli} {et~al.}(2008){Marulli}, {Bonoli}, {Branchini},
  {Moscardini}, \& {Springel}}]{marulli08}
{Marulli}, F., {Bonoli}, S., {Branchini}, E., {Moscardini}, L., \& {Springel},
  V. 2008, \mnras, 385, 1846

\bibitem[{{Matt} {et~al.}(2004){Matt}, {Bianchi}, {D'Ammando}, \&
  {Martocchia}}]{matt04}
{Matt}, G., {Bianchi}, S., {D'Ammando}, F., \& {Martocchia}, A. 2004, \aap,
  421, 473

\bibitem[{{Moran} {et~al.}(1992){Moran}, {Halpern}, {Bothun}, \&
  {Becker}}]{moran92}
{Moran}, E.~C., {Halpern}, J.~P., {Bothun}, G.~D., \& {Becker}, R.~H. 1992,
  \aj, 104, 990

\bibitem[{{Morris} \& {Ward}(1988)}]{mw88}
{Morris}, S.~L. \& {Ward}, M.~J. 1988, \mnras, 230, 639

\bibitem[{{Murphy} {et~al.}(2007){Murphy}, {Yaqoob}, \& {Terashima}}]{murphy07}
{Murphy}, K.~D., {Yaqoob}, T., \& {Terashima}, Y. 2007, \apj, 666, 96

\bibitem[{{Osterbrock}(1981{\natexlab{a}})}]{o81}
{Osterbrock}, D.~E. 1981{\natexlab{a}}, \apj, 249, 462

\bibitem[{{Osterbrock}(1981{\natexlab{b}})}]{o81iii}
{Osterbrock}, D.~E. 1981{\natexlab{b}}, \apj, 246, 696

\bibitem[{{Osterbrock} {et~al.}(1976){Osterbrock}, {Koski}, \&
  {Phillips}}]{o76}
{Osterbrock}, D.~E., {Koski}, A.~T., \& {Phillips}, M.~M. 1976, \apj, 206, 898

\bibitem[{{Osterbrock} \& {Miller}(1975)}]{om75}
{Osterbrock}, D.~E. \& {Miller}, J.~S. 1975, \apj, 197, 535

\bibitem[{{Osterbrock} \& {Pogge}(1985)}]{op85}
{Osterbrock}, D.~E. \& {Pogge}, R.~W. 1985, \apj, 297, 166

\bibitem[{{Panessa} {et~al.}(2006){Panessa}, {Bassani}, {Cappi}, {Dadina},
  {Barcons}, {Carrera}, {Ho}, \& {Iwasawa}}]{panessa06}
{Panessa}, F., {Bassani}, L., {Cappi}, M., {et~al.} 2006, \aap, 455, 173

\bibitem[{{Piconcelli} {et~al.}(2007{\natexlab{a}}){Piconcelli}, {Bianchi},
  {Guainazzi}, {Fiore}, \& {Chiaberge}}]{pico07ngc}
{Piconcelli}, E., {Bianchi}, S., {Guainazzi}, M., {Fiore}, F., \& {Chiaberge},
  M. 2007{\natexlab{a}}, \aap, 466, 855

\bibitem[{{Piconcelli} {et~al.}(2007{\natexlab{b}}){Piconcelli}, {Fiore},
  {Nicastro}, {Mathur}, {Brusa}, {Comastri}, \& {Puccetti}}]{pico07iras}
{Piconcelli}, E., {Fiore}, F., {Nicastro}, F., {et~al.} 2007{\natexlab{b}},
  \aap, 473, 85

\bibitem[{{Piconcelli} {et~al.}(2005){Piconcelli}, {Jimenez-Bail{\'o}n},
  {Guainazzi}, {Schartel}, {Rodr{\'{\i}}guez-Pascual}, \&
  {Santos-Lle{\'o}}}]{pico05}
{Piconcelli}, E., {Jimenez-Bail{\'o}n}, E., {Guainazzi}, M., {et~al.} 2005,
  \aap, 432, 15

\bibitem[{{Ponti} {et~al.}(2004){Ponti}, {Cappi}, {Dadina}, \&
  {Malaguti}}]{ponti04}
{Ponti}, G., {Cappi}, M., {Dadina}, M., \& {Malaguti}, G. 2004, \aap, 417, 451

\bibitem[{{Ptak} {et~al.}(2006){Ptak}, {Zakamska}, {Strauss}, {Krolik},
  {Heckman}, {Schneider}, \& {Brinkmann}}]{ptak06}
{Ptak}, A., {Zakamska}, N.~L., {Strauss}, M.~A., {et~al.} 2006, \apj, 637, 147

\bibitem[{{Puccetti} {et~al.}(2009){Puccetti}, {Vignali}, {Cappelluti},
  {Fiore}, {Zamorani}, {Aldcroft}, {Elvis}, {Gilli}, {Miyaji}, {Brunner},
  {Brusa}, {Civano}, {Comastri}, {Damiani}, {Fruscione}, {Finoguenov},
  {Koekemoer}, \& {Mainieri}}]{puccetti09}
{Puccetti}, S., {Vignali}, C., {Cappelluti}, N., {et~al.} 2009, ArXiv e-prints

\bibitem[{{Reyes} {et~al.}(2008){Reyes}, {Zakamska}, {Strauss}, {Green},
  {Krolik}, {Shen}, {Richards}, {Anderson}, \& {Schneider}}]{reyes08}
{Reyes}, R., {Zakamska}, N.~L., {Strauss}, M.~A., {et~al.} 2008, \aj, 136, 2373

\bibitem[{{Rigby} {et~al.}(2009){Rigby}, {Diamond-Stanic}, \&
  {Aniano}}]{rigby09}
{Rigby}, J.~R., {Diamond-Stanic}, A.~M., \& {Aniano}, G. 2009, \apj, 700, 1878

\bibitem[{{Risaliti} {et~al.}(2000){Risaliti}, {Gilli}, {Maiolino}, \&
  {Salvati}}]{guido00}
{Risaliti}, G., {Gilli}, R., {Maiolino}, R., \& {Salvati}, M. 2000, \aap, 357,
  13

\bibitem[{{Risaliti} {et~al.}(1999){Risaliti}, {Maiolino}, \&
  {Salvati}}]{guido99}
{Risaliti}, G., {Maiolino}, R., \& {Salvati}, M. 1999, \apj, 522, 157

\bibitem[{{Sambruna} {et~al.}(2007){Sambruna}, {Reeves}, \&
  {Braito}}]{sambruna07}
{Sambruna}, R.~M., {Reeves}, J.~N., \& {Braito}, V. 2007, \apj, 665, 1030

\bibitem[{{Schmidt} {et~al.}(1998){Schmidt}, {Hasinger}, {Gunn}, {Schneider},
  {Burg}, {Giacconi}, {Lehmann}, {MacKenty}, {Trumper}, \&
  {Zamorani}}]{schmidt98}
{Schmidt}, M., {Hasinger}, G., {Gunn}, J., {et~al.} 1998, \aap, 329, 495

\bibitem[{{Schmitt}(1998)}]{schmitt98}
{Schmitt}, H.~R. 1998, \apj, 506, 647

\bibitem[{{Scoville} {et~al.}(2007){Scoville}, {Aussel}, {Brusa}, {Capak},
  {Carollo}, {Elvis}, {Giavalisco}, {Guzzo}, {Hasinger}, {Impey}, {Kneib},
  {LeFevre}, {Lilly}, {Mobasher}, {Renzini}, {Rich}, {Sanders}, {Schinnerer},
  {Schminovich}, {Shopbell}, {Taniguchi}, \& {Tyson}}]{scoville07}
{Scoville}, N., {Aussel}, H., {Brusa}, M., {et~al.} 2007, \apjs, 172, 1

\bibitem[{{Shankar} {et~al.}(2004){Shankar}, {Salucci}, {Granato}, {De Zotti},
  \& {Danese}}]{shankar04}
{Shankar}, F., {Salucci}, P., {Granato}, G.~L., {De Zotti}, G., \& {Danese}, L.
  2004, \mnras, 354, 1020

\bibitem[{{Shields} \& {Filippenko}(1990)}]{sf90}
{Shields}, J.~C. \& {Filippenko}, A.~V. 1990, \aj, 100, 1034

\bibitem[{{Shinozaki} {et~al.}(2006){Shinozaki}, {Miyaji}, {Ishisaki}, {Ueda},
  \& {Ogasaka}}]{shinozaki06}
{Shinozaki}, K., {Miyaji}, T., {Ishisaki}, Y., {Ueda}, Y., \& {Ogasaka}, Y.
  2006, \aj, 131, 2843

\bibitem[{{Shu} {et~al.}(2007){Shu}, {Wang}, {Jiang}, {Fan}, \& {Wang}}]{shu07}
{Shu}, X.~W., {Wang}, J.~X., {Jiang}, P., {Fan}, L.~L., \& {Wang}, T.~G. 2007,
  \apj, 657, 167

\bibitem[{{Shuder} \& {Osterbrock}(1981)}]{so81}
{Shuder}, J.~M. \& {Osterbrock}, D.~E. 1981, \apj, 250, 55

\bibitem[{{Silverman} {et~al.}(2009){Silverman}, {Lamareille}, {Maier},
  {Lilly}, {Mainieri}, {Brusa}, {Cappelluti}, {Hasinger}, {Zamorani},
  {Scodeggio}, {Bolzonella}, {Contini}, {Carollo}, {Jahnke}, {Kneib}, {Le
  F{\`e}vre}, {Merloni}, {Bardelli}, {Bongiorno}, {Brunner}, {Caputi},
  {Civano}, {Comastri}, {Coppa}, {Cucciati}, {de la Torre}, {de Ravel},
  {Elvis}, {Finoguenov}, {Fiore}, {Franzetti}, {Garilli}, {Gilli}, {Iovino},
  {Kampczyk}, {Knobel}, {Kova{\v c}}, {Le Borgne}, {Le Brun}, {Mignoli},
  {Pello}, {Peng}, {Montero}, {Ricciardelli}, {Tanaka}, {Tasca}, {Tresse},
  {Vergani}, {Vignali}, {Zucca}, {Bottini}, {Cappi}, {Cassata}, {Fumana},
  {Griffiths}, {Kartaltepe}, {Koekemoer}, {Marinoni}, {McCracken}, {Memeo},
  {Meneux}, {Oesch}, {Porciani}, \& {Salvato}}]{silverman09}
{Silverman}, J.~D., {Lamareille}, F., {Maier}, C., {et~al.} 2009, \apj, 696,
  396

\bibitem[{{Spinelli} {et~al.}(2006){Spinelli}, {Storchi-Bergmann}, {Brandt}, \&
  {Calzetti}}]{spinelli06}
{Spinelli}, P.~F., {Storchi-Bergmann}, T., {Brandt}, C.~H., \& {Calzetti}, D.
  2006, \apjs, 166, 498

\bibitem[{{Storchi-Bergmann} {et~al.}(1995){Storchi-Bergmann}, {Kinney}, \&
  {Challis}}]{storchi95}
{Storchi-Bergmann}, T., {Kinney}, A.~L., \& {Challis}, P. 1995, \apjs, 98, 103

\bibitem[{{Thornton} {et~al.}(2008){Thornton}, {Barth}, {Ho}, {Rutledge}, \&
  {Greene}}]{thornton08}
{Thornton}, C.~E., {Barth}, A.~J., {Ho}, L.~C., {Rutledge}, R.~E., \& {Greene},
  J.~E. 2008, \apj, 686, 892

\bibitem[{{Tozzi} {et~al.}(2006){Tozzi}, {Gilli}, {Mainieri}, {Norman},
  {Risaliti}, {Rosati}, {Bergeron}, {Borgani}, {Giacconi}, {Hasinger},
  {Nonino}, {Streblyanska}, {Szokoly}, {Wang}, \& {Zheng}}]{tozzi06}
{Tozzi}, P., {Gilli}, R., {Mainieri}, V., {et~al.} 2006, \aap, 451, 457

\bibitem[{{Treister} {et~al.}(2009{\natexlab{a}}){Treister}, {Cardamone},
  {Schawinski}, {Urry}, {Gawiser}, {Virani}, {Lira}, {Kartaltepe}, {Damen},
  {Taylor}, {Le Floc'h}, {Justham}, \& {Koekemoer}}]{trei09_ct}
{Treister}, E., {Cardamone}, C.~N., {Schawinski}, K., {et~al.}
  2009{\natexlab{a}}, \apj, 706, 535

\bibitem[{{Treister} {et~al.}(2009{\natexlab{b}}){Treister}, {Urry}, \&
  {Virani}}]{tuv09}
{Treister}, E., {Urry}, C.~M., \& {Virani}, S. 2009{\natexlab{b}}, \apj, 696,
  110

\bibitem[{{Tueller} {et~al.}(2008){Tueller}, {Mushotzky}, {Barthelmy},
  {Cannizzo}, {Gehrels}, {Markwardt}, {Skinner}, \& {Winter}}]{tueller08}
{Tueller}, J., {Mushotzky}, R.~F., {Barthelmy}, S., {et~al.} 2008, \apj, 681,
  113

\bibitem[{{Urrutia} {et~al.}(2005){Urrutia}, {Lacy}, {Gregg}, \&
  {Becker}}]{urrutia05}
{Urrutia}, T., {Lacy}, M., {Gregg}, M.~D., \& {Becker}, R.~H. 2005, \apj, 627,
  75

\bibitem[{{Vanden Berk} {et~al.}(2001){Vanden Berk}, {Richards}, {Bauer},
  {Strauss}, {Schneider}, {Heckman}, {York}, {Hall}, {Fan}, {Knapp},
  {Anderson}, {Annis}, {Bahcall}, {Bernardi}, {Briggs}, {Brinkmann}, {Brunner},
  {Burles}, {Carey}, {Castander}, {Connolly}, {Crocker}, {Csabai}, {Doi},
  {Finkbeiner}, {Friedman}, {Frieman}, {Fukugita}, {Gunn}, {Hennessy},
  {Ivezi{\'c}}, {Kent}, {Kunszt}, {Lamb}, {Leger}, {Long}, {Loveday}, {Lupton},
  {Meiksin}, {Merelli}, {Munn}, {Newberg}, {Newcomb}, {Nichol}, {Owen}, {Pier},
  {Pope}, {Rockosi}, {Schlegel}, {Siegmund}, {Smee}, {Snir}, {Stoughton},
  {Stubbs}, {SubbaRao}, {Szalay}, {Szokoly}, {Tremonti}, {Uomoto}, {Waddell},
  {Yanny}, \& {Zheng}}]{vanden01}
{Vanden Berk}, D.~E., {Richards}, G.~T., {Bauer}, A., {et~al.} 2001, \aj, 122,
  549

\bibitem[{{Vasudevan} \& {Fabian}(2007)}]{vf07}
{Vasudevan}, R.~V. \& {Fabian}, A.~C. 2007, \mnras, 381, 1235

\bibitem[{{Vaughan} {et~al.}(2004){Vaughan}, {Fabian}, {Ballantyne}, {De Rosa},
  {Piro}, \& {Matt}}]{vaughan04}
{Vaughan}, S., {Fabian}, A.~C., {Ballantyne}, D.~R., {et~al.} 2004, \mnras,
  351, 193

\bibitem[{{Vaughan} {et~al.}(1999){Vaughan}, {Reeves}, {Warwick}, \&
  {Edelson}}]{vaughan99}
{Vaughan}, S., {Reeves}, J., {Warwick}, R., \& {Edelson}, R. 1999, \mnras, 309,
  113

\bibitem[{{Vignali} {et~al.}(2006){Vignali}, {Alexander}, \& {Comastri}}]{v06}
{Vignali}, C., {Alexander}, D.~M., \& {Comastri}, A. 2006, \mnras, 373, 321

\bibitem[{{Vignali} {et~al.}(2010){Vignali}, {Alexander}, {Gilli}, \&
  {Pozzi}}]{v10}
{Vignali}, C., {Alexander}, D.~M., {Gilli}, R., \& {Pozzi}, F. 2010, \mnras,
  404, 48 (V10)

\bibitem[{{Vignali} {et~al.}(2004){Vignali}, {Brandt}, {Boller}, {Fabian}, \&
  {Vaughan}}]{vignali04}
{Vignali}, C., {Brandt}, W.~N., {Boller}, T., {Fabian}, A.~C., \& {Vaughan}, S.
  2004, \mnras, 347, 854

\bibitem[{{Vignali} \& {Comastri}(2002)}]{vignali02}
{Vignali}, C. \& {Comastri}, A. 2002, \aap, 381, 834

\bibitem[{{Ward} {et~al.}(1980){Ward}, {Penston}, {Blades}, \&
  {Turtle}}]{ward80}
{Ward}, M., {Penston}, M.~V., {Blades}, J.~C., \& {Turtle}, A.~J. 1980, \mnras,
  193, 563

\bibitem[{{Wilkes} {et~al.}(2005){Wilkes}, {Pounds}, {Schmidt}, {Smith},
  {Cutri}, {Ghosh}, {Nelson}, \& {Hines}}]{wilkes05}
{Wilkes}, B.~J., {Pounds}, K.~A., {Schmidt}, G.~D., {et~al.} 2005, \apj, 634,
  183

\bibitem[{{Wilkes} {et~al.}(2002){Wilkes}, {Schmidt}, {Cutri}, {Ghosh},
  {Hines}, {Nelson}, \& {Smith}}]{wilkes02}
{Wilkes}, B.~J., {Schmidt}, G.~D., {Cutri}, R.~M., {et~al.} 2002, \apjl, 564,
  L65

\bibitem[{{Winkler}(1992)}]{winkler92}
{Winkler}, H. 1992, \mnras, 257, 677

\bibitem[{{Young} {et~al.}(2009){Young}, {Elvis}, \& {Risaliti}}]{y09}
{Young}, M., {Elvis}, M., \& {Risaliti}, G. 2009, \apjs, 183, 17 (Y09)

\bibitem[{{Zakamska} {et~al.}(2003){Zakamska}, {Strauss}, {Krolik}, {Collinge},
  {Hall}, {Hao}, {Heckman}, {Ivezi{\'c}}, {Richards}, {Schlegel}, {Schneider},
  {Strateva}, {Vanden Berk}, {Anderson}, \& {Brinkmann}}]{zak03}
{Zakamska}, N.~L., {Strauss}, M.~A., {Krolik}, J.~H., {et~al.} 2003, \aj, 126,
  2125

\end{thebibliography}

\end{document}